\documentclass[aps,pra,preprint,longbibliography]{revtex4-2}
\usepackage{graphicx}
\usepackage{float}
\usepackage{color}
\usepackage{amsmath,amssymb}
\usepackage[colorlinks=true, citecolor=red, linkcolor=blue,urlcolor=blue ]{hyperref}
\usepackage{appendix}
\usepackage{mathtools}
\usepackage{url}

\newcommand{\e}{\mathbf{e}_1}
\newcommand{\ee}{\mathbf{e}_2}
\newcommand{\eee}{\mathbf{e}_3}

\newcommand{\ex}{\mathbf{e}_x}
\newcommand{\ey}{\mathbf{e}_y}
\newcommand{\ez}{\mathbf{e}_z}

\newcommand{\xb}{\mathbf{x}}
\newcommand{\pb}{\mathbf{p}}
\newcommand{\rb}{\mathbf{r}}
\newcommand{\tb}{\hat{\mathbf{t}}}
\newcommand{\Fb}{\mathbf{F}}
\newcommand{\Tb}{\mathbf{T}}
\newcommand{\Ub}{\mathbf{U}}
\newcommand{\Omb}{\mathbf{\Omega}}
\newcommand{\db}{\mathbf{d}}
\newcommand{\dhat}{\hat{d}}
\newcommand{\dhatb}{\hat{\mathbf{d}}}
\newcommand{\Ib}{\mathbf{I}}
\newcommand{\Sb}{\mathbf{S}}
\newcommand{\Cb}{\mathbf{C}}
\newcommand{\fb}{\mathbf{f}}
\newcommand{\Sigb}{\mathbf{\Sigma}}
\newcommand{\ub}{\mathbf{u}}
\newcommand{\Rb}{\mathbf{R}}
\newcommand{\Qb}{\mathbf{Q}}
\newcommand{\Pb}{\mathbf{P}}
\newcommand{\mb}{\mathbf{m}}
\newcommand{\Gb}{\mathbf{G}}

\newcommand{\sint}[1]{\int_{-1}^{+1} #1 \mathrm{d}s}
\newcommand{\sdint}[1]{\int_{-1}^{+1} #1 \mathrm{d}s'}
\newcommand{\sddint}[1]{\int_{-1}^{+1} #1 \mathrm{d}s''}
\newcommand{\piNs}{(\pi N s)}

\newcommand{\mean}[1]{\langle #1 \rangle}
\newcommand{\var}[1]{\mathrm{var}(#1)}

\definecolor{pointcolour}{RGB}{255,128,0}
\definecolor{mygreen}{RGB}{153,255,51}
\definecolor{mydarkgreen}{RGB}{131,209,53}
\definecolor{mycyan}{RGB}{0,255,255}
\definecolor{orange}{RGB}{255,69,0}

\usepackage{calculator}

\newlength{\cutwidth}
\newlength{\cutheight}

\newcommand{\landscapetrim}[2]{
	\LENGTHSUBTRACT{29.7cm}{#1}{\cutwidth}
	\LENGTHSUBTRACT{21.0cm}{#2}{\cutheight}
}
\newcommand{\portraittrim}[2]{
	\LENGTHSUBTRACT{21.0cm}{#1}{\cutwidth}
	\LENGTHSUBTRACT{29.7cm}{#2}{\cutheight}
}

\begin{document}
	
\title{Asymptotic theory of hydrodynamic interactions between slender filaments}

\author{Maria T\u{a}tulea-Codrean}
\author{Eric Lauga}
\email[]{e.lauga@damtp.cam.ac.uk}
\affiliation{Department of Applied Mathematics and Theoretical Physics, University of Cambridge, Cambridge CB3 0WA, United Kingdom}

\date{\today}

\begin{abstract}
Hydrodynamic interactions (HIs) are important in biophysics research because they influence both the collective and the individual behaviour of microorganisms and self-propelled particles. For instance, HIs at the micro-swimmer level determine the attraction or repulsion between individuals, and hence their collective behaviour. Meanwhile, HIs between swimming appendages (e.g. cilia and flagella) influence the emergence of swimming gaits, synchronised bundles and metachronal waves. In this study, we address the issue of HIs between slender filaments separated by a distance larger than their contour length ($d>L$) by means of asymptotic calculations and numerical simulations. We first derive analytical expressions for the extended resistance matrix of two arbitrarily-shaped rigid filaments as a series expansion in inverse powers of $d/L>1$. The coefficients in our asymptotic series expansion are then evaluated using two well-established methods for slender filaments, resistive-force theory (RFT) and slender-body theory (SBT), and our asymptotic theory is verified using numerical simulations based on SBT for the case of two parallel helices. The theory captures the qualitative features of the interactions in the regime $d/L>1$, which opens the path to a deeper physical understanding of hydrodynamically governed phenomena such as inter-filament synchronisation and multiflagellar propulsion. To demonstrate the usefulness of our results, we next apply our theory to the case of two helices rotating side-by-side, where we quantify the dependence of all forces and torques on the distance and phase difference between them. Using our understanding of pairwise HIs, we then provide physical intuition for the case of a circular array of rotating helices. Our theoretical results will be useful for the study of HIs between bacterial flagella, nodal cilia, and slender microswimmers.
\end{abstract}

\maketitle

\section{Introduction}
The microscopic world is filled with examples of rigid structures that interact with each other as they move through fluids. In the biological context, these can range from very dense systems such as bacterial swarms \cite{Darnton2010}, where steric interactions are important, to regularly-spaced arrays of cilia, which can be coupled both hydrodynamically (through the fluid) \cite{Brumley2014} and elastically (through the cell membrane) \cite{Wan2016,Kanso2021}, down to dilute suspensions of planktonic bacteria and algae \cite{Ishikawa2009}, where only hydrodynamic interactions prevail. Outside biology, hydrodynamic interactions are important in the dynamics of sedimentation and the {\color{black} rheology of suspensions \cite{Shaqfeh1990,Mackaplow1996,Guazelli2011,Shelley2019}}, as well as the collective behaviour of synthetic active particles \cite{Ramaswamy2010,Marchetti2013}. For artificial devices such as diffusio- or electrophoretic swimmers, one must also consider long-range chemical interactions in addition to the hydrodynamics \cite{Sharifi2016,Varma2018,Varma2019,Saha2019}.

Hydrodynamic interactions (HIs) represent a particular interest for research because, due to their long-range nature, they can give rise to collective behaviour in systems with a large number of active, self-propelled particles \cite{Vicsek2012,Elgeti2015}. A popular approach for studying active matter is to coarse-grain the system and postulate phenomenological equations based on symmetries, but it remains important to capture the microscopic origin of interactions between the particles. Therefore, the study of HIs between a small number of suspended bodies is the necessary link between understanding the dynamics of a single body in an unbounded fluid and that of a large collection thereof.

On a microscopic length scale, the physics of the fluid is dominated by viscous dissipation, and inertia is negligible most of the time. Therefore, the interaction of micro-swimmers is usually a low Reynolds number problem, governed by the Stokes equations. Naturally, HIs are important in biology across all Reynolds numbers. For instance, they influence predator-prey interactions and sexual reproduction in small marine organisms such as copepods, which operate at low to intermediate Reynolds number \cite{Arezoo2016}. HIs are also very important in schools of fish (usually high Reynolds number), where they give rise to stable swimming formations and affect endurance and propulsive efficiency \cite{Weihs1973,Dai2018,Pan2020}. At intermediate and high Reynolds number, however, the problem of HIs is usually approached with experimental and computational tools. In contrast, in the low Reynolds number limit, the linearity of the Stokes equations allows for exact analytical solutions if the geometry is simple enough, e.g.~the interaction between two rigid spheres. 

For rigid spheres at low Reynolds number, exact analytical solutions were found for the flow field around two spheres of arbitrary size but specified orientation \cite{Jeffery1915,StimsonJeffery1926,Goddard2020}, as well as around two identical spheres with arbitrary orientation \cite{Goldman1966,Wakiya1967}. These exact solutions are possible either by exploiting a cylindrical symmetry in the problem \cite{Jeffery1915,StimsonJeffery1926}, or by using a bispherical coordinate system \cite{Goddard2020,Goldman1966,Wakiya1967}. These classical analytical results were later confirmed by computational studies \cite{Dabros1985,Kim1985,YoonKim1987}. In addition to the exact solutions, there are also approximate analytical solutions for the interaction of two spheres sufficiently far apart \cite{Felderhof1977,Cichocki1988}. These solutions are expressed as series expansions in inverse powers of the distance between the spheres, and have the advantage of circumventing bispherical coordinates. For more than two spheres, the interactions become more complicated, but researchers have studied this problem experimentally \cite{Jayaweera1964} and numerically \cite{Cichocki1994}, and have also made analytical progress in the form of a far-field theory \cite{Hocking1964}. 

For shapes more complex than a sphere, it is often necessary to approach the modelling problem with computational tools. In the biological context, full boundary-element method (BEM) simulations have been carried out to study the HIs between micromachines with spiral tails \cite{Nasseri1997}, uniflagellar bacteria swimming side by side \cite{Ishikawa2007}, and spherical colonies of algae swimming near boundaries \cite{Ishikawa2020}. Other computational studies have considered the interactions between more abstract types of swimmers such as dumbbell-type \cite{Gyrya2010} or squirmer-type pushers and pullers \cite{Goetze2010,Molina2013}. One important question to consider when talking about HIs between microorganisms is whether there is any net attraction or repulsion between the swimmers, and if they settle into stable swimming patterns. These questions are also motivated by experimental observations of swimming bacteria and volvocine algae \cite{Liao2007,Drescher2009}.  

In this study we focus on HIs between slender filaments at low Reynolds number, in order to tackle the interactions between swimming appendages such as cilia and flagella, rather than entire microorganisms. If HIs between microorganisms are important for the stability of swimming patterns in groups of swimmers, then the HIs between swimming appendages are essential to single-cell behaviour. This includes questions such as the speed and state of flagellar synchronisation \cite{Kim2004b,Reigh2012,Reigh2013,Brumley2014,Chakrabarti2019,Man2020}, the emergence of swimming gaits \cite{Wan2016} and metachronal waves \cite{Joanny2007mcw,Elgeti2013}, and the propulsive capacity of an organism with multiple appendages \cite{Elgeti2013,Nguyen2018}. Much previous work in this area is computational \cite{Kim2004b,Reigh2012,Reigh2013,Chakrabarti2019,Man2020,Nguyen2018,Elgeti2013}, but there has also been some analytical work on the HIs between nearby slender filaments \cite{Man2016}, as well as experimental work on HIs between the beating cilia of live algae \cite{Brumley2014}, and between rotating helices in macro-scale models of bacterial flagella \cite{Kim2003,Kim2004a}. 

After spheres, the next shapes that can be tackled analytically are slender filaments. This is because we now have well-developed theories for modelling the flows generated by moving filaments using a distribution of force singularities along the centreline of the slender body. One very successful analytical method is resistive-force theory (RFT) \cite{Hancock1953,Gray1955,Lighthill1996_helical}, which describes the anisotropic drag on a slender filament by a linear and local relationship between the force and velocity distributions along the centreline. Since it neglects non-local interactions along the filament, RFT is quantitatively accurate only for exponentially slender filaments, but it usually reproduces the qualitative features of the flow and it is analytically tractable, which leads to a deeper physical understanding. For more accurate quantitative results, one can use slender-body theory (SBT), which takes into account both local and non-local hydrodynamic effects \cite{Cox1970,Lighthill1976,Johnson1980}. While RFT is logarithmically correct, the errors in SBT are algebraically small.

In this investigation we apply the theoretical techniques commonly used for single filaments (RFT and SBT) to describe the HIs between two slender filaments {\color{black} separated by a distance, $d$, greater than the contour length of the filaments, $L$}. In a similar way to previous studies on spheres \cite{Felderhof1977,Cichocki1988}, we express the force distribution along each filament as a series expansion in inverse powers of {\color{black}$d/L>1$}. This uses principles from the method of reflections, where some contributions in the expansion correspond to hydrodynamic effects that have reflected back and forth between the filaments a number of times. {\color{black} The method of scattering has previously been employed in the theoretical study of suspensions of rods \cite{Shaqfeh1990,Mackaplow1996}, but these studies focus on the bulk rheology of a suspension of passive fibres, whereas our current purpose is to derive analytical expressions for the specific HIs between two active slender filaments. Furthermore, the present study can handle helical and other shapes of filaments, while the aforementioned work was limited to straight rods.}

Our final analytical results pertain specifically to rigid filaments, whose motion can be encapsulated in one mathematical object -- the resistance matrix. For multiple filaments, it is the extended resistance matrix (see also Ref.~\cite{Cichocki1988}) that relates the full dynamics (forces and torques on all the filaments) to the full kinematics (the linear and angular velocities of all the filaments). {\color{black} We expand our solution for the extended resistance matrix up to and including second-order corrections in $L/d<1$. This is motivated by our subsequent application to rotating helical pumps, where the net attraction or repulsion between the helices is only noticeable at second order. It is also at second order that the power of slender-filament methods like RFT and SBT comes into play. The first-order contribution of HIs is the same for slender filaments as it is for spheres or any rigid object that exerts a net force on the fluid. At second order, however, we have contributions not only from the flow that is reflected between the objects (which is the same for spheres), but also from expanding the shape of the filament centreline about its centre.} 

The paper is structured around three central parts -- the derivation, validation, and application of the theory for HIs between slender filaments at low Reynolds number. In Section \ref{sec:model} we derive analytical expressions for the extended resistance matrix of two arbitrarily-shaped rigid slender filaments, written as a series expansion up to second-order corrections in inverse distance. {\color{black} We then evaluate the coefficients in this series using both RFT and SBT, and in Section \ref{sec:validation} we validate the asymptotic theory against numerical simulations based on SBT}. Finally, in Section \ref{sec:application}, we apply both theory and simulations to the case of two helical pumps rotating side by side in an infinite fluid. We perform a thorough investigation of the forces and torques exerted by the helical pumps, and derive analytical expressions that capture the qualitative effects of HIs with varying distance and phase difference between the helices. Based on our understanding of pairwise HIs between helical pumps, we then provide a perspective on the HIs within a circular array of helical pumps, and we conclude this study in Section \ref{sec:discussion} by discussing our results in a wider context. 

\section{Asymptotic model for hydrodynamic interactions}
\label{sec:model}

In this section, we consider the HIs between two rigid slender filaments {\color{black} separated by a distance, $d$, greater than their contour length, $L$}. We quantify the dynamics of the interacting filaments through an extended resistance matrix, for which we derive a series expansion solution up to second-order corrections in {\color{black} $L/d<1$}.	
	
\subsection{Geometrical setup}

\begin{figure} 
	\landscapetrim{17cm}{9cm}
	\includegraphics[trim={{.5\cutwidth} {.5\cutheight} {.5\cutwidth} {.5\cutheight}},clip,width=17cm]{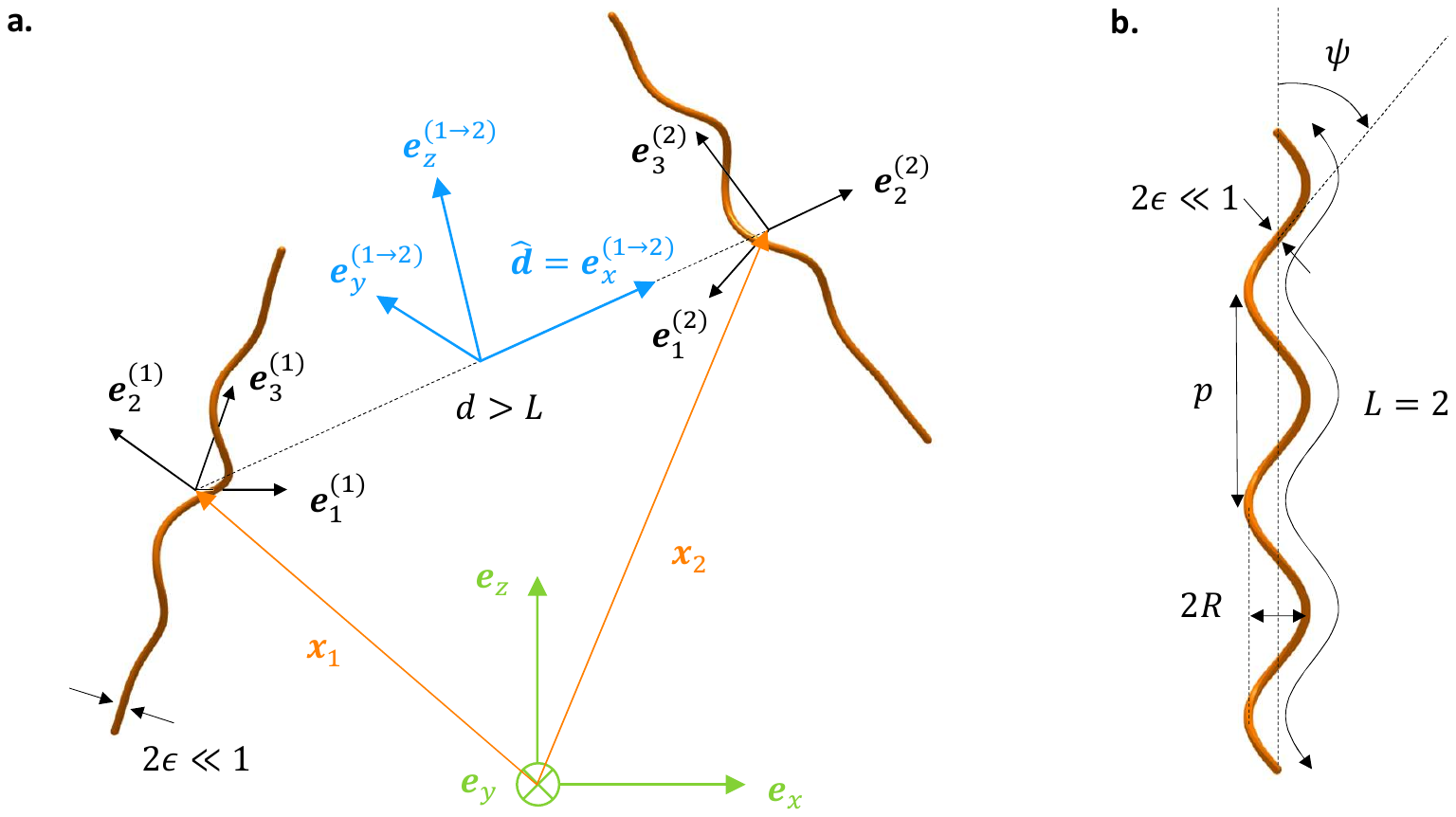}
	\caption{Geometrical setup of the problem. (a) Two rigid filaments of dimensionless contour length $L=2$ interact with each other hydrodynamically as they move through a viscous fluid. Our asymptotic theory is valid {\color{black} for sufficiently large inter-filament separation, $d > L$, and in the limit of small filament thickness, $\epsilon \ll 1$}. We identify three useful coordinate systems: the laboratory frame (green), the interaction frame for a pair of filaments (blue), and the body frame for an individual filament (black).  (b) Parameters describing the geometry of a helical filament, which we will use for the validation and application of our asymptotic theory.}
	\label{fig:setup}
\end{figure}

We begin by sketching the setup of our hydrodynamic problem and introducing the mathematical notation. In Fig.~\ref{fig:setup} (a) we illustrate the different coordinate systems used in this paper. First, there is the laboratory frame $\{\ex,\ey,\ez\}$ in usual Cartesian coordinates. 
Then there is a body frame $\{\e^{(k)},\ee^{(k)},\eee^{(k)}\}$ for each filament, labelled by $k$. Relative to the laboratory frame, we define the body frame vectors for a filament with orientation $\mathbf{p} = (\phi,\theta,\chi)$  to be 
\begin{eqnarray}
	\e &=& \cos\chi \left[ \cos\theta\left(\cos\phi \ex + \sin\phi \ey\right) -\sin\theta \ez \right] + \sin\chi \left[ -\sin\phi \ex + \cos\phi \ey \right], \label{eq:bodyframe-A} \\
	\ee &=& -\sin\chi \left[  \cos\theta\left(\cos\phi \ex + \sin\phi \ey\right) -\sin\theta \ez \right]  + \cos\chi \left[ -\sin\phi \ex + \cos\phi \ey \right] , \\
	\eee &=& \sin\theta\left(\cos\phi \ex + \sin\phi \ey\right) + \cos\theta \ez. \label{eq:bodyframe-Z}
\end{eqnarray}
Working outwards through the transformations applied to the laboratory frame vectors $\{\ex,\ey,\ez\}$, we see that the body frame $\{\e,\ee,\eee\}$ is obtained by a rotation through angle $\phi$ around the vertical, $\ez$, then a tilting by angle $\theta$ away from the vertical (i.e.~a rotation through angle $\theta$ around $-\sin\phi \ex + \cos\phi \ey$), and finally a rotation by angle $\chi$ around the axis $\eee$. Relative to the body frame, we write the position of the centreline and the unit tangent along an arbitrarily-shaped filament $k$ as 
\begin{eqnarray}
	\rb_k(s) &=& x^{(k)}_1(s) \e^{(k)} + x^{(k)}_2(s) \ee^{(k)} + x^{(k)}_3(s)\eee^{(k)}, \\
	\tb_k(s) &=& \frac{\partial x^{(k)}_1}{\partial s} \e^{(k)} + \frac{\partial x^{(k)}_2}{\partial s} \ee^{(k)} + \frac{\partial x^{(k)}_3}{\partial s}\eee^{(k)},
	\label{eq:filament-arbitrary}
\end{eqnarray}
where $s$ is the arc length along the filament. 

Finally there is a frame of interaction, $\{\ex^{(j \to k)},\ey^{(j \to k)},\ez^{(j \to k)}\}$, defined for every pair of filaments $j$ and $k$ such that the unit vector $\ex^{(j \to k)}$ points from the origin of the body frame of filament $j$ to that of filament $k$. This frame is useful for discussing interactions between three filaments or more, where there could be multiple pairwise interaction frames distinct from the absolute laboratory frame. However, in our discussion of interactions between two filaments, we may assume without loss of generality that the interaction frame is identical to the laboratory frame. 

Our asymptotic theory is written in terms of dimensionless quantities. We measure lengths in units of $\tilde{L}/2$ and viscosity in units of $\tilde{\mu}$, where $\tilde{L}$ is the integrated length of the filament and $\tilde{\mu}$ is the viscosity of the medium. This is equivalent to taking $L=2$ and $\mu=1$ in dimensionless terms. In these units, the cross-sectional radius of the filament, $\epsilon$, and the centre-to-centre distance between the filaments, $d$, must {satisfy \color{black}$\epsilon \ll 1 < d$} in order for our theory to hold. We also note that, in our notation, the arc length falls in the interval  $s\in (-1,+1)$, giving a total dimensionless length $L=2$ for the filament, and placing the midpoint of the filament at $s=0$. 

In Fig.~\ref{fig:setup} (b), we illustrate a filament geometry of particular interest - a helical filament with helical radius, $R$, and helical pitch, $p$. It is convenient to introduce the helix angle $\psi = \tan^{-1}(2\pi R/p)$ and the number of helical turns $N=L/\sqrt{(2\pi R)^2+p^2}$. In terms of these, the dimensionless radius of the helix is $R = \sin(\psi)/(\pi N)$ and the pitch is $p = 2\cos(\psi)/N$. We write the centreline of helix $k$ relative to the midpoint of the helical axis, $\xb_k$, as
\begin{equation}
\rb_k(s) = R \cos(\pi N s) \e^{(k)} + \sigma R \sin(\pi N s) \ee^{(k)} + s\cos\psi\eee^{(k)},
\label{eq:centreline}
\end{equation}
where $s\in (-1,+1)$ is the arc length along the helix and $\sigma=\pm 1$ is the chirality (negative for left-handed helices, positive for right-handed). We can also write the unit tangent vector along the centreline as
\begin{equation}
\tb_k(s) = -\sin\psi \sin(\pi N s) \e^{(k)} + \sigma \sin\psi \cos(\pi N s) \ee^{(k)} + \cos\psi\eee^{(k)}.
\label{eq:tangent}
\end{equation}
The calculations in Section \ref{sec:model} are valid for filaments of arbitrary shape, but in later sections we focus on helical filaments for the purposes of validating and applying our analytical results. 

\subsection{Hydrodynamic setup}

The goal is to find a relationship between the kinematics and the dynamics of the two filaments. This is generally quantified by an extended resistance matrix, which relates the forces and torques exerted by the filaments to their linear and angular velocities, such that
\begin{equation}
\begin{pmatrix}
\Fb_1 \\ \Tb_1 \\ \Fb_2 \\ \Tb_2 
\end{pmatrix} = \begin{pmatrix}
\Sb(\xb_1,\xb_2,\pb_1,\pb_2) & \Cb(\xb_1,\xb_2,\pb_1,\pb_2) \\
\Cb(\xb_2,\xb_1,\pb_2,\pb_1) & \Sb(\xb_2,\xb_1,\pb_2,\pb_1)
\end{pmatrix}
\begin{pmatrix}
\Ub_1 \\ \Omb_1 \\ \Ub_2 \\ \Omb_2 
\end{pmatrix},
\label{eq:defn-resistance-matrix}
\end{equation} 
where the matrix $\mathbf{S}$ stands for self-induced dynamics and the matrix $\mathbf{C}$ represents cross-interactions between the filaments. We have made it explicit that the resistance matrix depends on the positions, $\mathbf{x}_j$, and orientations, $\mathbf{p}_j$, of the two filaments. {\color{black} Note that even the matrix $\mathbf{S}$ for self-induced dynamics depends on the position of both filaments, because fluid disturbances induced by the motion of one filament will reflect off the second filament and travel back to the position where they originated.} Because $\mathbf{F}_j$ and $\mathbf{T}_j$ are the forces and torques exerted by the filaments on the fluid, the resistance matrix is positive definite and, by the reciprocal theorem, also symmetric. In particular, this means that $\mathbf{C}(\xb_2,\xb_1,\pb_2,\pb_1) = \mathbf{C}(\xb_1,\xb_2,\pb_1,\pb_2)^T$.

Without loss of generality for the two filament case, we may define the laboratory frame to be centred on the first filament, so that $\xb_1 = 0$. Thus, the resistance matrix only depends on the directed distance $\db = \xb_2 - \xb_1$ so that
\begin{eqnarray}
\begin{pmatrix}
\Fb_1 \\ \Tb_1
\end{pmatrix} &=& \phantom{-}\Sb(\db,\pb_1,\pb_2)\begin{pmatrix}
\Ub_1 \\ \Omb_1
\end{pmatrix} + \phantom{-}\Cb(\db,\pb_1,\pb_2)\begin{pmatrix}
\Ub_2 \\ \Omb_2
\end{pmatrix},
\label{eq:F_T_first_helix} 
\\
\begin{pmatrix}
\Fb_2 \\ \Tb_2
\end{pmatrix} &=& \Sb(-\db,\pb_2,\pb_1)\begin{pmatrix}
\Ub_2 \\ \Omb_2
\end{pmatrix} + \Cb(-\db,\pb_2,\pb_1)\begin{pmatrix}
\Ub_1 \\ \Omb_1
\end{pmatrix}.
\label{eq:F_T_second_helix}
\end{eqnarray}

If the filaments are slender ($\epsilon\ll 1$), then we may represent the dynamics of filament $k$ by a force density $\fb_k(s)$ along its centreline. {\color{black} We define an arclength-dependent drag tensor $\Sigb(s)$ which relates the force density to the relative velocity of the filament centreline through the expression
\begin{equation}
\fb_k(s) = 
\Sigb_k(s) \cdot \left[\ub(\rb_k(s))-\mathbf{u}_\infty(\rb_k(s))\right].
\label{eq:defn-force-density-RFT}
\end{equation}
In Section \ref{sec:evalcoeff} we will return to the drag tensor and explain how to evaluate it using resistive-force theory (RFT) and slender-body theory (SBT). Until then, the derivation of the asymptotic series expansion is independent of which method we use to characterise the drag on an individual filament.}

For a rigid filament, the velocity of the centreline is given by the rigid body motion
\begin{equation}
\ub(\rb_k(s)) = \Ub_k + \Omb_k\times\rb_k(s).
\label{eq:defn-u-vector-form}
\end{equation}
To make our notation more compact, we introduce a kinematics vector with six components made through the concatenation of the linear and angular velocities of the filament, i.e.~$(\Ub_k,\Omb_k)$. Then, using summation convention, we may write the velocity of the first filament's centreline as 
\begin{equation}
u_i(\rb_1(s)) = (\delta_{ij}+\varepsilon_{i,j-3,k}(\rb_1(s))_k) (\Ub_1,\Omb_1)_j,
\label{eq:defn-u-suffix-notation}
\end{equation}
where the index $j$ is summed over from $1$ to $6$, while the other free indices run from $1$ to $3$ as usual, and the Kronecker delta and Levi-Civita symbol are understood to be identically zero if any index falls outside the normal range $\{1,2,3\}$. 

Next, we consider the background flow at the position of the first filament, which is nothing more than the flow induced by the second filament. {\color{black} At distances much greater than the filament thickness, $\epsilon$, the dominant flow induced by the second filament is the cumulative effect of a distribution of Stokeslets placed along its centreline, and represented by the force density $\fb_2(s)$. Hence, we can express the background flow as 
\begin{equation}
\mathbf{u}_\infty(\rb_1(s)) = \frac{1}{8\pi\mu} \sdint{\frac{\Ib+\hat{\Rb}_d(s,s')\hat{\Rb}_d(s,s')}{|\Rb_d(s,s')|} \cdot\fb_2(s')},
\label{eq:defn-induced-flow}
\end{equation}
where $\Rb_d(s,s') = \db + \rb_2(s') - \rb_1(s)$ is the relative distance between a point $s'$ on the centreline of the second filament and a point $s$ on the centreline of the first filament.} Note that $\mu = 1$ in our dimensionless units, but was included for clarity. {\color{black} Higher-order singularities, such as the source dipoles included in computational studies \cite{Tornberg2004,Maxian2021}, decay at least as fast as the inverse cube of distance, and hence do not contribute to HIs at order $\mathcal{O}(d^{-2})$, which is as far as we go with the asymptotic series expansion in this paper.}

To obtain the total hydrodynamic force and torque exerted by the filament, we need to calculate force moments along the length of the filament, so that
\begin{equation}
\mathbf{F} = \int_{-1}^{+1} \mathbf{f}(s) \mathrm{d}s, \quad \mathbf{T} = \int_{-1}^{+1} \rb(s)\times\mathbf{f}(s) \mathrm{d}s.
\label{eq:F_T_vector_form}
\end{equation}
Using the compact notation introduced earlier, we can write an expression for the dynamics vector $(\Fb_1,\Tb_1)$ of the first filament as
\begin{equation}
(\Fb_1,\Tb_1)_i = \sint{(\delta_{ij}+\varepsilon_{i-3,kj}(\rb_1(s))_k)(\fb_1(s))_j},
\label{eq:defn-F-T-suffix-notation}
\end{equation}
where the index $i$ runs from $1$ to $6$, while the other indices are summed over from $1$ to $3$.

\subsection{Asymptotic series formulation}

Equations \eqref{eq:defn-force-density-RFT}-\eqref{eq:defn-induced-flow} define a coupled system of equations for the force densities on the two filaments, which we will solve {\color{black} in the regime $d > L =2$}. We write the force distribution along each filament as an asymptotic series expansion
\begin{equation}
\fb_k(s) = \fb_k^{(0)}(s) + d^{-1}\fb_k^{(1)}(s) + d^{-2}\fb_k^{(2)}(s) + \mathcal{O}(d^{-3}),
\label{eq:expn-f}
\end{equation}
with the ultimate goal of calculating series expansions for the self-induced and cross-interaction resistance matrices in Eq.~\eqref{eq:F_T_first_helix}. We can write these as
\begin{eqnarray}
\Sb(\db,\pb_1,\pb_2) &=& \Sb^{(0)}(\dhatb,\pb_1,\pb_2) + d^{-1}\Sb^{(1)}(\dhatb, \pb_1,\pb_2) + d^{-2}\Sb^{(2)}(\dhatb, \pb_1,\pb_2) + \mathcal{O}(d^{-3}), \label{eq:expn-S} \\
\Cb(\db,\pb_1,\pb_2) &=& \Cb^{(0)}(\dhatb,\pb_1,\pb_2) + d^{-1}\Cb^{(1)}(\dhatb, \pb_1,\pb_2) + d^{-2}\Cb^{(2)}(\dhatb, \pb_1,\pb_2) + \mathcal{O}(d^{-3}), \label{eq:expn-C}
\end{eqnarray}
where the matrices at each order only depend on the direction of separation, $\dhatb$, with all dependence on the magnitude of separation, $|\mathbf{d}|=d$, captured by the algebraic power of the given order. Because the leading order is given by the limit $d \to \infty$, where the filaments do not know of each other's presence, we deduce that 
\begin{equation}
\Sb^{(0)}(\dhatb,\pb_1,\pb_2) = \Sb^{(0)}(\pb_1), \quad \Cb^{(0)}(\dhatb,\pb_1,\pb_2) = \mathbf{0}.
\label{eq:result-C0}
\end{equation}

In order to solve Eq.~\eqref{eq:defn-force-density-RFT} as an asymptotic series, we need to expand the flow induced by the second filament in inverse powers of distance. {\color{black} The Stokeslets decay like $1/|\Rb_d|$, so we first write the magnitude of the relative distance as}
\begin{equation}
|\Rb_d| = d\left(1 + \frac{2\dhatb\cdot(\rb_2(s')-\rb_1(s))}{d} + \frac{|\rb_2(s')-\rb_1(s)|^2}{d^2} \right)^{1/2}.
\end{equation}
{\color{black}Because all points on the filament centreline lie within a sphere of diameter $L$ around the centre, we have $|\rb_2(s')-\rb_1(s)| < L < d$, so we can apply the binomial expansion to get}
\begin{eqnarray}
\frac{1}{|\Rb_d|} &=& \frac{1}{d} - \frac{\dhatb\cdot(\rb_2(s')-\rb_1(s))}{d^2} + \mathcal{O}(d^{-3}),\\
\hat{\Rb}_d &=& \dhatb + \frac{(\Ib - \dhatb\dhatb)\cdot(\rb_2(s')-\rb_1(s))}{d} + \mathcal{O}(d^{-2}).
\end{eqnarray}
{\color{black} Note that these binomial expansions is valid for any $d>L$, and higher accuracy can be obtained by including more terms in the series.} Therefore, we can expand the induced flow in Eq.~\eqref{eq:defn-induced-flow} as 
\begin{equation}
u_{\infty,i}(\rb_1(s)) = \sdint{\left(d^{-1} J_{ij}(\dhatb) + d^{-2}K_{ijp}(\dhatb)(\rb_2(s')-\rb_1(s))_p + \mathcal{O}(d^{-3}) \right)(\fb_2(s'))_j},
\label{eq:expn-induced-flow}
\end{equation}
where the second-rank tensor
\begin{equation}
J_{ij}(\dhatb) = \frac{\delta_{ij} + \dhat_i\dhat_j}{8\pi\mu}
\label{eq:defn-J}
\end{equation}
represents the leading-order Stokeslet induced by the second filament, and the third-rank tensor
\begin{equation}
K_{ijp}(\dhatb) = \frac{\dhat_i\delta_{jp} + \dhat_j\delta_{ip} - \dhat_p\delta_{ij} - 3\dhat_i\dhat_j\dhat_p}{8\pi\mu}
\label{eq:defn-K}
\end{equation}
represents higher-order moments of the force distribution along the second filament.

\subsection{Leading-order dynamics}
\label{sec:leading-order}

The induced flow, Eq.~\eqref{eq:expn-induced-flow}, makes no contributions to Eq.~\eqref{eq:defn-force-density-RFT} at $\mathcal{O}(1)$. By using Eq.~\eqref{eq:defn-u-suffix-notation} to express the rigid-body motion of the filament, we find that the leading-order force distribution is given by
\begin{equation}
(\fb_1^{(0)}(s))_i = (\Sigb_1(s))_{ij}(\delta_{jk}+\varepsilon_{j,k-3,l}(\rb_1(s))_l) (\Ub_1,\Omb_1)_k.
\label{eq:result-f0}
\end{equation}
Then, by using Eq.~\eqref{eq:defn-F-T-suffix-notation} to find the total force and torque exerted by the filament, and putting the result in the form of Eq.~\eqref{eq:F_T_first_helix}, we find that
\begin{equation}
S_{ij}^{(0)}(\pb_1) = \sint{(\delta_{ik}+\varepsilon_{i-3,lk}(\rb_1(s))_l) (\Sigb_1(s))_{km}(\delta_{mj}+\varepsilon_{j-3,nm}(\rb_1(s))_n)},
\label{eq:result-S0} 
\end{equation}
where the free indices $i$ and $j$ run from $1$ to $6$. but all others are summed over from $1$ to $3$. Note that the integral depends implicitly on the orientation $\pb_1$ of the filament through the filament centreline $\rb_1$ and {\color{black} the tensor} $\Sigb_1$.

The self-induced resistance matrix $\Sb^{(0)}(\pb_1)$ can be obtained, for any orientation $\pb_1$ of the filament, by applying a change of basis to the resistance matrix expressed in the body frame of the filament, which we denote by
\begin{equation}
\mathbf{S}_0 =
\begin{pmatrix}
\mathbf{A} & \textbf{B} \\ \mathbf{B}^T & \mathbf{D}
\end{pmatrix} \equiv \Sb^{(0)}(\mathbf{0}).
\label{eq:defn-S0}
\end{equation}
If $\Qb(\pb_1)$ is the orthogonal matrix whose columns are the unit vectors $\{\e^{(1)},\ee^{(1)},\eee^{(1)}\}$ defined in Eqs.~\eqref{eq:bodyframe-A}-\eqref{eq:bodyframe-Z}, then the self-induced resistance matrix for orientation $\pb_1$ is 
\begin{equation}
\Sb^{(0)}(\pb_1) = \begin{pmatrix}
\Qb(\pb_1) & \mathbf{0} \\ \mathbf{0} &\Qb(\pb_1)
\end{pmatrix} \begin{pmatrix}
\mathbf{A} & \textbf{B} \\ \mathbf{B}^T & \mathbf{D}
\end{pmatrix} \begin{pmatrix}
\Qb(\pb_1)^T & \mathbf{0} \\ \mathbf{0} & \Qb(\pb_1)^T
\end{pmatrix},
\label{eq:result-S0(p1)}
\end{equation}
where we applied the change of basis to each three-by-three block of the resistance matrix. 

\subsection{First-order correction}

Next, we analyse Eq.~\eqref{eq:defn-force-density-RFT} at $\mathcal{O}(d^{-1})$ using the expansion of the induced flow from Eq.~\eqref{eq:expn-induced-flow}. We find that the first-order correction to the force distribution is given by
\begin{equation}
(\fb_1^{(1)}(s))_i = -(\Sigb_1(s))_{ij}\sdint{J_{jk}(\dhatb)(\fb_2^{(0)}(s'))_k}.
\label{eq:expansion-f1}
\end{equation}
Then, substituting the leading-order force density from Eq.~\eqref{eq:result-f0}, we find that
\begin{equation}
(\fb_1^{(1)}(s))_i = -(\Sigb_1(s))_{ij}J_{jk}(\dhatb)\sdint{(\Sigb_2(s'))_{kl}(\delta_{ij}+\varepsilon_{i,j-3,k}(\rb_2(s'))_k)}(\Ub_2,\Omb_2)_l.
\label{eq:result-f1}
\end{equation}
Then, by using Eq.~\eqref{eq:defn-F-T-suffix-notation} to find the total force and torque exerted by the filament, and putting the result in the form of Eq.~\eqref{eq:F_T_first_helix}, we find that
\begin{equation}
S_{ij}^{(1)}(\dhatb,\pb_1,\pb_2) = 0,
\label{eq:result-S1}
\end{equation}
and
\begin{multline}
C_{ij}^{(1)}(\dhatb,\pb_1,\pb_2) =-\sint{(\Sigb_1(s))_{ik}(\delta_{kl}+\varepsilon_{k,l-3,m}(\rb_1(s))_m)} \\ \times J_{kn}(\dhatb)\sdint{(\Sigb_2(s'))_{np}(\delta_{pj}+\varepsilon_{p,j-3,q}(\rb_2(s'))_q)}.
\label{eq:dervn-C1}
\end{multline}
We recognise from Eq.~\eqref{eq:result-S0} that these integrals are the first three columns and rows of the leading-order matrix for the first and second filament, respectively, so we can write the leading-order cross-interaction matrix as
\begin{equation}
C_{ij}^{(1)}(\dhatb,\pb_1,\pb_2) =-S_{ik}^{(0)}(\pb_1)J_{kl}(\dhatb)S_{lj}^{(0)}(\pb_2),
\label{eq:result-C1}
\end{equation}
where the free indices $i$ and $j$ run from $1$ to $6$, but all others are summed over from $1$ to $3$. We can read this expression from right to left to understand its physical interpretation. At leading order, the second filament induces a Stokeslet flow of strength $(\Sb^{(0)}(\pb_2))_{lj}(\Ub_2,\Omb_2)_j$ (with $l\in\{1,2,3\}, j\in\{1,2,...,6\}$), which gets carried over to the position of the first filament by the Oseen tensor $J_{kl}(\dhatb)/d$. The first filament sees a uniform background flow at leading order and responds to it using its own self-induced resistance matrix $(\Sb^{(0)}(\pb_1))_{ik}$ (with $i\in\{1,2,...,6\},k\in\{1,2,3\}$), as if it was translating with a uniform velocity in the opposite direction to the background flow, hence the minus sign.

We note that directionality is lost at this order, because the tensor $J_{ij}(\dhatb)$, defined in Eq.~\eqref{eq:defn-J}, is invariant under the transformation $\dhatb \mapsto - \dhatb$. All that matters at this order is the distance $d$ between the two filaments. Furthermore, $\Cb^{(1)}(\dhatb,\pb_1,\pb_2)^T = \mathbf{C}^{(1)}(-\dhatb,\pb_2,\pb_1)$, so the reciprocal theorem is satisfied at this order.

The result can also be extended to non-identical filaments by incorporating information about the filament geometry. We can make this dependence explicit in our notation by writing $\mathbf{S}^{(0)}(\mathbf{p};\mathbf{g})$, where the vector parameter $\mathbf{g}$ encapsulates all information about the filament geometry. For the particular case of helical filaments, note from Eqs.~\eqref{eq:Acomponents-A}-\eqref{eq:Dcomponents-Z} that our dimensionless $S^{(0)}_{ij}$ depends explicitly on the helix angle $\psi$, the number of turns $N$, and implicitly on the slenderness parameter $\epsilon$ through the drag coefficients $c_\perp$ and $c_\parallel$, hence $\mathbf{g} = (\psi,N,\epsilon)$ for a helix.  Note also that, in our derivation of the dimensionless $\mathbf{S}(\mathbf{n};\mathbf{g})$ we had rescaled lengths by the filament length, so we would need to add this information back in if we wanted to consider filaments of different lengths.

Using tildes to denote dimensional quantities, we can write the leading-order self-induced resistance matrix as
\begin{equation}
\tilde{\mathbf{S}}^{(0)}(\mathbf{p};\mathbf{g},\tilde{L}) = \frac{\tilde{\mu}\tilde{L}}{2}\begin{pmatrix}
\mathbf{I} & 0 \\ 0 & \mathbf{I}\tilde{L}/2
\end{pmatrix} \begin{pmatrix}
\Qb(\pb)\mathbf{A}(\mathbf{g})\Qb(\pb)^T & \Qb(\pb)\textbf{B}(\mathbf{g})\Qb(\pb)^T \\ \Qb(\pb)\mathbf{B}(\mathbf{g})^T\Qb(\pb)^T & \Qb(\pb)\mathbf{D}(\mathbf{g})\Qb(\pb)^T
\end{pmatrix}
\begin{pmatrix}
\mathbf{I} & 0 \\ 0 & \mathbf{I}\tilde{L}/2
\end{pmatrix},
\label{eq:S0-general}
\end{equation}
and also the dimensional cross-interaction matrix as
\begin{equation}
\tilde{C}^{(1)}_{ij}(\db,\mathbf{p}_1,\mathbf{p}_2;\mathbf{g}_1,\mathbf{g}_2,\tilde{L}_1,\tilde{L}_2) = - \tilde{S}^{(0)}_{ip}(\mathbf{p}_1;\mathbf{g}_1,\tilde{L}_1)\frac{\left(\delta_{pq}+\hat{d}_p\hat{d}_q\right)}{8\pi \tilde{\mu} \tilde{d}}\tilde{S}^{(0)}_{qj}(\mathbf{p}_2;\mathbf{g}_2,\tilde{L}_2).
\label{eq:C1-general}
\end{equation}
The results in Eqs.~\eqref{eq:S0-general} and \eqref{eq:C1-general} describe in full generality the far-field HIs between two filaments of arbitrary shape and orientation up to order $\mathcal{O}(\tilde{d}^{-1})$.

\subsection{Second-order correction}
We now begin to analyse Eq.~\eqref{eq:defn-force-density-RFT} at $\mathcal{O}(d^{-2})$ using the expansion of the induced flow from Eq.~\eqref{eq:expn-induced-flow}. We find that the second-order correction to the force distribution is given by
\begin{multline}
(\fb_1^{(2)}(s))_i = -(\Sigb_1(s))_{ij}\sdint{J_{jk}(\dhatb)(\fb_2^{(1)}(s'))_k} \\
- (\Sigb_1(s))_{ij}\sdint{K_{jkp}(\dhatb)(\rb_2(s')-\rb_1(s))_p(\fb_2^{(0)}(s'))_k}.
\label{eq:expansion-f2}
\end{multline}
The first of these terms will contribute to the self-induced resistance matrix because $\fb_2^{(1)}$ is linear in the kinematics of the first filament, while the second of them will contribute to the cross-interaction matrix because $\fb_2^{(0)}$ is linear in the kinematics of the second filament. 

After substituting the first-order force density from Eq.~\eqref{eq:result-f1} into Eq.~\eqref{eq:expansion-f2}, we find that there is a contribution to $\fb_1^{(2)}(s)$ of the form
\begin{equation}
-(\Sigb_1(s))_{ij}\sdint{J_{jk}(\dhatb)(-\Sigb_2(s'))_{kl}J_{lm}(\dhatb)} \sddint{(\Sigb_1(s''))_{mn}(\delta_{np}+\varepsilon_{n,p-3,q}(\rb_1(s''))_q)}(\Ub_1,\Omb_1)_p.
\end{equation}
Then, using Eqs.~\eqref{eq:defn-F-T-suffix-notation} and \eqref{eq:F_T_first_helix} to bring the result to its final form, we deduce that
\begin{equation}
S_{ij}^{(2)}(\dhatb,\pb_1,\pb_2) = S_{ik}^{(0)}(\pb_1)J_{kl}(\dhatb) S_{lm}^{(0)}(\pb_2) J_{mn}(\dhatb) S_{nj}^{(0)}(\pb_1),
\label{eq:result-S2}
\end{equation}
where the free indices $i$ and $j$ run from $1$ to $6$, but all others are summed from $1$ to $3$. Note that this clearly satisfies the reciprocal theorem because both $\Sb^{(0)}$ and the Oseen tensor are symmetric.

Physically, the result in Eq.~\eqref{eq:result-S2} expresses the fact that the Stokeslet field produced by the first filament propagates with an $\mathcal{O}(d^{-1})$ decay to the position of the second filament, where it produces a disturbance in the force. The $\mathcal{O}(d^{-1})$ perturbation in the force exerted by the second filament gets reflected back to the first filament with the same $\mathcal{O}(d^{-1})$ decay. This generates an $\mathcal{O}(d^{-2})$ disturbance in the dynamics of the first filament that is self-induced (i.e.~proportional to its own kinematics).

Similarly, after substituting the leading-order force density from Eq.~\eqref{eq:result-f0} into Eq.~\eqref{eq:expansion-f2}, we find that there is a contribution to $\fb_1^{(2)}(s)$ of the form
\begin{multline}
-(\Sigb_1(s))_{ij}\sdint{K_{jkl}(\dhatb)(\rb_2(s'))_l (\Sigb_2(s'))_{km}(\delta_{mn}+\varepsilon_{m,n-3,p}(\rb_2(s'))_p)}(\Ub_2,\Omb_2)_n \\ +(\Sigb_1(s))_{ij}K_{jkl}(\dhatb)(\rb_1(s))_l\sdint{(\Sigb_2(s'))_{km}(\delta_{mn}+\varepsilon_{m,n-3,p}(\rb_2(s'))_p)}(\Ub_2,\Omb_2)_n.
\label{eq:contrib_C2}
\end{multline}
We introduce the notation
\begin{equation}
P_{ij}(\dhatb,\pb_2) = \sdint{K_{ikl}(\dhatb)(\rb_2(s'))_l (\Sigb_2(s'))_{km}(\delta_{mj}+\varepsilon_{m,j-3,n}(\rb_2(s'))_n)}
\label{eq:defn-P}
\end{equation}
for the second-rank tensor appearing in Eq.~\eqref{eq:contrib_C2}, and rewrite this contribution as
\begin{equation}
\left[-(\Sigb_1(s))_{ij}P_{jn}(\dhatb,\pb_2) +(\Sigb_1(s))_{ij}K_{jkl}(\dhatb)(\rb_1(s))_l S^{(0)}_{kn}(\pb_2)\right](\Ub_2,\Omb_2)_n
\end{equation}
with the help of Eq.~\eqref{eq:result-S0}. Finally, we integrate the force density as per Eq.~\eqref{eq:defn-F-T-suffix-notation} to find the correction to the total force and torque due to the kinematics of the second filament. Using the fact that $K_{jkl}(\dhatb) = K_{kjl}(\dhatb)$ (follows directly from the definition in Eq.~\eqref{eq:defn-K}), we deduce that the $\mathcal{O}(d^{-2})$ correction to the cross-interaction matrix is
\begin{equation}
C_{ij}^{(2)}(\dhatb,\pb_1,\pb_2) = -S_{ik}^{(0)}(\pb_1)P_{kj}(\dhatb,\pb_2) + P^T_{ik}(\dhatb,\pb_1)S_{kj}^{(0)}(\pb_2),
\label{eq:result-C2}
\end{equation}
where the free indices $i$ and $j$ run from $1$ to $6$, but $k$ is summed from $1$ to $3$. Note that this also satisfies the reciprocal theorem, according to which $\Cb(\dhatb,\pb_1,\pb_2)^T = \Cb(-\dhatb,\pb_2,\pb_1)$ because $P_{ij}(-\dhatb,\pb_2)=-P_{ij}(\dhatb,\pb_2)$ (follows directly from the definitions of $K_{ijp}$ and $P_{ij}$ in Eqs.~\eqref{eq:defn-K} and \eqref{eq:defn-P}, respectively).

The final result for $C_{ij}^{(2)}(\dhatb,\pb_1,\pb_2)$, given by Eq.~\eqref{eq:result-C2}, involves a new quantity that we have not calculated explicitly yet -- the tensor $P_{ij}$, defined in Eq.~\eqref{eq:defn-P}. In contrast, the expressions for $C_{ij}^{(1)}(\dhatb,\pb_1,\pb_2)$ and $S_{ij}^{(2)}(\dhatb,\pb_1,\pb_2)$ (Eqs.~\eqref{eq:result-C1} and \eqref{eq:result-S2}, respectively) have the advantage that they involve only the leading-order resistance matrices $S_{ij}^{(0)}(\pb_1)$ and $S_{ij}^{(0)}(\pb_2)$. These can be easily calculated {\color{black} from RFT or SBT} since they are nothing more than the resistance matrix for an isolated filament. Our final task is to show that the tensor $P_{ij}(\dhatb,\pb_1)$ can also be calculated easily from the leading-order resistance matrix $S_{ij}^{(0)}(\pb_1)$ and two minor follow-up calculations. 

\subsection{Force moments for second-order correction}
The tensor $P_{ij}$ defined in Eq.~\eqref{eq:defn-P} is constructed in a similar way to the last three rows of the leading-order resistance matrix from Eq.~\eqref{eq:result-S0}. If we introduce the quantity
\begin{equation}
M_{lkj}(\pb_2) = \sint{(\rb_2(s))_l (\Sigb_2(s))_{km}(\delta_{mj}+\varepsilon_{j-3,nm}(\rb_2(s))_n)},
\label{eq:defn-M}
\end{equation}
which represents force moments along the centreline of a filament with orientation $\pb_2$, then what we want to compute is
\begin{equation}
P_{ij}(\dhatb,\pb_2) = K_{ikl}(\dhatb)M_{lkj}(\pb_2),
\end{equation}
but we already have an expression for the last three rows ($4\leq i \leq 6$) of the resistance matrix
\begin{equation}
S_{ij}^{(0)}(\pb_2) = \varepsilon_{i-3,lk}M_{lkj}(\pb_2),
\end{equation}
in the laboratory frame, Eq.~\eqref{eq:result-S0(p1)}.

So far we have assumed that the laboratory and interaction frame are identical, and we have only talked about changing basis from the body frame to the laboratory frame, Eq.~\eqref{eq:result-S0(p1)}. This was convenient because $S_{ij}^{(0)}(\pb_2)$ has a simple representation in the body frame of the second filament, since the orientation of the filament is $\pb_2 = \mathbf{0}$ relative to this frame. But the natural frame in which to describe the tensor $K_{ikl}(\dhatb)$ is the interaction frame where $\dhatb = \ex^{(1\to2)}$, as shown in Fig.~\ref{fig:setup} (b). In this frame, the tensor $K_{ijp}(\dhatb)$ defined in Eq.~\eqref{eq:defn-K} has components
\begin{equation}
K_{1kl}(\ex^{(1\to2)}) = \frac{1}{8\pi}\begin{pmatrix}
-2 & 0 & 0 \\ 0 & 1 & 0 \\ 0 & 0 & 1
\end{pmatrix}, ~ K_{2kl}(\ex^{(1\to2)}) = \frac{1}{8\pi}\begin{pmatrix}
0 & 1 & 0 \\ -1 & 0 & 0 \\ 0 & 0 & 0
\end{pmatrix}, ~ K_{3kl}(\ex^{(1\to2)}) = \frac{1}{8\pi}\begin{pmatrix}
0 & 0 & 1 \\ 0 & 0 & 0 \\ -1 & 0 & 0
\end{pmatrix}.
\end{equation}
Hence, the tensor $P_{ij}(\dhatb,\pb_2)$ can be written in the interaction frame as
\begin{multline}
P_{ij}(\ex^{(1\to2)},\pb_2') = \frac{1}{8\pi}\delta_{i1}(-2M_{11j}(\pb_2')+M_{22j}(\pb_2')+M_{33j}(\pb_2')) \\ + \frac{1}{8\pi}\delta_{i2}(-M_{12j}(\pb_2')+M_{21j}(\pb_2')) + \frac{1}{8\pi}\delta_{i2}(-M_{13j}(\pb_2')+M_{31j}(\pb_2')),
\label{eq:dervn-Pij}
\end{multline}
whereas the last three rows ($4\leq i \leq 6$) of the resistance matrix are
\begin{multline}
S_{ij}^{(0)}(\pb_2') = \delta_{i4}(M_{23j}(\pb_2')-M_{32j}(\pb_2')) \\ + \delta_{i5}(-M_{13j}(\pb_2')+M_{31j}(\pb_2')) + \delta_{i6}(M_{12j}(\pb_2')-M_{21j}(\pb_2')).
\label{eq:dervn-Sij}
\end{multline}
Note that we have used the notation $\pb_2'$ to indicate the orientation of the filament relative to the interaction frame, so the tensors $\mathbf{M}(\pb_2')$ and $\Sb^{(0)}(\pb_2')$ are also to be expressed in these coordinates. By comparing the two expressions in Eqs.~\eqref{eq:dervn-Pij} and \eqref{eq:dervn-Sij}, we deduce that 
\begin{equation}
P_{2j}(\ex^{(1\to2)},\pb_2') = -\frac{S_{6j}^{(0)}(\pb_2')}{8\pi}, \quad P_{3j}(\ex^{(1\to2)},\pb_2') = \frac{S_{5j}^{(0)}(\pb_2')}{8\pi},
\label{eq:result-P-tworows}
\end{equation}
so we get the last two rows of $P_{ij}$ for free.

To complete the top row of $P_{ij}$ we simply need to calculate the quantity
\begin{equation}
P_{1j}(\ex^{(1\to2)},\pb_2') = \frac{1}{8\pi}(-2M_{11j}(\pb_2')+M_{22j}(\pb_2')+M_{33j}(\pb_2')),
\label{eq:dervn-Pij-1row}
\end{equation}
which is more easily calculated in the body frame of the filament and then transferred to the interaction frame by a change of basis.

{\color{black} Everything we have done so far is valid for filaments of arbitrary shape. Below, we go into more detail about the evaluation of the new row $P_{1j}$ for helical filaments, which will be used later for the validation and application of our theory. In the body frame of a helical filament, where $\pb_2' \to \mathbf{0}$, we denote the right-hand side of Eq.~\eqref{eq:dervn-Pij-1row} by}
\begin{equation}
(\mb_0)_j = -2M_{11j}(\mathbf{0})+M_{22j}(\mathbf{0})+M_{33j}(\mathbf{0}).
\label{eq:defn-m0}
\end{equation} 
{\color{black} The helical centreline introduced in Eq.~\eqref{eq:centreline} is symmetric under a rotation by angle $\pi$ around the unit vector $\e$. Due to this symmetry, the vector $\mb_0$ has vanishing components along the $\ee$ and $\eee$ directions, regardless of the method (RFT or SBT) by which we choose to evaluate it, meaning that}
\begin{equation}
(\mb_0)_{i} = (\mathcal{M}_{1} \e)_i, ~ (\mb_0)_{i+3} = (\mathcal{M}_{4} \e)_i,
\label{eq:dervn-m0}
\end{equation}
for index $i = 1,2,3$. Hence, when we move this result to the interaction frame of two helices, we obtain the final result for the matrix $\Pb(\ex^{(1\to2)},\pb_2')$
\begin{equation}
\Pb(\ex^{(1\to2)},\pb_2') = \frac{1}{8\pi} 
\begin{pmatrix}
\mathcal{M}_{1} \alpha(\pb_2') & \mathcal{M}_{1} \beta(\pb_2') & \mathcal{M}_{1} \gamma(\pb_2') & \mathcal{M}_{4} \alpha(\pb_2') & \mathcal{M}_{4} \beta(\pb_2') & \mathcal{M}_{4} \gamma(\pb_2') \\
-S_{61}^{(0)}(\pb_2') & -S_{62}^{(0)}(\pb_2') & -S_{63}^{(0)}(\pb_2') & -S_{64}^{(0)}(\pb_2') & -S_{65}^{(0)}(\pb_2') & -S_{66}^{(0)}(\pb_2') \\
S_{51}^{(0)}(\pb_2') & S_{52}^{(0)}(\pb_2') & S_{53}^{(0)}(\pb_2') & S_{54}^{(0)}(\pb_2') & S_{55}^{(0)}(\pb_2') & S_{56}^{(0)}(\pb_2')
\end{pmatrix},
\label{eq:result-Pmatrix}
\end{equation}
where $\alpha(\pb_2') = \e^{(2)}\cdot\ex^{(1\to2)}$, $\beta(\pb_2') = \e^{(2)}\cdot\ey^{(1\to2)}$ and $\gamma(\pb_2') = \e^{(2)}\cdot\ez^{(1\to2)}$ are the components of $\e^{(2)}$ relative to the interaction frame of filaments $1$ and $2$. If the interaction frame does not coincide with the laboratory frame (e.g.~if there are more than two filaments), this result would have to be moved to the laboratory frame by a change of basis on each three-by-three block. 

{\color{black} 
\subsection{Evaluating coefficients in the series expansion}
\label{sec:evalcoeff}
The first and second-order coefficients in the series expansion only require the leading-order resistance matrix, $\Sb^{(0)}$, and the force moment, $\mb_0$, which themselves only depend on the shape of the filament, $\rb(s)$, and the drag tensor, $\Sigb(s)$. We now explain how to evaluate these coefficients using both resistive-force theory (RFT) and slender-body theory (SBT). The former has the advantage of being analytically tractable but only logarithmically accurate, while the latter is algebraically correct but requires computations.

In RFT \cite{Hancock1953,Gray1955,Lighthill1996_helical}, the drag tensor depends only on the local tangent to the filament,}
\begin{equation}
\Sigb_{\mathrm{RFT}}(s) = c_\perp[\mathbf{I} -\tb(s)\tb(s)]+c_\parallel\tb(s)\tb(s),
\label{eq:defn-Sigma}
\end{equation} 
and quantifies the {\color{black}anisotropic} drag on the filament through the perpendicular, $c_\perp$, and parallel, $c_\parallel$, drag coefficients 
\begin{equation}
	c_\perp = \frac{4\pi\mu}{\ln(2/\epsilon)+1/2}, \quad c_\parallel = \frac{2\pi\mu}{\ln(2/\epsilon)-1/2}.
\end{equation} 
Note that, for clarity, we have included the dimensionless viscosity $\mu=1$ in the above definition of the drag coefficients. For the special case of a helical filament, we {\color{black}use RFT to derive} analytical expressions for $\Sb_0$ in Appendix \ref{app:RFT} and for $\mb_0$ in Appendix \ref{app:forcemoments_RFT}.

{\color{black} In SBT \cite{Cox1970,Lighthill1976,Johnson1980}, on the other hand, the relationship between force density and velocity is non-local, so we cannot express the drag tensor as a local object. The value of $\Sigb_{\mathrm{SBT}}(s)$ at each point $s$ along the centreline depends on the specifics of the motion relative to the shape of the filament. However, we do not need to know the general form of $\Sigb_{\mathrm{SBT}}(s)$ in order to evaluate the coefficients in our asymptotic series expansion using SBT. An inspection of Eqs.~\eqref{eq:result-S0} and \eqref{eq:defn-P} reveals that the drag tensor always appears contracted with the six modes of rigid-body motion that are available to our rigid filaments, in the form $\Sigma_{ik}(s)(\delta_{kj}+\varepsilon_{j-3,lk}r_l(s))$. Therefore, we only need to know the SBT drag tensor as it pertains to rigid-body motion,
\begin{equation}
\Sigb_{\mathrm{SBT}}(s)\cdot (\Ub + \Omb\times\rb(s)) \equiv \fb_{\mathrm{SBT}}(s;\Ub,\Omb),
\end{equation}
where $\fb_{\mathrm{SBT}}(s;\Ub,\Omb)$ is the SBT force density along a filament with kinematics $(\Ub,\Omb)$. By considering each mode of rigid-body motion individually, we can write
\begin{equation}
\Sigma_{ik}(s)(\delta_{kj}+\varepsilon_{j-3,lk}r_l(s)) \equiv (\fb^{(j)}_{\mathrm{SBT}}(s))_i,
\label{eq:defn-fSBT}
\end{equation}
where $\fb^{(j)}_{\mathrm{SBT}}(s)$ is now the force density computed from SBT for the $j$th mode of rigid body motion ($j=1,2,3$ for translations, $j=4,5,6$ for rotations).

From Eqs.~\eqref{eq:result-S0} and \eqref{eq:defn-fSBT}, we get the leading-order resistance matrix, $\Sb^{(0)}$, from SBT 
\begin{equation}
    (\Sb^{(0)}_{\mathrm{SBT}})_{ij} = \sint{(\delta_{ik}+\varepsilon_{i-3,lk}(\rb_1(s))_l)(\fb^{(j)}_{\mathrm{SBT}}(s))_k}.
\end{equation}
Similarly, from Eqs.~\eqref{eq:defn-M}, \eqref{eq:defn-m0} and \eqref{eq:defn-fSBT}, we find the SBT equivalent of $\mb_0$ as 
\begin{equation}
(\mb_0^{\mathrm{SBT}})_j = \sint{\rb(s) \cdot (\Ib -3\ex^{(1 \to 2)}\ex^{(1 \to 2)})\cdot \fb^{(j)}_{\mathrm{SBT}}(s)}.
\label{eq:m0-SBT}
\end{equation}

Evaluating the force density $\fb^{(j)}_{\mathrm{SBT}}(s)$ does require a numerical computation but for a rigid filament this only needs to be done once, in the body frame of the filament, and then modified with a change of basis if the filament changes orientation over time. The SBT computation consists of solving Eq.~\eqref{eq:COMP-method} numerically, exactly as described in Section \ref{sec:comp-method}, but without the interaction term $\mathcal{J}[\fb_2(s'),\db]$.

In the following sections, when we refer to the asymptotic theory with RFT or SBT coefficients, we mean that we have used the series expansion for the extended resistance matrix from Eqs.~\eqref{eq:expn-S} and \eqref{eq:expn-C}, with coefficients up to second order given by Eqs.~\eqref{eq:result-C0},\eqref{eq:result-S0}, \eqref{eq:result-S1},\eqref{eq:result-C1},\eqref{eq:result-S2} and \eqref{eq:result-C2}, but these coefficients have been evaluated either analytically with RFT or computationally with SBT. The RFT calculations for the matrix $\Sb^{(0)}$ and the vector $\mb_0$ are given in Appendices \ref{app:RFT} and \ref{app:forcemoments_RFT}, respectively, while the computational method for SBT is described in Section \ref{sec:comp-method} (except that the interaction term $\mathcal{J}$ is not included in the SBT computation for a single filament).}

\newpage
\section{Validation of asymptotic model}
\label{sec:validation}

We will now verify {\color{black} the asymptotic theory with RFT/SBT coefficients} against numerical simulations {\color{black} based on SBT}. In this section, we focus on filaments with a helical centreline, which are very common in microscopic scale flows (e.g.~the helical flagellar filaments of bacteria, helical microbots actuated by external magnetic fields, elongated microorganisms with a spiral body shape). 

\subsection{Computational method for hydrodynamic interactions}
\label{sec:comp-method}

In order to validate our asymptotic model, we implement Johnson's slender-body theory \cite{Johnson1980,thesisKoens} with additional interactions between the filaments \cite{Tornberg2004}. In our computational method, we replace Eq.~\eqref{eq:defn-force-density-RFT} with the following relationship between the force density and velocity along the filament centreline,
\begin{equation}
8\pi\mu\ub(\rb_1(s)) = \mathcal{L}[\fb_1(s)] + \mathcal{K}[\fb_1(s')] + \mathcal{J}[\fb_2(s'),\db],
\label{eq:COMP-method}
\end{equation}
where the first operator represents local effects
\begin{equation}
\mathcal{L}[\fb_1(s)] = \left[2\left(\ln\left(\frac{2}{\epsilon}\right)+\frac{1}{2}\right)\Ib + 2\left(\ln\left(\frac{2}{\epsilon}\right)-\frac{3}{2}\right)\tb_1(s)\tb_1(s)\right]\cdot \fb_1(s),
\label{eq:COMP-local}
\end{equation}
and the second operator represents non-local effects
\begin{multline}
\mathcal{K}[\fb_1(s')] = \sdint{\left[\frac{\Ib+\hat{\Rb}_0(s,s')\hat{\Rb}_0(s,s')}{|\Rb_0(s,s')|}-\frac{\Ib+\tb_1(s)\tb_1(s)}{|s'-s|}\right]\cdot \fb_1(s')} \\ 
+ \left(\Ib+\tb_1(s)\tb_1(s)\right)\cdot\sdint{\frac{\fb_1(s')-\fb_1(s)}{|s'-s|}},
\label{eq:COMP-nonlocal}
\end{multline}
where $\Rb_0(s,s') = \rb_1(s)-\rb_1(s')$, and we have split the terms in such a way that both integrals have a removable singularity at $s'=s$. Finally, the third operator represents interactions between the two filaments {\color{black} as previously modelled by Tornberg and Shelley \cite{Tornberg2004}},
\begin{equation}
\mathcal{J}[\fb_2(s'),\db] = \sdint{\left[\frac{\Ib+\hat{\Rb}_d(s,s')\hat{\Rb}_d(s,s')}{|\Rb_d(s,s')|}  + \frac{\epsilon^2}{2}\frac{\Ib-3\hat{\Rb}_d(s,s')\hat{\Rb}_d(s,s')}{|\Rb_d(s,s')|^3}\right]\cdot\fb_2(s')},
\label{eq:COMP-interaction}
\end{equation}
where $\Rb_d(s,s') = \db +\rb_2(s')-\rb_1(s)$. {\color{black} In our computational method, which was implemented for purposes beyond the present study, we choose to include the source dipole term that was left out of our asymptotic theory, Eq.~\eqref{eq:defn-induced-flow}, because it would have contributed to the asymptotic series expansion only at order $\mathcal{O}(d^{-3})$. Note that we have used the same prefactor of $1/2$ for the dipole term as in \cite{Tornberg2004}, while a more recent study based on the Rotne-Prager-Yamakawa kernel and matched asymptotics uses a larger prefactor of $e^3/24$ \cite{Maxian2021}.}

{\color{black}
We solve Eqs.~\eqref{eq:COMP-method}-\eqref{eq:COMP-interaction} numerically using a spectral method based on Legendre polynomials as in Ref.~\cite{thesisKoens}. Other studies have chosen to solve these integral equations by regularizing the integral operator $\mathcal{K}$ and approximating its arguments with piecewise polynomials \cite{Tornberg2004}, or more recently using a spectral method based on Chebyshev polynomials \cite{Maxian2021}. In the present study, the choice of Legendre polynomials as a set of basis functions is motivated by their being eigenfunctions of the second integral in the non-local operator $\mathcal{K}$, meaning that
\begin{equation}
    \sdint{\frac{P_n(s')-P_n(s)}{|s'-s|}} = E_n P_n(s),
\end{equation}
with eigenvalues $E_0=0$ and
\begin{equation}
    E_n = -2\sum_{j=1}^{n}\frac{1}{j},
\end{equation}
for $n>0$ \cite{thesisGotz}.

We discretize the force density and velocity along the filaments as 
\begin{equation}
    \ub(\rb_k(s)) = \sum_{n=0}^\infty \ub_k^{(n)}P_n(s), \quad \fb_k(s) = \sum_{n=0}^\infty \fb_k^{(n)}P_n(s),
\end{equation}
where the velocity coefficients $\ub_k^{(n)}$ are known from the prescribed kinematics, and the force coefficients $\fb_k^{(n)}$ must be solved for.  After projecting Eq.~\eqref{eq:COMP-method} onto the space of Legendre polynomials and making use of the orthogonality condition
\begin{equation}
    \sint{P_n(s)P_m(s)} = \frac{2\delta_{mn}}{2n+1},
\end{equation}
we recover the following system of equations relating the velocity and the force coefficients
\begin{multline}
    8\pi\mu \ub_1^{(n)} = \left[2\left(\ln\left(\frac{2}{\epsilon}\right)+\frac{1}{2}\right) + E_n \right] \fb_1^{(n)}   \\  + \frac{2n+1}{2} \sum_{m=0}^{\infty} \Bigg[ \left[2\left(\ln\left(\frac{2}{\epsilon}\right)-\frac{3}{2}\right) + E_m \right]\mathbf{M}_{\parallel}^{(n,m)}\fb_1^{(m)} + \mathbf{M}_{0}^{(n,m)}\fb_1^{(m)} + \mathbf{M}_{d}^{(n,m)}\fb_2^{(m)}\Bigg],
    \label{eq:COMP-method-projected}
\end{multline}
where the matrices $\mathbf{M}_{\parallel}^{(n,m)}$, $\mathbf{M}_{0}^{(n,m)}$ and $\mathbf{M}_{d}^{(n,m)}$ are given by
\begin{eqnarray}
\mathbf{M}_{\parallel}^{(n,m)} &=& \sint{\tb_1(s)\tb_1(s)P_n(s)P_m(s)}, \\
\mathbf{M}_{0}^{(n,m)} &=& \sint{\sdint{\left[\frac{\Ib+\hat{\Rb}_0(s,s')\hat{\Rb}_0(s,s')}{|\Rb_0(s,s')|}-\frac{\Ib+\tb_1(s)\tb_1(s)}{|s'-s|}\right]P_n(s)P_m(s')}}, \\
\mathbf{M}_{d}^{(n,m)} &=& \sint{\sdint{\left[\frac{\Ib+\hat{\Rb}_d(s,s')\hat{\Rb}_d(s,s')}{|\Rb_d(s,s')|}  + \frac{\epsilon^2}{2}\frac{\Ib-3\hat{\Rb}_d\hat{\Rb}_d}{|\Rb_d(s,s')|^3}\right]P_n(s)P_m(s')}}.
\end{eqnarray}
The second of these matrices involves a removable singularity at $s'=s$, but the quadrature integration methods readily available in MATLAB can evaluate this integral accurately so long as the singular points lie on the boundaries of the integration domain. Therefore, when computing the matrices $\mathbf{M}_{0}^{(n,m)}$ in MATLAB we split the double integral into two parts - $s\in[-1,+1]$, $s'\in[-1,s]$ and $s\in[-1,+1]$, $s'\in[s,+1]$. 

The infinite system of linear equations from Eq.~\eqref{eq:COMP-method-projected} is truncated to $m \leq N_{\mathrm{Legendre}}$ modes and inverted numerically, in order to find the force density coefficients $\fb_1^{(k)}$ in terms of the velocity coefficients $\ub_1^{(k)}$, which themselves are linearly dependent on the filament kinematics $(\Ub_k,\Omb_k)$. The force density is then integrated along the filaments to find the extended resistance matrix that relates filament kinematics and dynamics. We implement this algorithm in MATLAB and validate it using the tests described in Appendix \ref{app:comptests}.}

For each set of parameters $(N,\psi,\epsilon)$ describing the geometry of the helical filament, we vary the number of Legendre modes in our truncation until the numerical solution for an isolated helix settles to within 1\% error. We then make the reasonable assumption that the number of Legendre modes determined from this single-helix self-convergence test is sufficient to obtain the same level of accuracy in our double-helix simulations as well. In general, we find that the required number of Legendre modes increases with the number of helical turns of the filament, because we must be able to capture variations in the force density and filament velocity which have the same wavenumber as the filament centreline. For most simulations presented in this study it was sufficient to use $N_{\mathrm{Legendre}} = 15$, because the helices have a small number of helical turns. 

\subsection{Relative errors}

In the absence of an exact solution, we use the numerical solution from SBT as a reference value against which to {\color{black} validate} our asymptotic model. {\color{black} In the previous section, we derived a series expansion for the extended resistance, $\mathbf{R}$, in the form
\begin{equation}
\mathbf{R} = \mathbf{R}^{(0)} + d^{-1}\mathbf{R}^{(1)} + d^{-2}\mathbf{R}^{(2)} + \mathcal{O}(d^{-3}),
\label{eq:expn-R}
\end{equation}
up to and including second-order terms. We wish to compare this expansion of the resistance matrix with the numerical solution, $\tilde{\mathbf{R}}$, of the fully-coupled integral equations described in Section \ref{sec:comp-method}. However, we cannot compare the matrices $\mathbf{R}$ and $\tilde{\mathbf{R}}$ component-wise, because this would depend on the basis in which we represent the matrices. One can always choose a vector basis in which some component of the ``true" solution $\tilde{\mathbf{R}}$ is zero, relative to which our approximate solution $\mathbf{R}$ would have an infinite relative error. Therefore, we need to think of the extended resistance matrices as linear operators between the space of filament kinematics and the space of filament dynamics, and define an error for the operator as a whole in a way that is basis-independent. A standard way to do this is to use an operator norm.}

Suppose we have some given kinematics $\mathbf{x}$ (two linear and two angular velocities, so a vector with twelve components) and we want to compute the dynamics $\mathbf{y}$. Then the error in $\mathbf{y}$ is $\Delta \mathbf{y} =  \mathbf{R}\mathbf{x} - \tilde{\mathbf{R}}\mathbf{x}$. We define the ``relative error" in the dynamics to be 
\begin{equation}
E_{\mathrm{dyn}} \equiv \sup_{\mathbf{x}}\left\{ \frac{||\tilde{\mathbf{R}}\mathbf{x} - \mathbf{R}\mathbf{x}||_p}{||\tilde{\mathbf{R}}\mathbf{x}||_p} \right\} = \sup_{\mathbf{y}}\left\{ \frac{||(\mathbf{I} - \mathbf{R}\tilde{\mathbf{R}}^{-1})\mathbf{y}||_p}{||\mathbf{y}||_p} \right\},
\label{eq:rel_error_dynamics}
\end{equation}
in other words the operator norm of $\mathbf{I} - \mathbf{R}\tilde{\mathbf{R}}^{-1}$. {\color{black} Note that taking the supremum over the entire space of filament kinematics is important, so that the value we compute for the relative error is not dependent on an arbitrary choice of filament kinematics.}

Similarly, we can define the relative error in the kinematics as
\begin{equation}
E_{\mathrm{kin}} \equiv  \sup_{\mathbf{y}}\left\{ \frac{||\tilde{\mathbf{R}}^{-1}\mathbf{y} - \mathbf{R}^{-1}\mathbf{y}||_p}{||\tilde{\mathbf{R}}^{-1}\mathbf{y}||_p} \right\} = \sup_{\mathbf{x}}\left\{ \frac{||(\mathbf{I} - \mathbf{R}^{-1}\tilde{\mathbf{R}})\mathbf{x}||_p}{||\mathbf{y}||_p} \right\},
\label{eq:rel_error_kinematics}
\end{equation}
so the operator norm of $\mathbf{I} - \mathbf{R}^{-1}\tilde{\mathbf{R}}$. Here again, {\color{black} taking the supremum is important, so that the relative error we compute does not depend on an arbitrary choice of filament dynamics}.

\begin{figure}	
	\landscapetrim{17cm}{10cm}
	\includegraphics[trim={{.5\cutwidth} {.5\cutheight} {.5\cutwidth} {.5\cutheight}},clip,width=17cm]{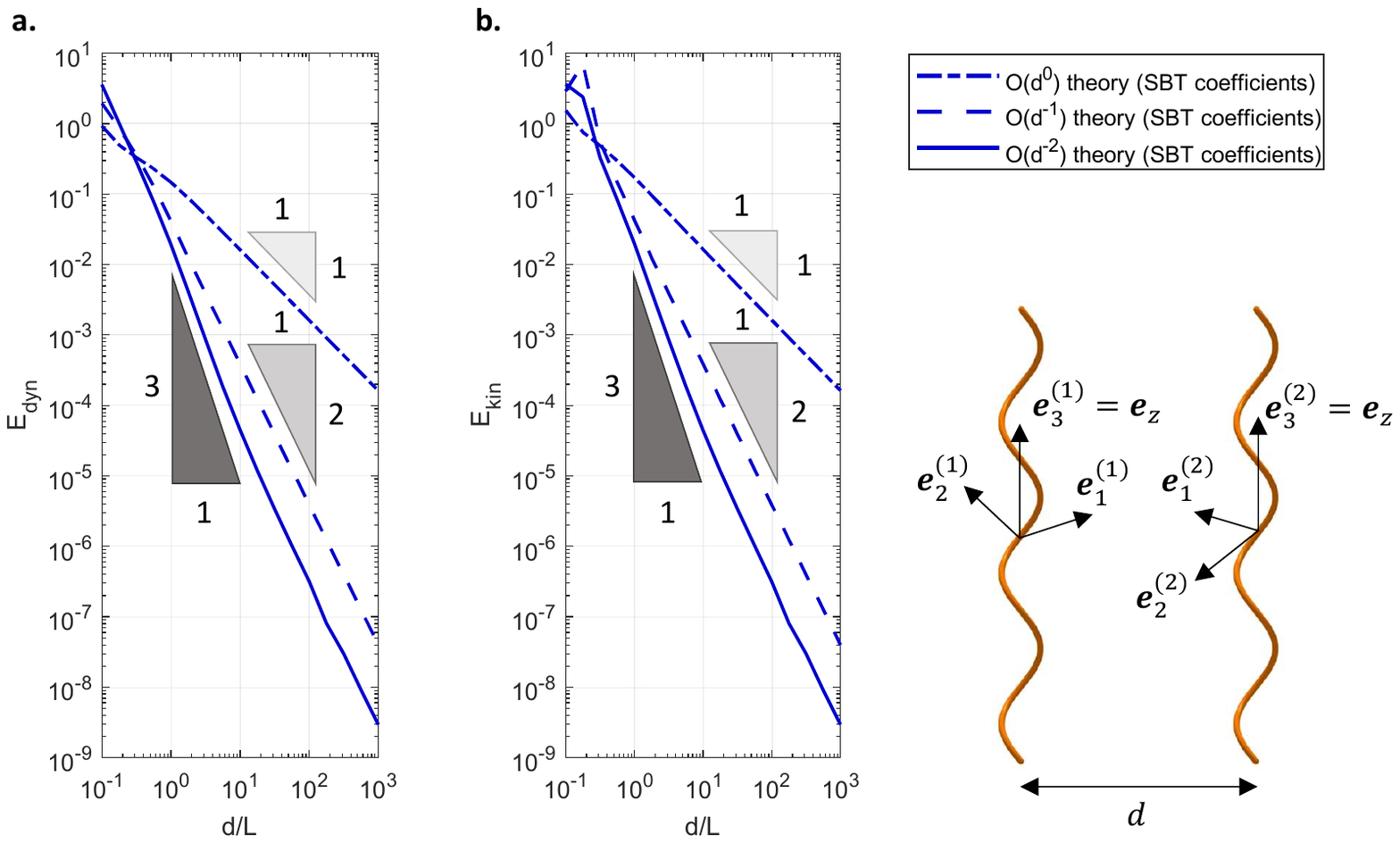}
	\caption{Relative error in (a) helix dynamics and (b) helix kinematics, as defined in Eqs.~\eqref{eq:rel_error_dynamics} and \eqref{eq:rel_error_kinematics} respectively, with $p=2$. {\color{black} As we increase the helix separation, $d$, the asymptotic theory with SBT coefficients} converges to the numerical solution, and the error decays as expected {\color{black} with each higher order included in the theory}. Parameter specification: helices have configurations $(\theta_1,\chi_1,\phi_1) = (0,0,\pi/6)$ and $(\theta_2,\chi_2,\phi_2) = (0,0,2\pi/3)$, and $N=2.75$ helical turns. Helix angle, $\psi = 0.5$ rad, and filament slenderness, $\epsilon = 10^{-2}$, are representative of bacterial flagella.}
	\label{fig:matrix_errors}
\end{figure}

In Fig.~\ref{fig:matrix_errors} (a) and (b) we compare the relative errors, {\color{black} defined with a $p=2$ norm}, for different {\color{black} orders in our asymptotic theory with SBT coefficients. If our asymptotic series expansion up to $\mathcal{O}(d^{-m})$ terms was calculated correctly, then we would expect the relative error to decay like $ d^{-(m+1)}$, the order of the first neglected terms. This is confirmed by the slopes of our log-log plots, which validate our asymptotic series expansion up to $\mathcal{O}(d^{-2})$. Note that the comparison is only meaningful between the computations and the asymptotic theory with SBT coefficients. This is an unavoidable consequence of our choice to implement the computational method based on SBT. The asymptotic theory with RFT coefficients differs at leading order from the numerical solution based on SBT, and so we would not be able to observe convergence unless we implemented a different computational method based on RFT. The results presented in Fig.~\ref{fig:matrix_errors} (a) and (b) serve to validate the asymptotic series expansion in itself, regardless of the method (RFT or SBT) by which we choose to calculate the leading-order resistance matrix, $\Sb^{(0)}$, and the force moment, $\mb_0$. } 

{\color{black} Furthermore, by examining the size of the relative error, we deduce that the asymptotic theory can be useful for any $d>L$, which is the regime of validity for our binomial expansion of the Oseen tensor. When the filaments are parallel and orthogonal to the line that connects their centres, we observe that our asymptotic theory with SBT coefficients can achieve 99\% accuracy for $d/L > 1.4$. This accuracy is achieved by the asymptotic solution up to and including $\mathcal{O}(d^{-2})$ terms. Higher accuracy could be obtained either by including more terms in the asymptotic series expansion, or by increasing the distance between the filaments. Based on further results presented in this study, where we also vary the phase difference between filaments, we believe this accuracy estimate to be representative of any parallel configuration of two filaments with this particular helical geometry. A broader numerical investigation would be necessary to determine the accuracy of our method for rigid filaments of arbitrary geometry and non-parallel configurations.}

\subsection{Time evolution of forces and torques}

{\color{black} The main purpose of the asymptotic theory presented in this paper is to provide a systematic method to calculate analytically the specific HIs between two filaments. When carrying out calculations by hand, we are interested in finding relative patterns more than in calculating accurate absolute values, which is the purpose of numerical schemes. With this perspective in mind, we propose to validate the asymptotic theory with RFT coefficients by looking at the time variation of hydrodynamically-induced forces and torques. We consider the case of two slender helices rotating in parallel with the same angular velocity}.

\begin{figure}
\centering
	\portraittrim{17cm}{20.9cm}
	\includegraphics[trim={{.5\cutwidth} {.5\cutheight} {.5\cutwidth} {.5\cutheight}},clip,width=14cm]{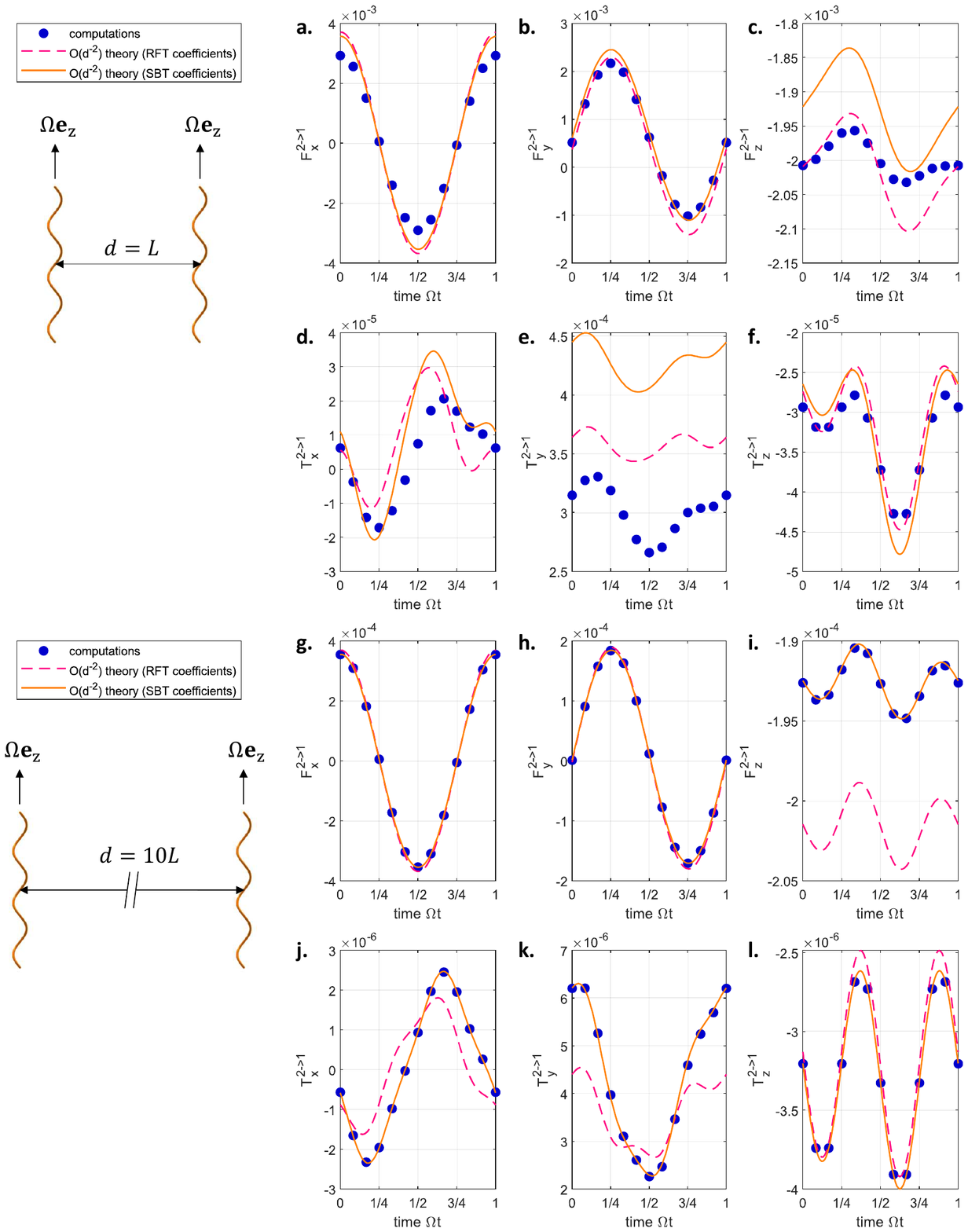}
	\caption{Comparison between {\color{black} computations and the asymptotic theory with RFT/SBT coefficients}, by means of the time evolution of forces and torques induced by the second (rightmost) filament on the first (leftmost). The helices are vertical {\color{black} ($\theta=0$)} and rotating with constant angular velocity $\Omega\ez$. We fix the phase difference $\Delta\phi = \pi/2$ between them, and a horizontal distance equal to the integrated filament length (a-f) or ten times larger (g-l). The helix angle, $\psi = 0.5043$ rad, and filament slenderness, $\epsilon =0.0038$, were chosen as representative of bacterial flagella. The helices have $N=2.5$ helical turns.}
	\label{fig:results_time_evolution}
\end{figure}

\begin{figure}
\centering
	\portraittrim{17cm}{20.9cm}
	\includegraphics[trim={{.5\cutwidth} {.5\cutheight} {.5\cutwidth} {.5\cutheight}},clip,width=14cm]{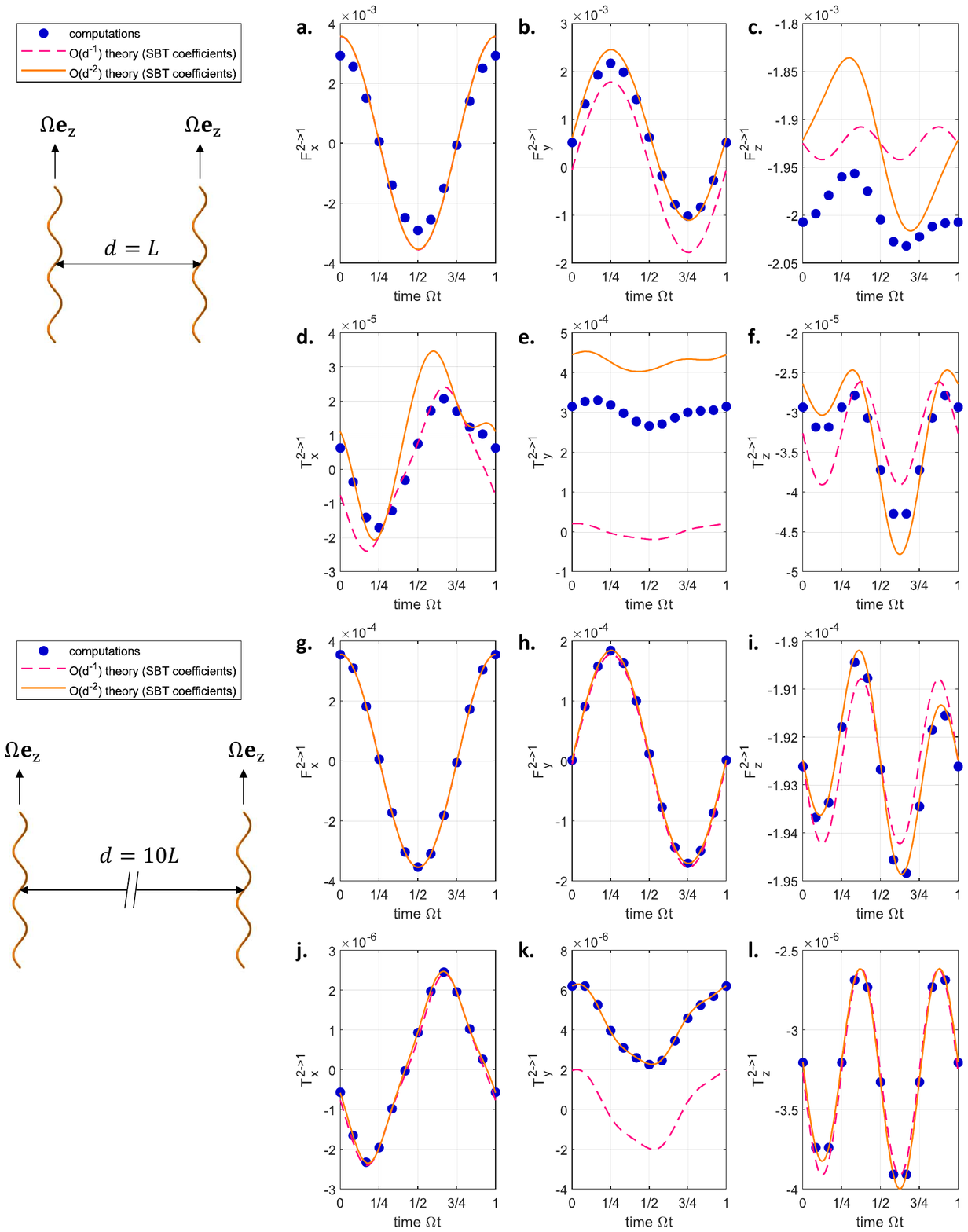}
	\caption{Comparison between {\color{black} computations and the asymptotic theory with SBT coefficients} to $\mathcal{O}(d^{-1})$ and $\mathcal{O}(d^{-2})$, by means of the time evolution of forces and torques induced by the second (rightmost) filament on the first (leftmost). The helices are vertical  {\color{black} ($\theta=0$)} and rotating with constant angular velocity $\Omega\ez$. We impose the phase difference $\Delta\phi = \pi/2$ between them, and a horizontal distance equal to the integrated filament length (a-f) or ten times larger (g-l). The helix angle, $\psi = 0.5043$ rad, and filament slenderness, $\epsilon =0.0038$, were chosen as representative of bacterial flagella. The helices have $N=2.5$ helical turns.}
	\label{fig:results_compareorders}
\end{figure}

Back in Fig.~\ref{fig:matrix_errors}, we examined the relative error for a fixed orientation of the helices, and we varied the distance between the filaments to see how the error decays - a quantitative {\color{black} validation} of our asymptotic model. In Fig.~\ref{fig:results_time_evolution}, however, we fix the distance between the helical filaments and we let time flow, and the orientation of the filaments along with it, to look for patterns over time - a qualitative {\color{black} validation} of our asymptotic model. {\color{black} Because the helices are vertical, their body-fixed axis $\eee$ is parallel to the laboratory frame $\ez$. Hence, the phase angle $\phi$ around $\ez$ and the spin angle $\chi$ around $\eee$, as defined in Eqs.~\eqref{eq:bodyframe-A}-\eqref{eq:bodyframe-Z}, are interchangeable. Without loss of generality, we can describe the configuration of the filaments from Figs.~\ref{fig:results_time_evolution} and \ref{fig:results_compareorders} as $(\theta_1,\chi_1,\phi_1) = (0,0,\Omega t)$ and $(\theta_2,\chi_2,\phi_2) = (0,0,\Omega t+\Delta\phi)$.}

{\color{black} Our asymptotic theory with both RFT and SBT coefficients} captures the qualitative features of the interaction even for smaller helix separations Fig.~\ref{fig:results_time_evolution} (a)-(f), with the agreement becoming quantitative at larger separations Fig.~\ref{fig:results_time_evolution} (g)-(l). This indicates that our {\color{black}asymptotic series expansion can be used to derive meaningful analytical expressions for the HIs between filaments separated by a distance greater than their contour length, as later demonstrated in Section \ref{sec:application}.}

We also provide a direct comparison between the {\color{black} asymptotic theory with SBT coefficients} at  $\mathcal{O}(d^{-1})$ and $\mathcal{O}(d^{-2})$, in Fig.~\ref{fig:results_compareorders}. These plots provide clearer visual evidence that higher-order corrections improve the fidelity of the asymptotic solution, as opposed to Fig.~\ref{fig:matrix_errors} where the evidence {\color{black} spanned a wider range of kinematic conditions, but was presented in a more condensed format}.

\section{Application to helical pumps}
\label{sec:application}

To demonstrate the usefulness of our asymptotic theory, we now apply and extend our analytical calculations to the interaction of rotating helical pumps. This particular application of our theory is motivated by previous theoretical and experimental studies of helical micropumps \cite{Darnton2004,Kim2008,Martindale2017,Dauparas2018,Buchmann2018}. Experimentally, these systems often take the form of bacterial carpets or forests, where the bacteria are stuck to a substrate while their helical flagellar filaments are free to rotate and pump fluid around.

\subsection{Problem specification}

\begin{figure}
	\landscapetrim{17cm}{15cm}
	\includegraphics[trim={{.5\cutwidth} {.5\cutheight} {.5\cutwidth} {.5\cutheight}},clip,width=17cm]{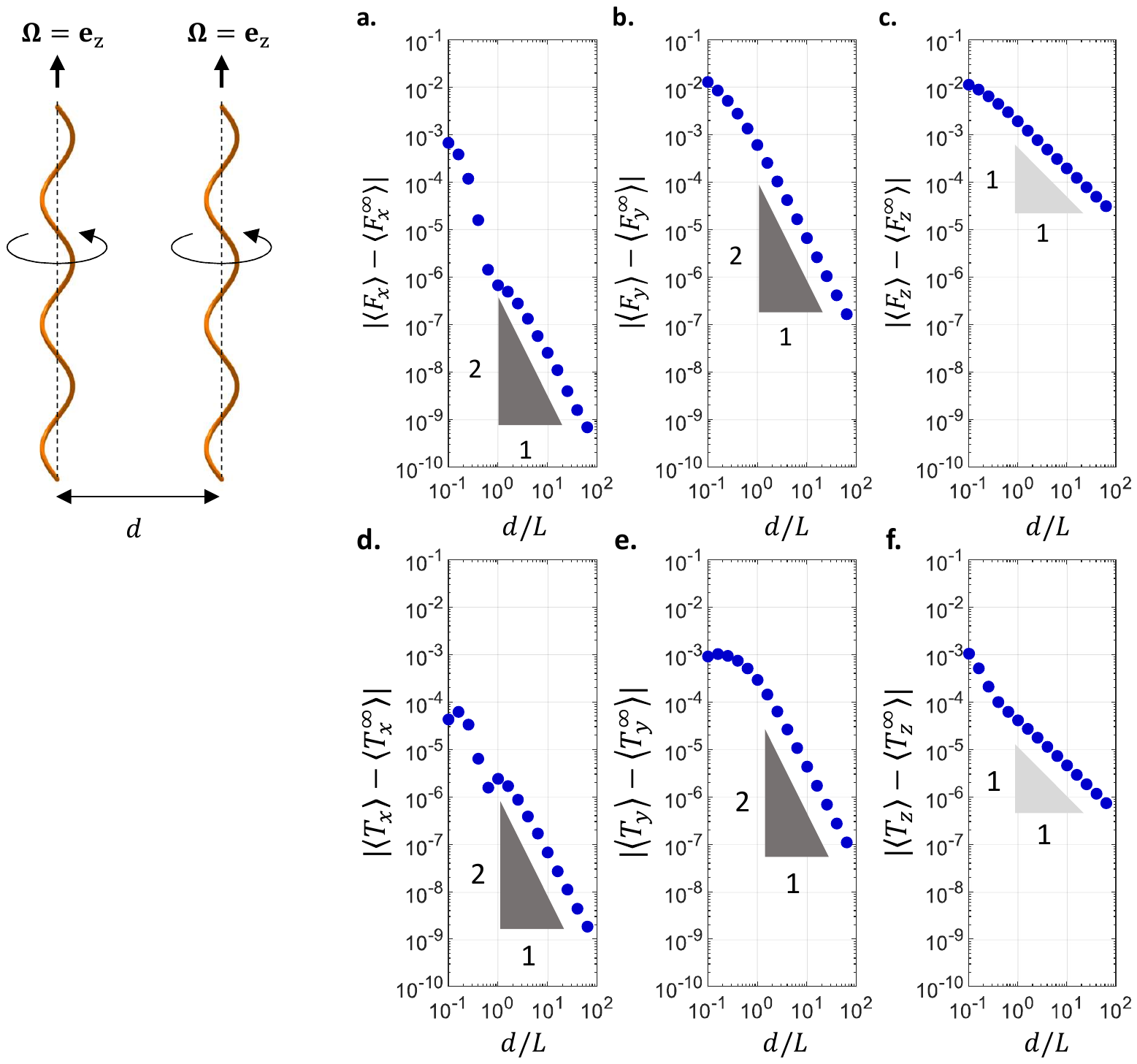}
	\caption{Average forces and torques exerted by the leftmost helix due to the presence of a second parallel helix rotating at a distance $d$ to the right, with fixed phase difference $\Delta\phi = \pi/4$. The data points come from SBT simulations including HIs. The power law triangles indicate that the average forces and torques along the axis of the helix (c,f) are an $\mathcal{O}(d^{-1})$ effect, while the other forces and torques (a,b,d,e) are an $\mathcal{O}(d^{-2})$ effect. Simulation parameters: $\psi = 0.5043$ rad, $\epsilon =0.0038$, $N=2.5$ helical turns.}
	\label{fig:mean_FT}
\end{figure}

We consider two parallel identical helices, rotating with constant angular velocity $\tilde{\Omega}$, as illustrated in Fig.~\ref{fig:mean_FT}. {\color{black} We may choose the laboratory frame so that the filaments are parallel to the $z$-axis and, therefore, the tilt angle $\theta$ is identically zero. When $\theta = 0$, the angles $\phi$ and $\chi$ can be used interchangeably to refer to the rotation of the filament about its own axis, because the body-fixed axis $\eee$ is parallel to $\ez$. Without loss of generality, we describe the configuration of the filaments using the angle $\chi=0$ and a varying phase $\phi$.} Because they are driven at constant angular velocity, the helices maintain a fixed phase difference $\phi_2-\phi_1 = \Delta\phi$. If we rescale time by $\tilde{\Omega}^{-1}$, such that $\Omega=1$ in dimensionless terms, then 
\begin{equation}
\phi_1 = t, \quad \phi_2=t + \Delta\phi.
\end{equation}
Since the helices are held in place, they exert a net force on the fluid, which is pumped in the positive $z$ direction for left-handed helices rotating clockwise. 

To characterise the net long-term effect of the helical pumps, we need to consider the time-averaged forces and torques exerted by the rotating filaments on the fluid, so we define the mean
\begin{equation}
\mean{Y} = \frac{1}{2\pi}\int_{0}^{2\pi}Y(t) \mathrm{d}t,
\end{equation}
for any time-varying quantity $Y$ that we are interested in. We may also want to look at the oscillations of this quantity around its mean value, so we define the variance over time as
\begin{equation}
\var{Y} = \frac{1}{2\pi}\int_{0}^{2\pi}(Y(t)-\mean{Y})^2\mathrm{d}t.
\end{equation}

Because our focus is on the HIs between helical pumps, we need to compare the effect of a helical pump when it is part of an ensemble, to what it otherwise would be if the helical pump was operating on its own. If $Y(t;d)$ is a force or torque exerted by a helical pump when there is second helical pump operating at distance $d$ away, then we define
\begin{equation}
Y_\infty (t) = \lim_{d\to\infty}Y(t;d),
\end{equation}
which is the force or torque that the same helical pump would exert in isolation. For our asymptotic theory, this corresponds to the leading-order terms in Section \ref{sec:leading-order}. For our computational method, this corresponds to the numerical solution of Eq.~\eqref{eq:COMP-method} without the interaction term $\mathcal{J}[\fb_2(s'),\db]$.

In the next sections, we will look at differences of the form $\mean{Y} - \mean{Y_\infty}$ to understand if HIs increase or decrease the net effect of the helical pumps on the fluid, and differences of the form $\var{Y} - \var{Y_\infty}$ to investigate whether HIs make the pumping fluctuate more or less over time. 

\subsection{Computational results}

\begin{figure}
	\landscapetrim{17cm}{15cm}
	\includegraphics[trim={{.5\cutwidth} {.5\cutheight} {.5\cutwidth} {.5\cutheight}},clip,width=17cm]{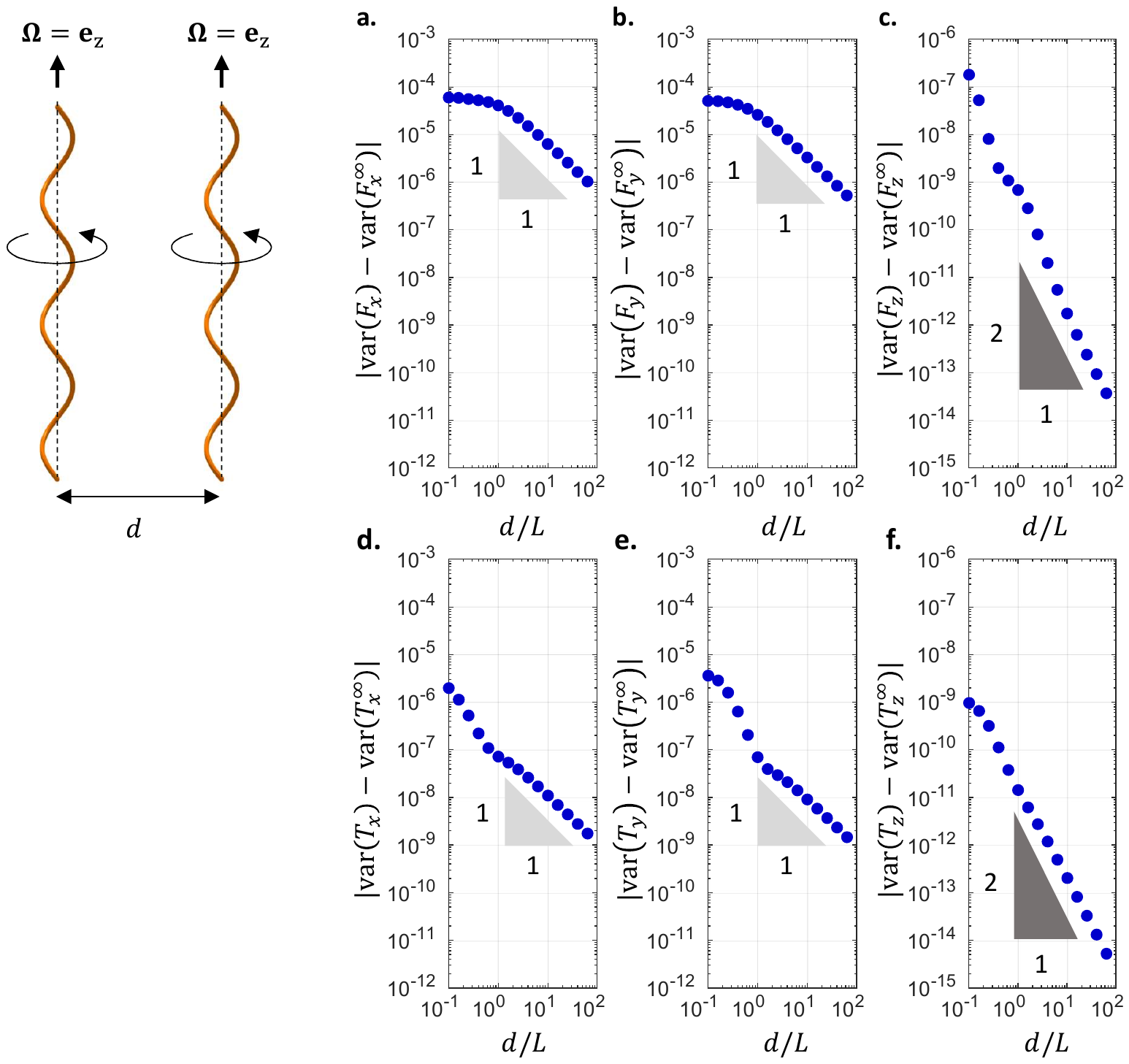}
	\caption{Variance over time in the forces and torques exerted by the leftmost helix due to the presence of a second parallel helix rotating at a distance $d$ to the right, with fixed phase difference $\Delta\phi = \pi/4$. The data points come from SBT simulations including HIs. The power law triangles indicate that the variances in force and torque along the axis of the helix (c,f) are an $\mathcal{O}(d^{-2})$ effect, while the other forces and torques (a,b,d,e) are an $\mathcal{O}(d^{-1})$ effect. Simulation parameters: $\psi = 0.5043$ rad, $\epsilon =0.0038$, $N=2.5$ helical turns.}
	\label{fig:var_FT}
\end{figure}

In our simulations,  we sample the forces and torques exerted by two helical pumps at twelve regular intervals over one period of rotation, i.e.~$0 \leq \Omega t\leq 2\pi$. The time-averaged forces and torques obtained in this way are shown in Fig.~\ref{fig:mean_FT}, while their variances over time are shown in Fig.~\ref{fig:var_FT}, both for a given phase difference $\Delta\phi = \pi/4$ and varying inter-filament distance. The geometry of the helices was chosen to be representative of bacterial flagella: helix angle, $\psi = 0.5043$ rad, filament slenderness, $\epsilon =0.0038$, and $N=2.5$ helical turns. 

We will now seek to interpret the trends observed in these computations using our asymptotic theory. Specifically, we want to understand why the interaction between the filaments alters the time average of $F_z$ and $T_z$ by $\mathcal{O}(d^{-1})$, but their fluctuation over time by $\mathcal{O}(d^{-2})$. Meanwhile, for the forces and torques in the $x$ and $y$ direction, we want to understand why the time average changes by $\mathcal{O}(d^{-2})$ due to inter-filament interaction, but their fluctuation over time changes by $\mathcal{O}(d^{-1})$.

\subsection{Asymptotic theory}
We start by computing the intrinsic resistance matrix $\mathbf{S}^{(0)}(0,0,\phi)$ for a vertical helix with arbitrary phase $\phi$, which we will denote from now on simply as $\mathbf{S}^{(0)}(\phi)$. We need to apply the change of basis from Eqs.~\eqref{eq:result-S0(p1)} with the orthogonal matrix
\begin{equation}
\mathbf{Q}(0,0,\phi) = \begin{pmatrix}
\cos\phi & -\sin\phi & 0 \\
\sin\phi & \cos\phi & 0 \\
0 & 0 & 1
\end{pmatrix}.
\end{equation}
Because the filament is symmetric under a rotation by angle $\pi$ around the first vector ($\e$) in the body frame basis, the resistance matrix expressed in the body frame has the structure
\begin{equation}
\mathbf{S}_0 = \begin{pmatrix}
A_{11} & 0 & 0 & B_{11} & 0 & 0 \\
0 & A_{22} & A_{23} & 0 & B_{22} & B_{23}\\
0 & A_{32} & A_{33} & 0 & B_{32} & B_{33}\\
B_{11} & 0 & 0 & D_{11} & 0 & 0 \\
0 & B_{22} & B_{32} & 0 & D_{22} & D_{23}\\
0 & B_{23} & B_{33} & 0 & D_{32} & D_{33}\\
\end{pmatrix},
\end{equation}
noting that $A_{23} = A_{32}$ and $D_{23} = D_{32}$ because the resistance matrix is symmetric. Hence, after a rotation by angle $\phi$, the matrix can be written as
\begin{equation}
\mathbf{S}^{(0)}(\phi) = \begin{pmatrix}
\mathbf{A}(\phi) & \mathbf{B}(\phi) \\
\mathbf{B}(\phi)^T & \mathbf{D}(\phi) 
\end{pmatrix},
\label{eq:S_structure}
\end{equation}
where the matrices $\mathbf{A}(\phi)$, $\mathbf{B}(\phi)$ and $\mathbf{D}(\phi)$ have the same structure with respect to $\phi$, that is
\begin{equation}
\mathbf{A}(\phi) = \begin{pmatrix}
A_0 + \Delta A \cos(2\phi) & \Delta A \sin(2\phi) & -A_{23}\sin(\phi) \\
\Delta A \sin(2\phi) & A_0 - \Delta A \cos(2\phi) & A_{23}\cos(\phi) \\
-A_{32}\sin(\phi) & A_{32}\cos(\phi) & A_{33}
\end{pmatrix},
\label{eq:Aphi_structure}
\end{equation}
where we define $A_0 = (A_{11} + A_{22})/2$ and $\Delta A = (A_{11}-A_{22})/2$, and similarly for $\mathbf{B}(\phi)$ and $\mathbf{D}(\phi)$ but with $A_{ij} \mapsto B_{ij}$ and $A_{ij} \mapsto D_{ij}$ respectively.  

Without loss of generality, we may choose our laboratory frame to coincide with the interaction frame of the two filaments, so the directed distance between the two helices is $\mathbf{d} = d\ex$. From Eqs.~\eqref{eq:defn-J} and \eqref{eq:result-C1}, we can write
\begin{equation}
C^{(1)}_{ij} (\phi_1,\phi_2) = -\frac{1}{8\pi}\left(2S^{(0)}_{i1}(\phi_1)S^{(0)}_{1j}(\phi_2)+S^{(0)}_{i2}(\phi_1)S^{(0)}_{2j}(\phi_2)+S^{(0)}_{i3}(\phi_1)S^{(0)}_{3j}(\phi_2) \right),
\label{eq:appln-C1}
\end{equation}
and then replace the expressions for the elements of $\mathbf{S}(\phi)$ from Eqs.~\eqref{eq:S_structure}-\eqref{eq:Aphi_structure}. 

Furthermore, from Eq.~\eqref{eq:result-Pmatrix} we derive the matrix 
\begin{equation}
\Pb(\phi) = \begin{pmatrix}
\mathbf{G}(\phi) & \mathbf{H}(\phi) 
\end{pmatrix},
\label{eq:P_structure}
\end{equation}
where the matrices $\mathbf{G}(\phi)$ and $\mathbf{H}(\phi)$ have the same structure with respect to the phase $\phi$. Because $\e = \cos\phi\ex + \sin\phi \ey$, we have
\begin{equation}
\Gb(\phi) = \frac{1}{8\pi}\begin{pmatrix}
\mathcal{M}_1 \cos\phi & \mathcal{M}_1 \sin\phi & 0 \\
B_{23}\sin(\phi) & -B_{23}\cos(\phi) & -B_{33} \\
\Delta B \sin(2\phi) & B_0 - \Delta B \cos(2\phi) & B_{32}\cos(\phi) 
\end{pmatrix}, 
\label{eq:Gphi_structure}
\end{equation}
and similarly for $\mathbf{H}(\phi)$ but with $B_{ij} \mapsto D_{ij}$ and $\mathcal{M}_1 \mapsto \mathcal{M}_4$.

We are now ready to evaluate the mean forces and torques, and their fluctuations over time, for the specific case of constant rotation about the helical axis $\eee = \ez$. The two helical pumps rotate with constant angular velocities $\Omb_1 = \ez$ and $\Omb_2 = \ez$, since $\Omega = 1$ in our chosen units of time. Therefore, the forces and torques exerted by the first filament are
\begin{equation}
\begin{pmatrix}
\Fb_1 \\ \Tb_1
\end{pmatrix}_{\hspace{-.15cm}i} = S^{(0)}_{i6}(t) + \frac{C^{(1)}_{i6}(t,t+\Delta\phi)}{d} + \frac{S^{(2)}_{i6}(t,t+\Delta\phi) + C^{(2)}_{i6}(t,t+\Delta\phi)}{d^2} + \mathcal{O}(d^{-3}),
\label{eq:appln-FT}
\end{equation}
where we have substituted the phases $\phi_1 = t,~\phi_2 = t + \Delta\phi$.

{\color{black}
\subsection{Forces and torques parallel to axis of rotation}}

\begin{figure}
	\landscapetrim{17.25cm}{16cm}
	\includegraphics[trim={{.5\cutwidth} {.5\cutheight} {.5\cutwidth} {.5\cutheight}},clip,width=17cm]{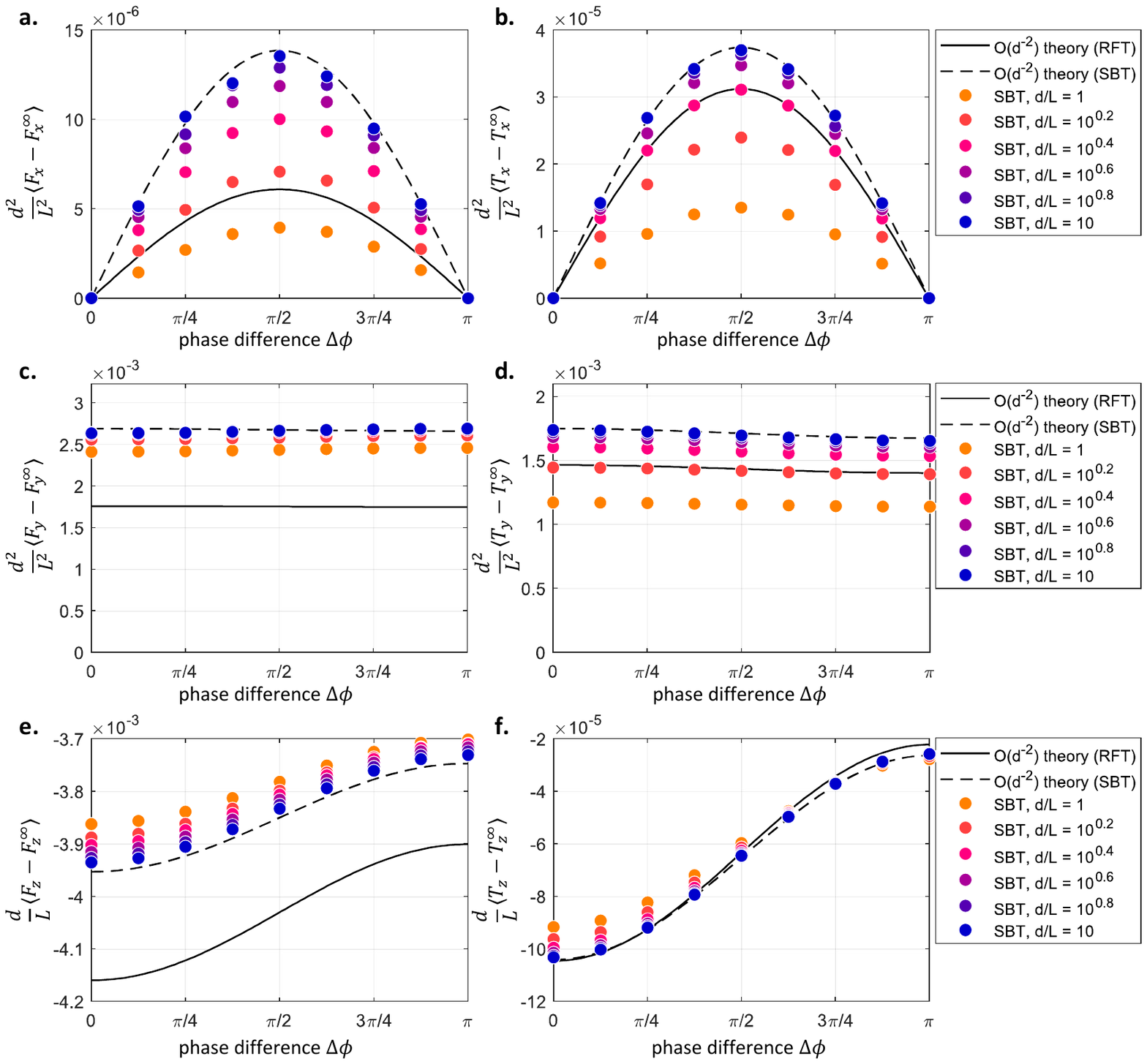}
	\caption{Average forces (a,c,e) and torques (b,d,f) due to HIs between the helices, as a function of the phase difference between filaments. The helix angle, $\psi = 0.5043$ rad, and filament slenderness, $\epsilon =0.0038$, were chosen as representative of bacterial flagella. The helices have $N=2.5$ helical turns.}
	\label{fig:helices_time_average}
\end{figure}

We begin by looking at the force exerted by the leftmost filament along its helical axis, $\eee = \ez$.  From Eqs.~\eqref{eq:S_structure},\eqref{eq:Aphi_structure} and \eqref{eq:appln-FT}, we see that
\begin{equation}
F_z(t) = B_{33}+d^{-1}C^{(1)}_{36}(t,t+\Delta\phi) + \mathcal{O}(d^{-2}),
\end{equation}
which is constant at leading order with $\mean{F_z^\infty} = B_{33}$. The first-order correction, given by Eqs.~\eqref{eq:S_structure},\eqref{eq:Aphi_structure} and \eqref{eq:appln-C1}, will be
\begin{equation}
C^{(1)}_{36} = -\frac{1}{8\pi }\left[A_{33}B_{33}+A_{23}B_{23}\left(2\sin(t)\sin(t+\Delta\phi)+\cos(t)\cos(t+\Delta\phi)\right)\right], \label{eq:C36}
\end{equation} 
which has a non-zero time-average. Hence, the mean thrust provided by the helical pump is
\begin{equation}
\mean{F_z} - \mean{F_z^\infty} = -\frac{1}{8\pi d} \left(A_{33}B_{33}+\frac{3}{2}A_{23}B_{23}\cos(\Delta\phi)\right) + \mathcal{O}(d^{-2}), 
\label{eq:result-meanFz}
\end{equation}
so indeed the interaction between the filaments changes the mean thrust by $\mathcal{O}(d^{-1})$, as seen in the computations. {\color{black} Note that the result in Eq.~\eqref{eq:result-meanFz} is independent of the method (RFT or SBT) by which we choose to evaluate the coefficients $A_{33}, B_{33}, A_{23}$ and $B_{23}$. In Fig.~\ref{fig:helices_time_average} (e), we examine how the $\mathcal{O}(d^{-1})$ change in thrust depends on the phase difference between the filament}s. The {\color{black} asymptotic theory with SBT coefficients} provides perfect quantitative agreement in the limit of large $d$, while the {\color{black} asymptotic theory with RFT coefficients} has an approximate error of 5\% but captures all qualitative features.

\begin{figure}
	\landscapetrim{17.25cm}{16cm}
	\includegraphics[trim={{.5\cutwidth} {.5\cutheight} {.5\cutwidth} {.5\cutheight}},clip,width=17cm]{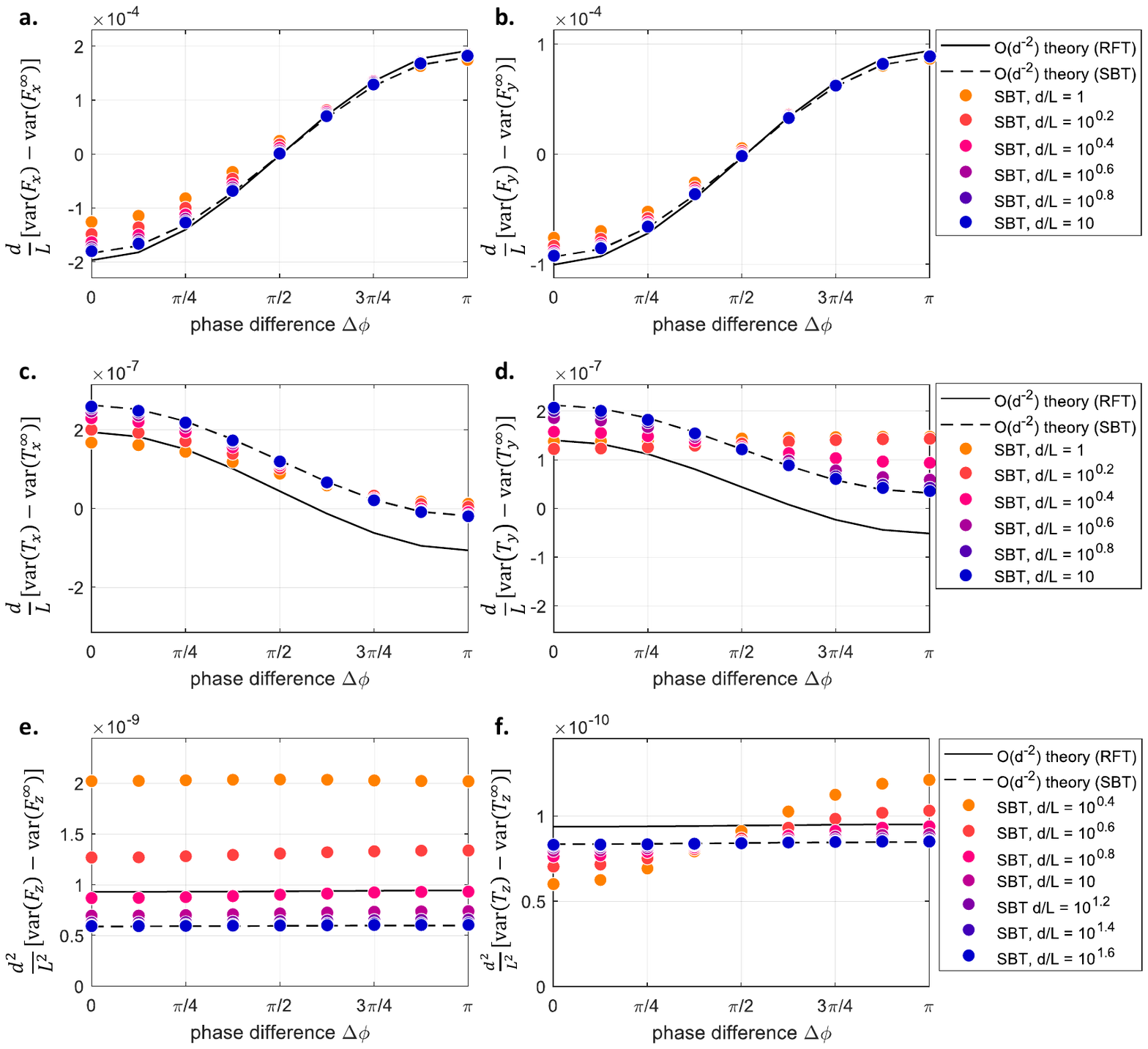}
	\caption{Variance in forces (a,b,e) and torques (c,d,f) due to HIs between the helices, as a function of the phase difference between filaments. The helix angle, $\psi = 0.5043$ rad, and filament slenderness, $\epsilon =0.0038$, were chosen as representative of bacterial flagella. The helices have $N=2.5$ helical turns.}
	\label{fig:helices_variances_over_time}
\end{figure}

Because $F_z$ is constant at leading order, i.e.~$\var{F_z^\infty} = 0$, its variance over time will be given by 
\begin{equation}
\var{F_z} - \var{F_z^\infty} = \frac{1}{ d^2}\left(\mean{C^{(1)}_{36}(t,t+\Delta\phi)^2}  - \mean{C^{(1)}_{36}(t,t+\Delta\phi)}^2 \right) + \mathcal{O}(d^{-3}),
\end{equation}
which is indeed an $\mathcal{O}(d^{-2})$ effect as seen in computations. This is shown in Fig.~\ref{fig:helices_variances_over_time} (e), where we look at how this $\mathcal{O}(d^{-2})$ effect depends on the phase difference between the filaments. Once again, the {\color{black} asymptotic theory with SBT coefficients} provides quantitative agreement, while the {\color{black}theory with RFT coefficients captures the  correct shape and order of magnitude}.

Moving on to the torque exerted by the leftmost filament along its helical axis, we can derive in a similar way expressions for the time-average
\begin{equation}
\mean{T_z} - \mean{T_z^\infty} = -\frac{1}{8\pi d} \left(B_{33}^2+\frac{3}{2}B_{23}^2\cos(\Delta\phi)\right) + \mathcal{O}(d^{-2}), 
\label{eq:result-meanTz}
\end{equation}
and the fluctuation over time 
\begin{equation}
\var{T_z} - \var{T_z^\infty} = \frac{1}{ d^2}\left(\mean{C^{(1)}_{66}(t,t+\Delta\phi)^2}  - \mean{C^{(1)}_{66}(t,t+\Delta\phi)}^2 \right) + \mathcal{O}(d^{-3}),
\label{eq:result-varTz}
\end{equation}
which are compare against computations in Figs.~\ref{fig:helices_time_average} (f) and \ref{fig:helices_variances_over_time} (f), respectively.

\vspace{\baselineskip}
{\color{black}
\subsection{Forces and torques perpendicular to axis of rotation}}

Next, we evaluate the forces and torques perpendicular to the filament axis, starting with $F_x$. From Eqs.~\eqref{eq:S_structure},\eqref{eq:Aphi_structure} and \eqref{eq:appln-FT}, we see that
\begin{multline}
F_x(t) = -B_{23}\sin(t) + d^{-1}C^{(1)}_{16}(t,t+\Delta\phi) + \\ d^{-2}(S^{(2)}_{16}(t,t+\Delta\phi) + C^{(2)}_{16}(t,t+\Delta\phi)) +  \mathcal{O}(d^{-3}),
\end{multline}
which averages out to zero at leading order,i.e.~$\mean{F_x^\infty} = 0$. The first-order correction, 
\begin{multline}
C^{(1)}_{16} = -\frac{1}{8\pi }\left[-A_{23}B_{33}\sin(t) - 2A_0 B_{23}\sin(t+\Delta\phi) \right. \\ - \left. \Delta A B_{23}\left(2\cos(2t)\sin(t+\Delta\phi)-\sin(2t)\cos(t+\Delta\phi)\right) \right], \label{eq:C16}
\end{multline}
also averages out to zero, so the mean of $F_x$ is an $\mathcal{O}(d^{-2})$ effect  as seen in Fig.~\ref{fig:mean_FT} (a). Using Eqs.~\eqref{eq:result-S2},\eqref{eq:S_structure} and \eqref{eq:Aphi_structure}, we obtain that
\begin{equation}
\langle S^{(2)}_{16}(t,t+\Delta\phi) \rangle = 0.
\end{equation}
Then, by using Eqs.~\eqref{eq:result-C2},\eqref{eq:S_structure},\eqref{eq:Aphi_structure},\eqref{eq:P_structure} and \eqref{eq:Gphi_structure}, we get that
\begin{equation}
\langle C^{(2)}_{16}(t,t+\Delta\phi) \rangle = -\frac{1}{16\pi}\left(A_{23}D_{23} + B_{23}^2 + B_{23}\mathcal{M}_1\right)\sin(\Delta\phi),
\label{eq:C2_16}
\end{equation} 
and hence
\begin{equation}
\mean{F_x} - \mean{F_x^\infty} = -\frac{1}{16\pi d^2}\left(A_{23}D_{23} + B_{23}^2 + B_{23}\mathcal{M}_1\right)\sin(\Delta\phi).
\label{eq:result-meanFx}
\end{equation}
Because the time-average of $F_x$ is only $\mathcal{O}(d^{-2})$, we deduce that the variance over time is 
\begin{equation}
\var{F_x} = \mean{(-B_{23}\sin(t)+d^{-1}C^{(1)}_{16}(t,t+\Delta\phi) + \mathcal{O}(d^{-2}))^2}.
\end{equation}
Because $F_x$ oscillates at leading order with variance $\var{F_x^\infty} = A_{23}^2/22$, we deduce that the variance due to HIs is given by
\begin{equation}
\var{F_x} - \var{F_x^\infty}= -\frac{2B_{23}}{d}\mean{\sin(t)C^{(1)}_{16}(t,t+\Delta\phi)} + \mathcal{O}(d^{-2}),
\end{equation}
so indeed an $\mathcal{O}(d^{-1})$ effect as seen in Fig.~\ref{fig:var_FT} (a). Using Eq.~\eqref{eq:C16}, we arrive at the final result
\begin{equation}
\var{F_x} - \var{F_x^\infty}= -\frac{B_{23}}{8\pi d}\left(A_{23}B_{33} + 2A_0 B_{23}\cos(\Delta\phi) + \frac{1}{2} \Delta A B_{23}\cos(\Delta\phi) \right) + \mathcal{O}(d^{-2}).
\label{eq:result-varFx}
\end{equation}
The analytical expressions from Eqs.~\eqref{eq:result-meanFx} and \eqref{eq:result-varFx} are compared against computational results in Fig.~\ref{fig:helices_time_average} (a) and \ref{fig:helices_variances_over_time} (a), respectively. As above, we have quantitative agreement between computations and the {\color{black} asymptotic theory with SBT coefficients} in the limit $d\to\infty$, and qualitative agreement with the {\color{black} asymptotic theory with RFT coefficients}.

Just as we have done for $F_x$, we may compute the time-average of the other transverse forces and torques to  $\mathcal{O}(d^{-2})$,
\begin{eqnarray}
\mean{F_y} - \mean{F_y^\infty} &=& \frac{1}{16\pi d^2}\left(2(A_0 D_{33}+B_0 B_{33}) - (A_{23}D_{23} + B_{23}^2 + B_{23}\mathcal{M}_1)\cos(\Delta\phi)\right),
\label{eq:result-meanFy} \\ 
\mean{T_x} - \mean{T_x^\infty} &=& -\frac{1}{16\pi d^2}\left(B_{23}D_{23} + B_{23}D_{23} + B_{23}\mathcal{M}_4\right)\sin(\Delta\phi),
\label{eq:result-meanTx} \\
\mean{T_y} - \mean{T_y^\infty} &=& \frac{1}{16\pi d^2}\left(2(B_0 D_{33}+D_0 B_{33}) - (B_{23}D_{23} + B_{23}D_{23} + B_{23}\mathcal{M}_4)\cos(\Delta\phi)\right).
\label{eq:result-meanTy}
\end{eqnarray}
Similarly, we can derive the fluctuations over time to  $\mathcal{O}(d^{-1})$,
\begin{eqnarray}
\var{F_y} - \var{F_y^\infty} &=& -\frac{B_{23}}{8\pi d}\left(A_{23}B_{33} + \phantom{2}A_0 B_{23}\cos(\Delta\phi) - \frac{3}{2} \Delta A B_{23}\cos(\Delta\phi)\right),
\label{eq:result-varFy} \\
\var{T_x} - \var{T_x^\infty} &=& -\frac{D_{23}}{8\pi d}\left(B_{32}B_{33} + 2B_0 B_{23}\cos(\Delta\phi) + \frac{1}{2} \Delta B B_{23}\cos(\Delta\phi)\right),
\label{eq:result-varTx} \\
\var{T_y} - \var{T_y^\infty} &=& -\frac{D_{23}}{8\pi d}\left(B_{32}B_{33} + \phantom{2}B_0 B_{23}\cos(\Delta\phi) - \frac{3}{2} \Delta B B_{23}\cos(\Delta\phi)\right).
\label{eq:result-varTy}
\end{eqnarray}
The analytical expressions from Eqs.~\eqref{eq:result-meanFy}-\eqref{eq:result-varTy} are compared against computational results in Fig.~\ref{fig:helices_time_average} (b)-(d) and \ref{fig:helices_variances_over_time} (b)-(d).

\vspace{\baselineskip}
{\color{black}
\subsection{Deducing the dynamics of the second filament}}
\label{sec:deducing-second}

We remind the reader that the forces and torques plotted in Fig.~\ref{fig:helices_time_average} are those exerted \textit{on} the fluid \textit{by} the leftmost filament - see Fig.~\ref{fig:interpretation} (a). Relative to this, the rightmost filament is in the positive $x$ direction, and accordingly we have taken $\dhatb=\ex$ in our calculation of second-order corrections from Eqs.~\eqref{eq:result-meanFx}, \eqref{eq:result-meanFy}-\eqref{eq:result-meanTy}. To obtain the forces and torques exerted by the rightmost filament, we can rotate our coordinate system by an angle $\pi$ about the $z$-axis. First of all, this swaps the filaments around and, hence, reverses the sign of the phase difference. It also changes the signs of all $x$ and $y$ components, but not the $z$ components. {\color{black} Hence, the average dynamics of the second filament satisfy the relations $-\Gamma^{(2)}_{x,y}(\Delta\phi) = \Gamma^{(1)}_{x,y}(-\Delta\phi)$ and  $ \Gamma^{(2)}_{z}(\Delta\phi) = \Gamma^{(1)}_{z}(-\Delta\phi)$, where $\Gamma^{(k)}$ is a placeholder for the time-averaged force or torque exerted by the $k$th filament on the fluid.}

Because $\langle F_x \rangle$ and $\langle T_x \rangle$ depend on the sine of the phase difference (see Eqs.~\eqref{eq:result-meanFx} and \eqref{eq:result-meanTx}), the rightmost helix exerts the same average force $\langle F_x \rangle$ and torque $\langle T_x \rangle$ as the leftmost helix. Meanwhile, for $\langle F_y \rangle$ and $\langle T_y \rangle$, which depend on the cosine of the phase difference (see Eqs.~\eqref{eq:result-meanFy} and \eqref{eq:result-meanTy}), the rightmost helix exerts an equal and opposite average force and torque to the leftmost helix. Finally, the average $\langle F_z \rangle$ and $\langle T_z \rangle$ are the same for the two helices, because the two quantities depend on the cosine of the phase difference (see Eqs.~\eqref{eq:result-meanFz} and \eqref{eq:result-meanTz}), and the sign of $z$ components has not changed {\color{black} due to the rotation}.

\vspace{\baselineskip}
\subsection{Interpretation of results}
\label{sec:application-interpretation}
{\color{black}

We now provide some physical interpretation for the earlier computational results.

\subsubsection*{Deficit in pumping force}}

\begin{figure}
	\landscapetrim{17cm}{13cm}
	\includegraphics[trim={{.5\cutwidth} {.5\cutheight} {.5\cutwidth} {.5\cutheight}},clip,width=17cm]{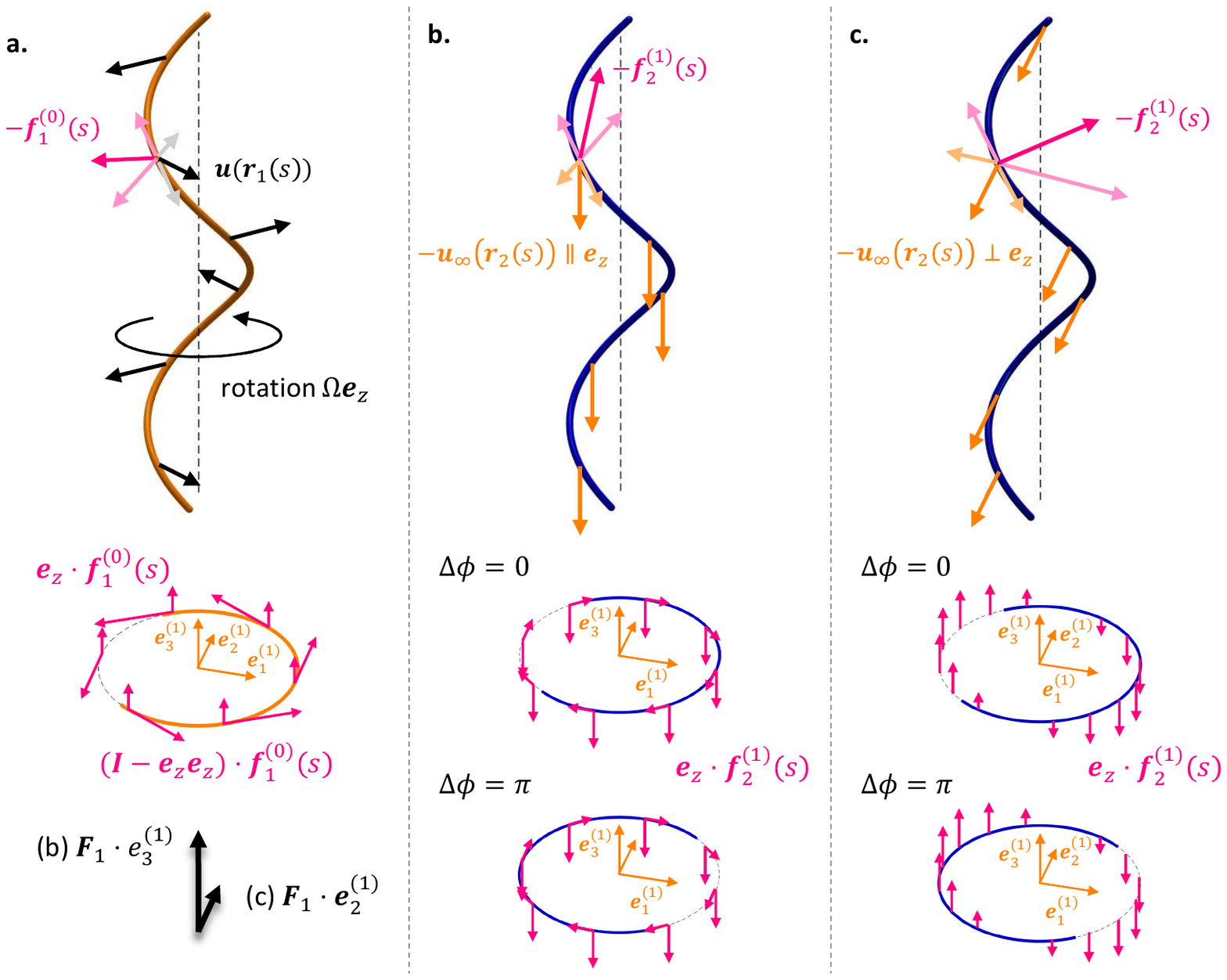}
	\caption{\color{black} (Not to scale) Physical mechanism for the reduction in pumping force due to HIs. Top panels illustrate the local velocity of the filament relative to the surrounding fluid. Lower panels show the periodic force density along the filament, rendered at points along a horizontal projection of the centreline. The total force and torque exerted by the helical pump are obtained by integrating the force density around the circle as many times as needed.  (a) Due to the anisotropic drag on the slender filament, a rotating helix exerts a net force along its axis of rotation, $\eee^{(1)}$. If the helix does not have an integer number of turns, there is also a net component of the force along the $\ee^{(1)}$ direction, due to a ``surplus" of filament on one side (indicated by a thick orange arc on the circular projection of the centreline). (b) Changes to the force density along the second filament due to the $\eee^{(1)}$ component of the force exerted by the first filament on the fluid. (c) Likewise for the $\ee^{(1)}$ component of the force.}
	\label{fig:physicalmechanism}
\end{figure}

{\color{black} Since the main purpose of the helical pumps is to push fluid along their axes, we start by explaining how HIs affect the vertical \color{black} pumping force, $\langle F_z \rangle$. The leading-order dynamics of a rotating helical pump are illustrated in Fig.~\ref{fig:physicalmechanism} (a) using a local description of the problem (i.e.~no end effects). The local velocity of the centreline relative to the fluid is shown at various points along the filament. At one of these points we decompose the velocity into the directions tangent and perpendicular to the filament. Because the perpendicular drag coefficient on a slender rod is higher, by roughly a factor of two, than the parallel drag coefficient, this gives rise to a leading-order viscous drag on the filament, $-\fb_1^{(0)}(s)$, that has a negative vertical component. Below the three-dimensional picture of the filament, we draw the projection of the filament centreline onto the horizontal plane. At each point on this circular projection, we show the corresponding force density exerted by the filament on the fluid, $\fb_1^{(0)}(s)$, decomposed into vertical and horizontal components. Notice that the force density simply rotates about the axis $\eee^{(1)}=\ez$ as we rotate around the circle, due to the rotational symmetry of the system. The total force and torque exerted by the helical pump are obtained by integrating the force density along the entirety the filament, or equivalently by integrating around the circular projection as many times as needed. For left-handed helices rotating counter-clockwise, the vertical components of the force density are positive, so the helical pump exerts a net positive force in the $\eee^{(1)}$ direction. The fluid is pumped vertically upwards. By integrating the horizontal components of the force density, we also obtain a net counter-clockwise torque that must be applied to the helical pump to keep it rotating. Furthermore, if the helical filament does not have an integer number of turns, there will be a surplus of filament on one side, indicated by a thick orange line on the circular projection. This means that the helical pump also exerts a net horizontal force on the fluid along the $\ee^{(1)}$ direction. 

In Fig.~\ref{fig:physicalmechanism} (b) and (c) we explain how the $\eee^{(1)}$ and the $\ee^{(1)}$ components of the leading-order force exerted by the first helical pump, respectively, affect the pumping force exerted by the second helical pump. Firstly, the $\eee^{(1)}$ component of the pumping force exerted by one helical pump on the fluid leads to an upward vertical flow at the position of the other helical pump. This flow is uniform to leading-order in the distance between the filaments. Therefore, the second filament appears to be moving in the negative vertical direction relative to the fluid, with velocity $-\ub_\infty(\rb_2(s))$, as indicated at various points along the filament in Fig.~\ref{fig:physicalmechanism} (b). Following the same procedure as above, we can determine the local force density along the second filament and depict it along the horizontal projection of the centreline. The first-order change in the force density, $\fb_2^{(1)}(s)$, has negative vertical components, because the second filament appears to be moving downward with respect to the background flow. When integrated along the filament, this leads to a deficit in pumping force due to the HIs between the helical pumps. This is confirmed by the negative sign in Fig.~\ref{fig:helices_time_average} (e). Note that this effect is independent of the phase difference between the filaments, because the force density has a constant vertical component along the entire filament, due to rotational symmetry. By integrating the horizontal components of the force density, we also deduce that HIs lead to a deficit in the torque exerted by the helical pumps, as seen in Fig.~\ref{fig:helices_time_average} (f) as well. Hence, less power is needed to actuate two helical pumps with the same angular velocity, if they are rotating in parallel.

Secondly, the $\ee^{(1)}$ component of the leading-order force exerted by the first helical pump generates a horizontal flow at the position of the second helical pump, which is again depicted at various points along the filament in Fig.~\ref{fig:physicalmechanism} (c). Because the flow is horizontal, we no longer have rotational symmetry so the force density is variable along the filament. Note that we only depict the vertical components of the force in the lower panels of Fig.~\ref{fig:physicalmechanism} (c), to avoid overcrowding the diagram. Unlike Figs.~\ref{fig:physicalmechanism} (a) and (b), where the force density simply rotates around the vertical axis as we go around the centreline, in Fig.~\ref{fig:physicalmechanism} (c) we observe that the vertical component of the force density depends on the alignment of the tangent vector and the direction of the flow. Where the velocity of the filament relative to the background flow, $-\ub_\infty(\rb_2(s))$, has a positive (or negative) component in the direction of the local tangent, the force density has a positive (or negative) vertical component. Hence, this particular contribution of HIs to the pumping force will depend on the phase difference between the two helical pumps. If the two are in-phase, $\Delta\phi =0$ and $\ee^{(2)} = \ee^{(1)}$, there is a surplus of negative vertical force as we integrate along the centreline.  If the pumps are anti-phase, $\Delta\phi =\pi$ and $\ee^{(2)} = -\ee^{(1)}$, there is a surplus of positive vertical force instead. This dependence on the phase difference is confirmed by Fig.~\ref{fig:helices_time_average} (e), where the deficit in pumping force is greater when the filaments are in-phase than anti-phase.

It is important to emphasise that the dominant effect here comes from the flow discussed in Fig.~\ref{fig:physicalmechanism} (b), which is a result of integrating a constant force along the entire length of the filament. The effect described in Fig.~\ref{fig:physicalmechanism} (c) is a correction that comes from integrating forces along just a fraction of the filament, if the helix deviates from an integer number of turns. Regardless of the phase difference between the helical pumps, each of them will pump fluid with less force when they are interacting, because each filament tries to push fluid that has already been entrained by the other pump. The deficit is greatest when the filaments are in-phase, because they entrain the fluid in the same direction both vertically and horizontally, whereas filaments that are anti-phase will work against each other in the horizontal plane (Fig.~\ref{fig:physicalmechanism} (c)).}

{\color{black}
\subsubsection*{Fluctuations over time}}

Another question to consider is whether HIs dampen or enhance fluctuations in the dynamics of the helical pumps. The results in Fig.~\ref{fig:helices_variances_over_time} suggest that HIs tend to increase the variances over time for most forces and torques. The only exceptions we observe, for this set of parameters, are the forces $F_x$ and $F_y$ when $|\Delta\phi|<\pi/2$ and the torque $T_x$ in a small interval around $\Delta\phi=\pi$. 

{\color{black}
\subsubsection*{Attraction vs.~repulsion}}

We have so far considered the average forces and torques exerted by the filaments on the fluid while they are held in place, except for rotating about the vertical axis. It is also important to consider what would happen to the helices if they were not held in place, but free to move in response to the forces and torques exerted on them by the fluid. Note that the time averages we previously computed assumed that the helices remain vertical. However, we may still use these results to get a sense for what happens in the early stages, when the axes of the helices are still close to vertical. 

In Fig.~\ref{fig:interpretation} (b) and (c) we show the horizontal components of the average force exerted by the fluid on {\color{black} two left-handed filaments rotating counter-clockwise. The relative directions of the forces and torques on the two helices were established in Section \hyperref[sec:deducing-second]{IV F}}. The first observation is that, at second order, there is no net attraction or repulsion between the helices.  Previous theoretical work had ruled out the possibility of attraction or repulsion between two helices rotating with zero phase difference, based on symmetry arguments \cite{Kim2004b}. Our findings add to that observation by excluding any net attraction or repulsion between helices rotating with any phase difference, so long as they are parallel. Instead, we discover a net migration to one side, because the two filaments experience the same force along the $x$ direction -- Fig.~\ref{fig:interpretation} (b). The direction of migration depends on the sine of the phase difference, so it is not a consistent behaviour. On the other hand, the helices will be swirled around by the fluid in the counter-clockwise direction, because they experience equal and opposite forces along the $y$ direction -- Fig.~\ref{fig:interpretation} (c). The direction of the swirl is consistent with the individual rotation of the helices, and this effect is persistent across all phase differences, as demonstrated by Fig.~\ref{fig:helices_time_average} (c).

Note from Fig.~\ref{fig:helices_time_average} (a)-(d) that the sign of $\langle T_x \rangle$ is the same as $\langle F_x \rangle$, likewise for $\langle T_y \rangle$ and $\langle F_y \rangle$. Hence, the arrows in Fig.~\ref{fig:interpretation} (b) and (c) could equally well represent the horizontal components of the torques exerted by the fluid on the filaments. The key observation here is that, due to equal and opposite average torques along $y$, the helices would initially experience a splaying out effect where the fluid pushes the tips of the helical pumps apart (the tips being {\color{black} the ends pointing in the same direction as the angular velocity}) and brings their bases together.

\subsection{Outlook: circular array of helical pumps}

\begin{figure}
	\landscapetrim{17cm}{8cm}
	\includegraphics[trim={{.5\cutwidth} {.5\cutheight} {.5\cutwidth} {.5\cutheight}},clip,width=17cm]{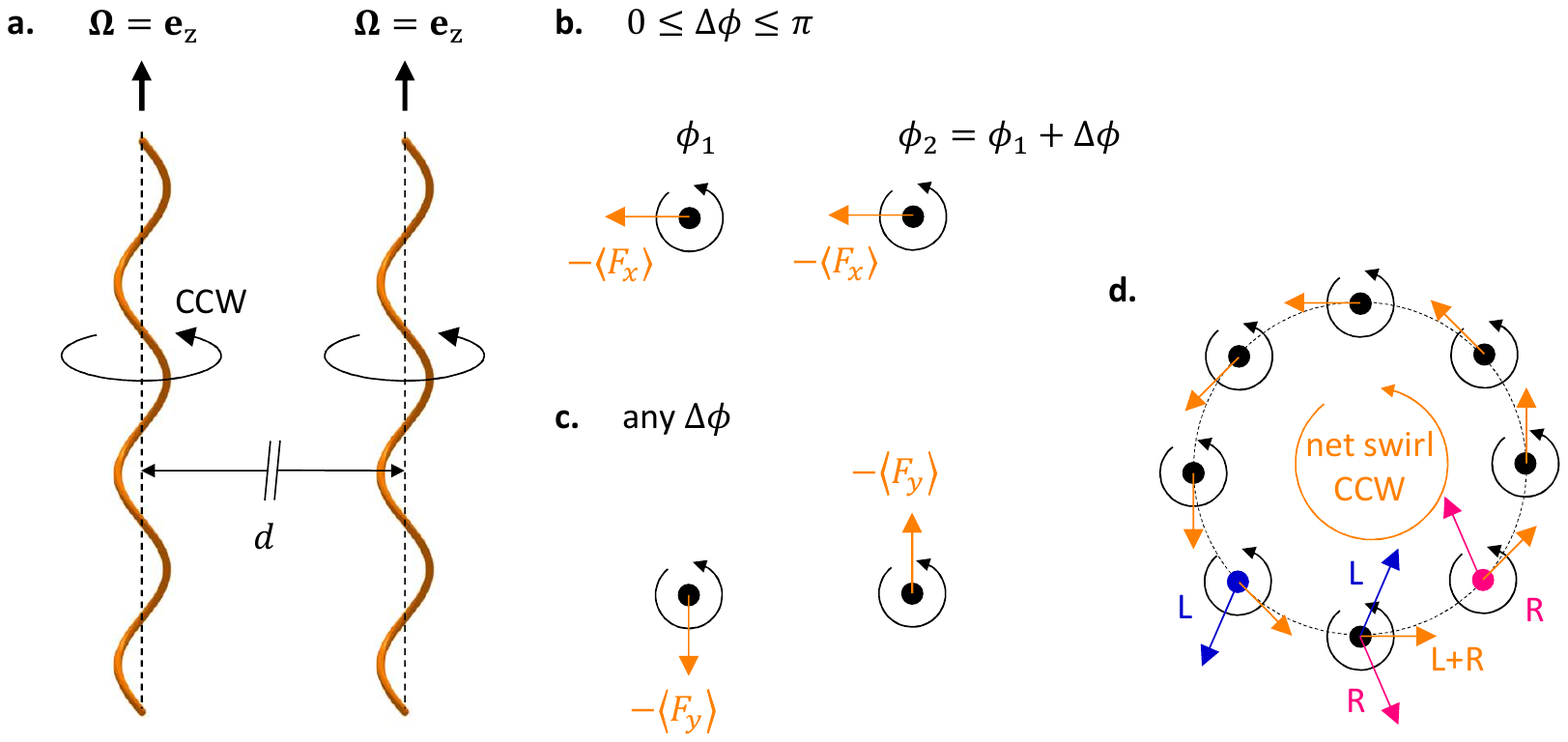}
	\caption{Basic principles of HIs between helical pumps. (a) Minimal setup with two helical pumps rotating with constant angular velocity around their axes. (b) There is no net attraction or repulsion between the two rotating helices  (cf.~symmetry arguments for zero phase difference in Ref.~\cite{Kim2004b}), but rather a sideways migration whose sign depends on the phase difference. (c) There is a persistent (i.e.~independent of phase difference) swirling effect in the same direction as the rotation of the helices. (d) A ring of helical pumps would initially experience counter-clockwise swirling (due to the forces $-\langle F_y \rangle$ exerted by the fluid) and outward splaying of the tips (due to the torques $-\langle T_y \rangle$ exerted by the fluid).}
	\label{fig:interpretation}
\end{figure}

Once we understand the basic principles of pairwise HIs between helical pumps, it is natural to consider ensembles with more than two helical pumps. The simplest example is a ring of regularly spaced helical pumps, illustrated from the top in Fig.~\ref{fig:interpretation} (d). For simplicity, let us consider a ring of sufficiently large radius that the dominant HIs come from the nearest neighbours only. We expect the dominant contribution to the horizontal force to come from  $\langle F_y \rangle$, which is two orders of magnitude larger than $\langle F_x \rangle$ -- cf.~Fig~\ref{fig:helices_time_average} (a) and (c). The effects of $\langle F_y \rangle$ are also consistent, compared to $\langle F_x \rangle$ which depends strongly on the phase difference. In conclusion, we need to focus on the force components perpendicular to the distance between nearest neighbours, depicted in Fig.~\ref{fig:interpretation} (c). 

By adding the contributions from the left nearest neighbour (L) and the right nearest neighbour (R), we find that the net effect is a force along the circumference of the ring. Therefore, the ring of helical pumps experiences a tendency towards counter-clockwise swirling about the centre. If instead of forces we consider the torques $\langle T_y \rangle$, which are likewise dominant over $\langle T_x \rangle$, we find once again that there is a net torque along the circumference of the circle. This means that the tips of the helical pumps have a tendency to spread out and away from the centre of the ring. Note that the sign of these two hydrodynamic effects (swirling and splaying) would stay the same if we include more than nearest neighbour interactions, due to the symmetry of the system. 

\section{Discussion}
\label{sec:discussion}

In this paper, we have considered the problem of HIs between slender filaments in viscous fluids. We have approached the topic theoretically, focusing on the case of two interacting rigid filaments whose dynamics can be described by an extended resistance matrix, Eq.~\eqref{eq:defn-resistance-matrix}. We have solved for the extended resistance matrix and the force distribution along two arbitrarily-shaped filaments as series expansions in inverse powers of the distance between the filaments, up to second-order corrections. Our asymptotic results from Section \ref{sec:model} are valid {\color{black}in the limit of small aspect ratio, $\epsilon\ll 1$, and in the regime, $d>L$, where the inter-filament separation is greater than the contour length of the filament.} {\color{black}Although HIs decrease in magnitude with increasing distance between the filaments, they continue to play a leading-order role important to physical mechanisms such as synchronisation and self-organisation. This provides a strong motivation for developing an analytical theory of HIs to advance our fundamental understanding of such phenomena. While other studies have dealt with the limit $d\ll L$, here we have chosen to focus on the regime $d>L$, which can provide just as many valuable physical insights.}

{\color{black} We have evaluated the coefficients in the asymptotic series expansion using both resistive-force theory (RFT) and slender-body theory (SBT), and validated our asymptotic theory against numerical simulations in Section \ref{sec:validation}.} In the final part, Section \ref{sec:application}, motivated by bacterial microfluidic pumps \cite{Darnton2004,Kim2008,Martindale2017,Dauparas2018}, we have demonstrated the usefulness of our asymptotic theory by applying it to the interaction of two rotating helical pumps. Here, we have identified the dependence of forces and torques on the distance and phase difference between the helices, which is illustrated in Figs.~\ref{fig:helices_time_average} and \ref{fig:helices_variances_over_time} and made explicit in Eqs.~\eqref{eq:result-meanFz}-\eqref{eq:result-varTz}, \eqref{eq:result-meanFx}, \eqref{eq:result-varFx}-\eqref{eq:result-varTy}. The analytical expressions are also implicitly dependent on the helix geometry through the components $A_{ij}, B_{ij}, D_{ij}$ of the single-helix resistance matrix, which are given in Appendix \ref{app:RFT}, and the force moments $\mathcal{M}_i$ from Appendix \ref{app:forcemoments_RFT}. 

Our theory provides us with new physical understanding of the HIs between helical pumps. We find that the pumping force exerted by each rotating helix is reduced due to HIs, and the reduction is greatest when the helical pumps are rotating in phase with each other. Similarly, the torque required to rotate the two helical pumps is lowest when they are in-phase and greatest when they are antiphase, as the helices are working against each other in the latter case. Because we include second-order corrections in our calculation of the average forces and torques acting on the helical pumps, we are able to determine that there is no net attraction or repulsion between the filaments, but rather a sideways migration whose sign depends on the phase difference. However, we identify two persistent hydrodynamic effects which are independent of the phase difference: a swirl in the direction of rotation of the helices and a splaying out at the tips of the helical pumps (i.e.~the ends pointing in the same direction as the angular velocity). We believe that these effects are consistent with the behaviour observed by Kim and co-authors {\color{black}in the initial stage (i.e. when the filaments are still nearly parallel) of their} macroscopic-scale experiments of flagellar bundling \cite{Kim2003}, despite the fact that our theory is intended for \textcolor{black}{$d > L$} while the experiments were carried out in the \textcolor{black}{$d < L$} regime. {\color{black} This suggests that there may be fundamental similarities in the HIs between helical filaments across different regimes of separation. Without further investigation, it is not possible to quantify in which ways the HIs between bacterial flagella within a bundle ($d<L$) are qualitatively different from the HIs between flagellar filaments that are further apart ($d>L$). Our theory provides a starting point to investigate these questions further, analytically.}

{\color{black} The primary purpose of our asymptotic theory is to provide a method to calculate, analytically, the specific HIs between two rigid filaments, as opposed to previous theoretical studies which focus on the bulk properties of suspensions of fibers \cite{Shaqfeh1990,Mackaplow1996}. The asymptotic theory with RFT coefficients is suitable for this purpose, since all the coefficients have closed-form solutions provided in Appendices \ref{app:RFT} and \ref{app:forcemoments_RFT}. The asymptotic theory with SBT coefficients can provide a quantitative improvement on some of these results, since SBT calculates the force density along the filament with algebraic accuracy, but the ultimate goal of the asymptotic theory is to capture the qualitative features of HIs such as the dependence on filament geometry and relative configuration. 

A secondary use of the asymptotic theory could be to speed up the simulation of long time-evolution problems governed by HIs or, in special cases, to provide a way to integrate the equations of motion by hand. The reduction in computation time would come from removing the need to recompute the interaction term $\mathcal{J}$ (see Section \ref{sec:comp-method}) at each time step, as the relative orientation of the two filaments changes. Our asymptotic series expansion provides expression for the HIs between filaments in terms of the resistance matrix of a single filament, which can be precomputed (either by evaluating the analytical expressions from RFT, or by numerically solving the integral equations of SBT for a single filament) and updated at each time step using a rigid-body rotation to reflect changes in filament orientation. This relies on the filaments being rigid so that the shape of their centreline does not change over time. {\color{black} However, we reiterate that the main purpose of our asymptotic theory is to provide a way to evaluate the HIs between filaments analytically, and not to challenge well-established computational methods.} For the simulation of flexible fibers, there exist specialised computational methods that can handle large numbers of filaments with HIs efficiently \cite{Tornberg2004,Maxian2021}.}

One advantage of the current asymptotic theory is the compactness of the final results in Eqs.~\eqref{eq:result-C1}, \eqref{eq:result-S2}, and \eqref{eq:result-C2}, which means they can be used to develop analytical models for certain hydrodynamic phenomena that have only been studied computationally until now. Another advantage is that the results of Eqs.~\eqref{eq:result-C1}, \eqref{eq:result-S2}, and \eqref{eq:result-C2} are valid for arbitrary filament shapes, in contrast to other theories of HIs which require a small-amplitude assumption for the shape of the filament. 

However, no theory is without its limitations. {\color{black} One important restriction is that, within the current setup, our asymptotic theory can only handle filaments in an infinite fluid domain. Further work would be needed to account for external surface such as the cell body of the organism to which the filaments might be attached.} {\color{black} Just as important is the fact that our asymptotic theory, in its current state,} can only fully describe the interaction of rigid filaments. A possible extension is to refine the series expansions for the force distributions from Eqs.~\eqref{eq:expansion-f1} and \eqref{eq:expansion-f2}, which are valid for any type of filament, in order to obtain a comprehensive theory for HIs between flexible filaments as well. We also note that we have neglected HIs due to moment distributions along the centrelines of the filaments. This is because such contributions would scale like $\epsilon^2/d^2$ and would always be smaller than the second-order corrections from the force distributions, which scale like $\log(\epsilon)/d^2$ and are the final terms included in our asymptotic theory. 

We have also considered the interactions between multiple slender filaments but only in a qualitative way, when discussing the physics of HIs in a circular array of helical pumps. Our asymptotic theory can be easily extended to include HIs between more than two filaments, because it is based on the method of reflections. With this approach, $j$th-order corrections to the extended resistance matrix come from hydrodynamic effects that have reflected $j$ times between the filament that induces the flow and the filament that feels its effect. The only complication comes from the fact that, in a collection of $N>2$ filaments, there is no single expansion parameter. Instead, there are $\frac{1}{2}N(N-1)$ pairwise distances between the filaments. Hence, the order in which corrections appear in the series expansion must be considered carefully, {\color{black}unless the filaments are so far apart that it is sufficient to consider first-order corrections due to pairwise interactions}.

There are many possible applications for the theoretical results presented in this paper, beyond the case of helical pumps discussed in Section \ref{sec:application}. Our asymptotic theory can be used to investigate the collective swimming of elongated microorganisms like the \textit{Spirochaetes} and \textit{Spiroplasma}, as well as some artificial micro-swimmers (e.g.~helical micromachines actuated by an external magnetic field). {\color{black} Amongst all moving appendages in the microscopic world, the closest to being rigid are the bacterial flagellum and nodal cilia, which makes them more suitable for applications of our asymptotic theory. Although the distance between flagellar filaments within a bundle is less than their contour length, there are other situations in which bacterial flagella interact on a larger length scale, making these problems directly relevant to our asymptotic theory. Examples include the HIs between filaments at either pole of an amphitrichous bacterium or filaments belonging to different cells in a sparse bacterial carpet or swarm.} Following an extension of our theory to the case of flexible filaments, as discussed before, one could also examine the HIs between eukaryotic cilia and flagella, or between fluctuating polymeric filaments in the cytoplasm, such as actin filaments and microtubules. Another, more technical, avenue for future research will be to bridge the gap between near-field {\color{black}($d\ll L$)} theories of HIs \cite{Man2016} {\color{black} and the present study ($d > L$)}.

\section*{Acknowledgements} 

We gratefully acknowledge funding from the George and Lillian Schiff Fund through the University of Cambridge (studentship supporting M.T.C.) and the European Research Council under the European Union's Horizon 2020 research and innovation programme (grant agreement 682754 to E.L.).

\appendix
\section{Calculating the leading-order resistance matrix from RFT}
\label{app:RFT}
We calculate the leading-order resistance matrix from Eq.~\eqref{eq:result-S0(p1)} using the resistive-force theory (RFT) representation of the force density from Eq.~\eqref{eq:defn-Sigma}. In the body frame of the filament, i.e.~relative to basis vectors $\{\e,\ee,\eee\}$, the local resistance tensor $\Sigb(s)$ (defined in Eq.~\eqref{eq:defn-Sigma}) can be written as
\begin{eqnarray}
\mathbf{\Sigma}_{11} &=& c_\perp + (c_\parallel-c_\perp)\sin^2\psi\sin^2(\pi Ns), \label{eq:Sigma-A}\\
\mathbf{\Sigma}_{22} &=& c_\perp + (c_\parallel-c_\perp)\sin^2\psi\cos^2(\pi Ns), \\
\mathbf{\Sigma}_{33} &=& c_\perp + (c_\parallel-c_\perp)\cos^2\psi, \\
\mathbf{\Sigma}_{12} &=& -\sigma(c_\parallel-c_\perp)\sin^2\psi\sin(\pi Ns)\cos(\pi Ns) = \mathbf{\Sigma}_{21}, \\ 
\mathbf{\Sigma}_{13} &=& -(c_\parallel-c_\perp)\sin\psi\cos\psi\sin(\pi Ns) = \mathbf{\Sigma}_{31}, \\
\mathbf{\Sigma}_{23} &=& \sigma(c_\parallel-c_\perp)\sin\psi\cos\psi\cos(\pi Ns) = \mathbf{\Sigma}_{32}, \label{eq:Sigma-Z}
\end{eqnarray}
where we have used the components of the tangent vector from Eq.~\eqref{eq:tangent}.

From Eqs.~\eqref{eq:result-S0} and \eqref{eq:defn-S0} we write down an equivalent definition of matrix $\mathbf{A}$ as 
\begin{equation}
A_{ij} = \int_{-1}^{+1} \Sigma_{ij}(s) \mathrm{d}s. \label{eq:Aij_defn}
\end{equation}
Using the integrals in Appendix \ref{app:useful-integrals}, Eqs.~\eqref{eq:app:cos},\eqref{eq:app:sin2},\eqref{eq:app:cos2} and \eqref{eq:app:odd}, we determine the components of matrix $\mathbf{A}$, which describes the force exerted by a translating filament,
\begin{eqnarray}
A_{11} &=& \sint{\Sigma_{11}} = 2c_\perp + (c_\parallel-c_\perp)\sin^2\psi\left(1-\frac{\sin(2\pi N)}{2\pi N}\right), \label{eq:Acomponents-A}\\ 
A_{22} &=& \sint{\Sigma_{22}} = 2c_\perp + (c_\parallel-c_\perp)\sin^2\psi\left(1+\frac{\sin(2\pi N)}{2\pi N}\right), \\
A_{33} &=& \sint{\Sigma_{33}} = 2(\cos^2\psi c_\parallel + \sin^2\psi c_\perp), \label{eq:A33}\\
A_{12} &=& \sint{\Sigma_{12}} = 0 = A_{21}, \\
A_{13} &=& \sint{\Sigma_{13}} = 0 = A_{31}, \\
A_{23} &=& \sint{\Sigma_{23}} = \sigma(c_\parallel-c_\perp)\sin\psi\cos\psi \frac{2\sin(\pi N)}{\pi N} = A_{32}. \label{eq:A23}
\end{eqnarray}

Similarly, from Eqs.~\eqref{eq:result-S0} and \eqref{eq:defn-S0} we write down an equivalent definition of matrix $\mathbf{B}$ as 
\begin{equation}
B_{ij} = \int_{-1}^{+1} \varepsilon_{jkl} r_k(s) \Sigma_{il}(s) \mathrm{d}s. \label{eq:Bij_defn}
\end{equation}
Using the integrals from Appendix \ref{app:useful-integrals}, Eqs.~\eqref{eq:app:cos},\eqref{eq:app:sin2},\eqref{eq:app:cos2},\eqref{eq:app:ssin},\eqref{eq:app:ssincos} and \eqref{eq:app:odd}, we determine the components of matrix $\mathbf{B}$, which describes the force exerted by a rotating filament (or, alternatively, the torque exerted by a translating filament)
\begin{multline}
B_{11} = \sint{r_2 \Sigma_{13}-r_3 \Sigma_{12}} \\
= -\sigma R(c_\parallel-c_\perp)\sin\psi\cos\psi\sint{\sin^2(\pi N s)} 
+
\sigma (c_\parallel-c_\perp)\sin^2\psi\cos\psi\sint{s\sin(\pi Ns)\cos(\pi Ns)} \\
= \sigma (c_\parallel-c_\perp)\sin^2\psi\cos\psi\left(-\frac{1}{\pi N}-\frac{\cos(2\pi N)}{2\pi N}+\frac{3 \sin(2 \pi N)}{(2\pi N)^2}\right), \label{eq:B11}
\end{multline}
\begin{multline}
B_{12} = \sint{r_3 \Sigma_{11}-r_1 \Sigma_{13}} \\
= (c_\parallel-c_\perp)\sin^2\psi\cos\psi\sint{s\sin^2(\pi Ns)} + R(c_\parallel-c_\perp)\sin\psi\cos\psi\sint{\sin(\pi N s)\cos(\pi N s) } = 0,
\end{multline}
\begin{multline}
B_{13} = \sint{r_1 \Sigma_{12}-r_2 \Sigma_{11}} 
= -\sigma R(c_\parallel-c_\perp)\sin^2\psi\sint{\sin(\pi Ns)} = 0,
\end{multline}
\begin{multline}
B_{21} = \sint{r_2 \Sigma_{23}-r_3 \Sigma_{22}} \\
= R(c_\parallel-c_\perp)\sin\psi\cos\psi\sint{\sin(\pi N s)\cos(\pi N s)} - (c_\parallel-c_\perp)\sin^2\psi\cos\psi\sint{s\cos^2(\pi Ns)} = 0,
\end{multline}
\begin{multline}
B_{22} = \sint{r_3 \Sigma_{21}-r_1 \Sigma_{23}} \\
= -\sigma(c_\parallel-c_\perp)\sin^2\psi\cos\psi\sint{s \sin(\pi Ns)\cos(\pi Ns)} - \sigma R(c_\parallel-c_\perp)\sin\psi\cos\psi\sint{\cos^2(\pi N s) } \\
= -\sigma(c_\parallel-c_\perp)\sin^2\psi\cos\psi\left(\frac{1}{\pi N}-\frac{\cos(2\pi N)}{2\pi N}+\frac{3 \sin(2 \pi N)}{(2\pi N)^2}\right),
\end{multline}
\begin{multline}
B_{23} = \sint{r_1 \Sigma_{22}-r_2 \Sigma_{21}} \\
= Rc_\perp\sint{\cos(\pi N s) } + R(c_\parallel-c_\perp)\sin^2\psi\sint{\cos(\pi N s)} \\
= \frac{2\sin(\pi N)\sin(\psi)}{(\pi N)^2}(\cos^2\psi c_\perp + \sin^2\psi c_\parallel), \label{eq:B23}
\end{multline}
\begin{multline}
B_{31} = \sint{r_2 \Sigma_{33}-r_3 \Sigma_{32}} \\
= \sigma R(\cos^2\psi c_\parallel + \sin^2\psi c_\perp)\sint{\sin(\pi N s)} - \sigma(c_\parallel-c_\perp)\sin\psi\cos^2\psi\sint{s\cos(\pi N s)} = 0,
\end{multline}
\begin{multline}
B_{32} = \sint{r_3 \Sigma_{31}-r_1 \Sigma_{33}} \\
= -(c_\parallel-c_\perp)\sin\psi\cos^2\psi\sint{s\sin(\pi N s)} - R(\cos^2\psi c_\parallel + \sin^2\psi c_\perp)\sint{\cos(\pi N s) } \\
= \frac{2\sin(\pi N)\sin(\psi)}{(\pi N)^2}\left(\cos(2\psi)c_\perp - (1+\cos(2\psi))c_\parallel\right)+\frac{\sin(2\psi)\cos\psi\cos(\pi N)}{\pi N}(c_\parallel-c_\perp),
\end{multline}
\begin{multline}
B_{33} = \sint{r_1 \Sigma_{32}-r_2 \Sigma_{31}} \\
= \sigma R(c_\parallel-c_\perp)\sin\psi\cos\psi\sint{\cos^2(\pi N s) } + \sigma R(c_\parallel-c_\perp)\sin\psi\cos\psi\sint{\sin^2(\pi N s)} \\
= \sigma \frac{\sin(\psi)\sin(2\psi)}{\pi N}(c_\parallel-c_\perp). \label{eq:B33}
\end{multline}
Note that we have used the identity $R=\sin\psi/(\pi N)$ to simplify the answers.

Finally, from Eqs.~\eqref{eq:result-S0} and \eqref{eq:defn-S0} we write down an equivalent definition of matrix $\mathbf{D}$ as 
\begin{equation}
D_{ij} = \int_{-1}^{+1} \varepsilon_{ikl} \varepsilon_{jmn} r_k(s)\Sigma_{ln}(s) r_m(s) \mathrm{d}s. \label{eq:Dij_defn}
\end{equation}
Using the integrals from Appendix \ref{app:useful-integrals}, Eqs.~\eqref{eq:app:cos},\eqref{eq:app:sin2},\eqref{eq:app:cos2},\eqref{eq:app:sin2cos},\eqref{eq:app:cos3},\eqref{eq:app:ssin},\eqref{eq:app:ssincos},\eqref{eq:app:s2sin2},\eqref{eq:app:s2cos2}, and \eqref{eq:app:odd}, we determine the components of matrix $\mathbf{D}$, which describes the torque exerted by a rotating filament:
\begin{multline}
D_{11} = \sint{(r_2^2\Sigma_{33} + r_3^2\Sigma_{22} - 2r_2r_3\Sigma_{23})} 
\\ 
= (\cos^2\psi c_\parallel + \sin^2\psi c_\perp)R^2\sint{\sin^2(\pi N s)} \\ + c_\perp \cos^2\psi \sint{s^2} 
+ (c_\parallel-c_\perp)\sin^2\psi\cos^2\psi \sint{s^2\cos^2(\pi N s)} 
\\ - 2(c_\parallel-c_\perp)R \sin\psi\cos^2\psi \sint{s \sin(\pi N s) \cos(\pi N s)}
\\ 
= \frac{\sin^2\psi}{(\pi N)^2}\left(\sin^2\psi c_\perp + \cos^2\psi c_\parallel\right) + \frac{2}{3}\cos^2\psi\left[\left(1-\frac{1}{2}\sin^2\psi\right)c_\perp + \frac{1}{2}\sin^2\psi c_\parallel\right]\\
- \sin^2\psi \left[ c_\perp \frac{4 \sin(2\pi N)}{(2\pi N)^3} + (c_\parallel - c_\perp)\cos^2\psi\left(-\frac{\sin(2\pi N)}{2\pi N}-\frac{6\cos(2\pi N)}{(2\pi N)^2}+\frac{10\sin(2\pi N)}{(2\pi N)^3}\right) \right], \label{eq:Dcomponents-A}
\end{multline}
\begin{multline}
D_{22} = \sint{(r_1^2\Sigma_{33} + r_3^2\Sigma_{11} - 2r_1r_3\Sigma_{13})}\\
= (\cos^2\psi c_\parallel + \sin^2\psi c_\perp)R^2\sint{\cos^2(\pi N s)} 
\\ + c_\perp \cos^2\psi \sint{s^2} 
+ (c_\parallel-c_\perp)\sin^2\psi\cos^2\psi \sint{s^2\sin^2(\pi N s)} 
\\ + 2(c_\parallel-c_\perp)R \sin\psi\cos^2\psi \sint{s \sin(\pi N s) \cos(\pi N s)}
\\ 
= \frac{\sin^2\psi}{(\pi N)^2}\left(\sin^2\psi c_\perp + \cos^2\psi c_\parallel\right) + \frac{2}{3}\cos^2\psi\left[\left(1-\frac{1}{2}\sin^2\psi\right)c_\perp + \frac{1}{2}\sin^2\psi c_\parallel\right]\\
+ \sin^2\psi \left[ c_\perp \frac{4 \sin(2\pi N)}{(2\pi N)^3} + (c_\parallel - c_\perp)\cos^2\psi\left(-\frac{\sin(2\pi N)}{2\pi N}-\frac{6\cos(2\pi N)}{(2\pi N)^2}+\frac{10\sin(2\pi N)}{(2\pi N)^3}\right) \right],
\end{multline}
\begin{multline}
D_{33} = \sint{(r_1^2\Sigma_{22} + r_2^2\Sigma_{11} - 2r_1r_2\Sigma_{12})}\\
= R^2c_\perp \sint{(\sin^2\piNs + \cos^2\piNs)}
\\+R^2(c_\parallel-c_\perp)\sin^2\psi\sint{(\cos^4(\pi N s)+\sin^4(\pi N s)+2\sin^2(\pi N s)\cos^2(\pi N s))} \\= \frac{2\sin^2\psi}{(\pi N)^2}(\cos^2\psi c_\perp + \sin^2\psi c_\parallel), \label{eq:D33}
\end{multline}
\begin{multline}
D_{12} = D_{21} = \sint{(r_2r_3\Sigma_{13}+r_1r_3\Sigma_{23}-r_1r_2\Sigma_{33}-r_3^2\Sigma_{12})} 
\\ = -\sigma (c_\parallel-c_\perp) R \sin\psi\cos^2\psi \sint{s \sin^2\piNs} 
\\+ \sigma (c_\parallel-c_\perp) R \sin\psi\cos^2\psi \sint{s \cos^2\piNs}
\\ - \sigma (\cos^2\psi c_\perp + \sin^2\psi c_\parallel) R^2 \sint{\sin\piNs \cos\piNs}
\\+ \sigma (c_\parallel-c_\perp)\sin^2\psi\cos^2\psi\sint{s^2\sin(\pi Ns)\cos(\pi Ns)}
= 0,
\end{multline}
\begin{multline}
D_{13} = D_{31} = \sint{(r_2r_3\Sigma_{12}+r_1r_2\Sigma_{23}-r_1r_3\Sigma_{22}-r_2^2\Sigma_{13})} 
\\ = - (c_\parallel-c_\perp) R \sin^2\psi\cos\psi \sint{s \sin^2\piNs\cos\piNs} 
\\+ (c_\parallel-c_\perp) R^2 \sin\psi\cos\psi \sint{\sin\piNs \cos^2\piNs}
\\ - c_\perp R\cos\psi \sint{s \cos\piNs} - (c_\parallel-c_\perp) R\sin^2\psi\cos\psi \sint{s \cos^3\piNs}
\\ + (c_\parallel-c_\perp) R^2 \sin\psi\cos\psi \sint{\sin^3\piNs}
= 0,
\end{multline}
\begin{multline}
D_{23} = D_{32} = \sint{(r_1r_3\Sigma_{12}+r_1r_2\Sigma_{13}-r_2r_3\Sigma_{11}-r_1^2\Sigma_{23})} \\ = - \sigma(c_\parallel-c_\perp) R \sin^2\psi\cos\psi \sint{s \sin\piNs\cos^2\piNs} 
\\-\sigma (c_\parallel-c_\perp) R^2 \sin\psi\cos\psi \sint{\sin^2\piNs \cos\piNs}
\\ - \sigma c_\perp R\cos\psi \sint{s \sin\piNs} - \sigma(c_\parallel-c_\perp) R\sin^2\psi\cos\psi \sint{s \sin^3\piNs}
\\ -\sigma (c_\parallel-c_\perp) R^2 \sin\psi\cos\psi \sint{\cos^3\piNs}
\\ = -\frac{\sigma\sin(2\psi)}{(\pi N)^2}\left[\left(\cos(2\psi)c_\perp + (1-\cos(2\psi))c_\parallel\right)\frac{\sin(\pi N)}{\pi N} - \left(\cos^2\psi c_\perp + \sin^2\psi c_\parallel\right)\cos(\pi N)\right]. \label{eq:Dcomponents-Z}
\end{multline}

\section{Calculating force moments from RFT}
\label{app:forcemoments_RFT}

In this section we use RFT to calculate analytical expressions for the vector of force moments $\mb_0$ defined in Eqs.~\eqref{eq:defn-M} and \eqref{eq:defn-m0}. In the body frame of a filament with centreline $\rb(s)$ and local resistance tensor $\Sigb(s)$, we have
\begin{eqnarray}
(\mb_0)_j &=& (-2\delta_{k1}\delta_{l1} + \delta_{k2}\delta_{l2} + \delta_{k3}\delta_{l3})\sint{r_l(s)\Sigma_{kj}(s)}, \quad (j=1,2,3) \\
(\mb_0)_j&=& (-2\delta_{k1}\delta_{l1} + \delta_{k2}\delta_{l2} + \delta_{k3}\delta_{l3})\sint{r_l(s)\Sigma_{km}(s)\varepsilon_{j-3,nm}r_n(s)}. \quad (j=4,5,6)
\end{eqnarray}

The first component is 
\begin{equation}
(\mb_0)_1 = \sint{-2r_1\Sigma_{11}+r_2\Sigma_{21}+r_3\Sigma_{31}}.
\end{equation}
Substituting the components of $\rb(s)$ from Eq.~\eqref{eq:centreline} and the values of $\Sigb(s)$ from Eqs.~\eqref{eq:Sigma-A}-\eqref{eq:Sigma-Z}, we find that
\begin{multline}
(\mb_0)_1 = -2Rc_\perp\sint{\cos(\pi N s)}-3R(c_\parallel-c_\perp)\sin^2\psi\sint{\cos(\pi Ns)\sin^2(\pi Ns)} \\
-(c_\parallel-c_\perp)\sin\psi\cos^2\psi\sint{s\sin(\pi Ns)}.
\end{multline}
Using the integrals in Appendix \ref{app:useful-integrals}, Eqs.~\eqref{eq:app:cos},\eqref{eq:app:sin2cos} and \eqref{eq:app:ssin}, we determine that
\begin{multline}
(\mb_0)_1 = -2Rc_\perp\frac{2\sin(\pi N)}{\pi N} - 3R(c_\parallel-c_\perp)\sin^2\psi\frac{2\sin^3(\pi N)}{3\pi N} \\
-(c_\parallel-c_\perp)\sin\psi\cos^2\psi\left(-\frac{2\cos(\pi N)}{\pi N} + \frac{2\sin(\pi N)}{(\pi N)^2}\right).
\end{multline}
Finally, with the substitution $R=\sin\psi/\pi N$ and notation $(\mb_0)_1 = \mathcal{M}_1$, we get
\begin{multline}
\mathcal{M}_1 = 2c_\perp\sin\psi\left[\cos^2\psi\left(-\frac{\cos(\pi N)}{\pi N} + \frac{\sin(\pi N)}{(\pi N)^2}\right) + \sin^2\psi\frac{\sin^3(\pi N)}{(\pi N)^2}-\frac{2\sin(\pi N)}{(\pi N)^2}\right] \\
-2c_\parallel\sin\psi\left[\cos^2\psi\left(-\frac{\cos(\pi N)}{\pi N} + \frac{\sin(\pi N)}{(\pi N)^2}\right) + \sin^2\psi\frac{\sin^3(\pi N)}{(\pi N)^2}\right].
\label{eq:result:M1}
\end{multline}

The next two components are zero
\begin{eqnarray}
(\mb_0)_2 &=& \sint{-2r_1\Sigma_{12}+r_2\Sigma_{22}+r_3\Sigma_{32}} = 0, \\
(\mb_0)_3 &=& \sint{-2r_1\Sigma_{13}+r_2\Sigma_{23}+r_3\Sigma_{33}} = 0,
\end{eqnarray}
because each term in the integrals is an odd function of $s$.  

The fourth component is
\begin{equation}
(\mb_0)_4 = \sint{-2r_1(\Sigma_{13}r_2-\Sigma_{12}r_3)+r_2(\Sigma_{23}r_2-\Sigma_{22}r_3)+r_3(\Sigma_{33}r_2-\Sigma_{32}r_3)} = 0.
\end{equation}
We will group terms according to their $s$ dependence as
\begin{multline}
(\mb_0)_4 = \sint{(-2r_1r_2\Sigma_{13} + r_2^2\Sigma_{23})} + \sint{(2r_1r_3\Sigma_{12} + r_2r_3(\Sigma_{33}-\Sigma_{22}))} - \sint{r_3^2 \Sigma_{23}}.
\end{multline}
After substituting the components of $\rb(s)$ from Eq.~\eqref{eq:centreline} and the values of $\Sigb(s)$ from Eqs.~\eqref{eq:Sigma-A}-\eqref{eq:Sigma-Z}, this becomes
\begin{multline}
(\mb_0)_4 = \sigma (c_\parallel-c_\perp)\left[3R^2\sin\psi\cos\psi\sint{\cos(\pi Ns)\sin^2(\pi Ns)} \right. \\
+ R\cos^3\psi\sint{s\sin(\pi N s)} -3R\sin^2\psi\cos\psi\sint{s\sin(\pi N s)\cos^2(\pi N s)} \\
\left. -\sin\psi\cos^3\psi\sint{s^2\cos(\pi N s)} \right].
\end{multline}
Using the integrals in Appendix \ref{app:useful-integrals}, Eqs.~\eqref{eq:app:sin2cos},\eqref{eq:app:ssin},\eqref{eq:app:ssincos2} and \eqref{eq:app:s2cos}, we determine that
\begin{multline}
(\mb_0)_4 = \sigma (c_\parallel-c_\perp)\left[
3R^2\sin\psi\cos\psi\frac{2\sin^3(\pi N)}{3\pi N} \right. \\
+ R\cos^3\psi\left( -\frac{2\cos(\pi N)}{\pi N} + \frac{2\sin(\pi N)}{(\pi N)^2} \right) \\
-3R\sin^2\psi\cos\psi\left( -\frac{2\cos^3(\pi N)}{3\pi N} + \frac{2\sin(\pi N)}{3(\pi N)^2} - \frac{2\sin^3(\pi N)}{9(\pi N)^2} \right)\\
\left. -\sin\psi\cos^3\psi\left(\frac{2\sin(\pi N)}{\pi N} + \frac{4\cos(\pi N)}{(\pi N)^2} -\frac{4\sin(\pi N)}{(\pi N)^3} \right) \right].
\end{multline}
Finally, with the substitution $R=\sin\psi/\pi N$ and notation $(\mb_0)_4 = \mathcal{M}_4$, we get
\begin{multline}
\mathcal{M}_4 = \sigma (c_\parallel-c_\perp)\sin(2\psi)
\left[
\cos^2\psi\left(
- \frac{\sin(\pi N)}{\pi N} - \frac{3\cos(\pi N)}{(\pi N)^2} +\frac{3\sin(\pi N)}{(\pi N)^3}
\right)\right.
\\
\left. 
+ \sin^2\psi\left(
\frac{\cos^3(\pi N)}{(\pi N)^2} + \frac{4\sin^3(\pi N)}{3(\pi N)^3} - \frac{\sin(\pi N)}{(\pi N)^3}
\right)\right].
\label{eq:result:M4}
\end{multline}

The next two components are also zero
\begin{eqnarray}
(\mb_0)_5 &=& \sint{-2r_1(\Sigma_{11}r_3-\Sigma_{13}r_1)+r_2(\Sigma_{21}r_3-\Sigma_{23}r_1)+r_3(\Sigma_{31}r_3-\Sigma_{33}r_1)} = 0, \\
(\mb_0)_6 &=& \sint{-2r_1(\Sigma_{12}r_1-\Sigma_{11}r_2)+r_2(\Sigma_{22}r_1-\Sigma_{21}r_2)+r_3(\Sigma_{32}r_1-\Sigma_{31}r_2)} = 0, 
\end{eqnarray}
because each term in the integrals is an odd function of $s$.
\vfill

\section{Useful integrals}
\label{app:useful-integrals}
We provide some useful integrals for the RFT calculations in Appendices \ref{app:RFT} and \ref{app:forcemoments_RFT}.
\begin{eqnarray}
&~& \int_{-1}^{+1} \cos(\pi N s)  \mathrm{d}s = \frac{2\sin(\pi N)}{\pi N}, 
\label{eq:app:cos}\\
&~& \int_{-1}^{+1} \sin^2(\pi N s)  \mathrm{d}s = 1-\frac{\sin(2\pi N)}{2\pi N}, 
\label{eq:app:sin2}\\
&~& \int_{-1}^{+1} \cos^2(\pi N s)  \mathrm{d}s = 1+\frac{\sin(2\pi N)}{2\pi N}, 
\label{eq:app:cos2}\\
&~& \sint{\sin^2(\pi N s)\cos(\pi N s)} = \frac{2\sin^3(\pi N)}{3\pi N},
\label{eq:app:sin2cos}\\
&~& \sint{\cos^3(\pi N s)} = \frac{2\sin(\pi N)}{\pi N} -  \frac{2\sin^3(\pi N)}{3\pi N}, 
\label{eq:app:cos3}\\
&~& \int_{-1}^{+1} s\sin(\pi N s)  \mathrm{d}s = -\frac{2\cos(\pi N)}{\pi N} + \frac{2\sin(\pi N)}{(\pi N)^2}, 
\label{eq:app:ssin}\\
&~& \int_{-1}^{+1} s\sin(\pi N s)\cos(\pi N s)  \mathrm{d}s = -\frac{\cos(2\pi N)}{2\pi N} + \frac{\sin(2\pi N)}{(2\pi N)^2},  
\label{eq:app:ssincos}\\
&~& \int_{-1}^{+1} s\sin(\pi N s)\cos^2(\pi N s)  \mathrm{d}s =  -\frac{2\cos^3(\pi N)}{3\pi N} + \frac{2\sin(\pi N)}{3(\pi N)^2} - \frac{2\sin^3(\pi N)}{9(\pi N)^2},
\label{eq:app:ssincos2}\\
&~& \int_{-1}^{+1} s^2\cos(\pi N s)  \mathrm{d}s = \frac{2\sin(\pi N)}{\pi N} + \frac{4\cos(\pi N)}{(\pi N)^2} -\frac{4\sin(\pi N)}{(\pi N)^3},
\label{eq:app:s2cos} \\
&~& \int_{-1}^{+1} s^2\sin^2(\pi N s)  \mathrm{d}s = \frac{1}{3}-\left[\frac{\sin(2\pi N)}{2\pi N}+\frac{2\cos(2\pi N)}{(2\pi N)^2} - \frac{2\sin(2\pi N)}{(2\pi N)^3}\right], 
\label{eq:app:s2sin2}\\
&~& \int_{-1}^{+1} s^2\cos^2(\pi N s)  \mathrm{d}s = \frac{1}{3}+\left[\frac{\sin(2\pi N)}{2\pi N}+\frac{2\cos(2\pi N)}{(2\pi N)^2} - \frac{2\sin(2\pi N)}{(2\pi N)^3}\right]. 
\label{eq:app:s2cos2}
\end{eqnarray}

In addition to these, we point out that
\begin{equation}
\sint{s^j\sin^k(\pi N s) \cos^l(\pi N s)} = 0,
\label{eq:app:odd}
\end{equation}
for any non-negative integer powers $j,k,l$ so long as $j+k\equiv 1 ~(\mathrm{mod} 2)$, because the integrand is an odd function of $s$. 
\vfill

\section{Validation of computational method}
\label{app:comptests}

\begin{figure}
	\landscapetrim{17cm}{12cm}
	\includegraphics[trim={{.5\cutwidth} {.5\cutheight} {.5\cutwidth} {.5\cutheight}},clip,width=17cm]{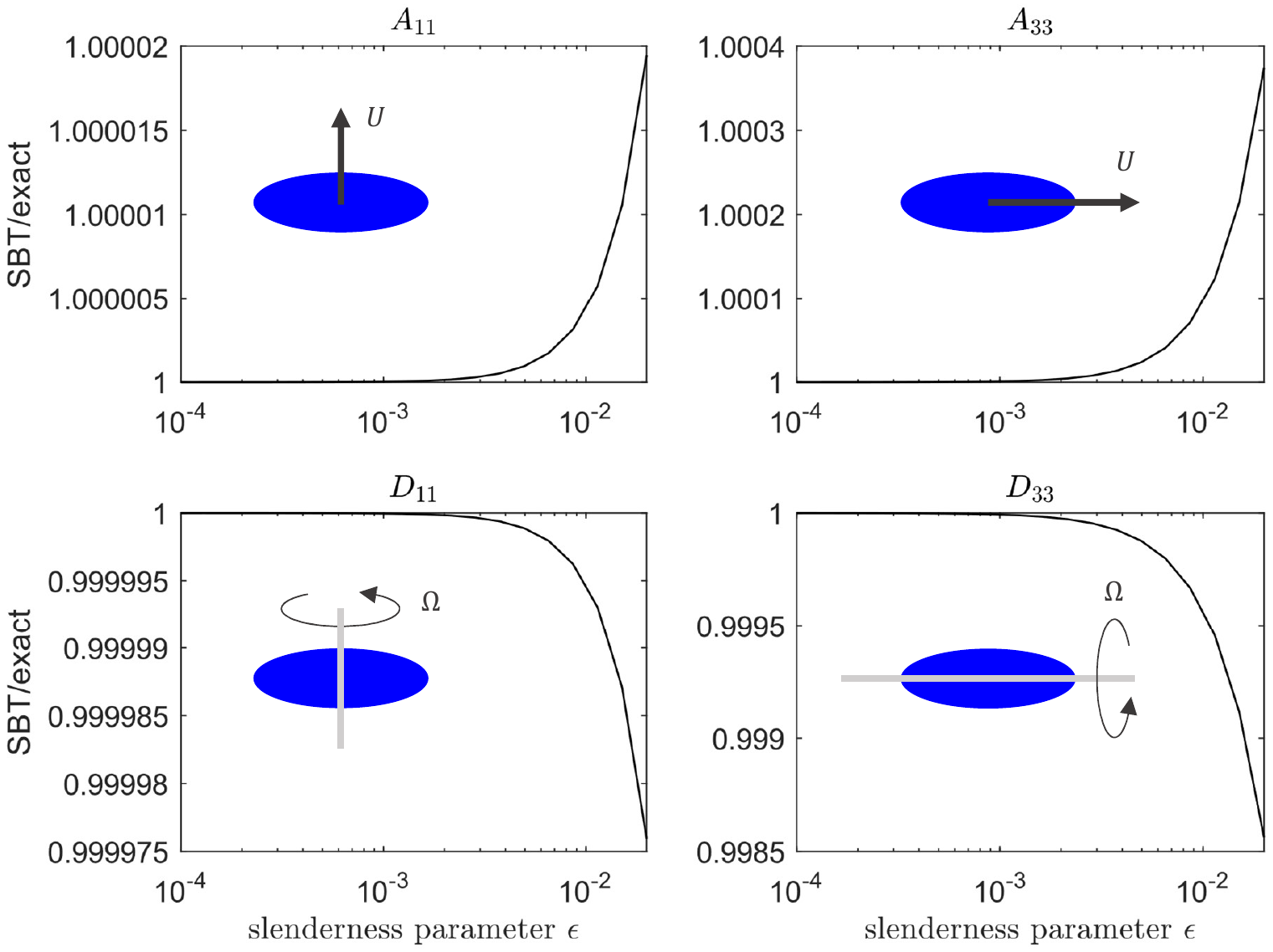}
	\caption{Tests for a prolate spheroid. Verifying our implementation of SBT against the exact solution for spheroids \cite{Chwang1975}. In our implementation of SBT, we truncate the numerical  solution to $N_{\mathrm{Legendre}} = 5$ Legendre polynomial modes.}
	\label{fig:app-spheroid}
\end{figure}

\begin{figure}
	\landscapetrim{17cm}{6cm}
	\includegraphics[trim={{.5\cutwidth} {.5\cutheight} {.5\cutwidth} {.5\cutheight}},clip,width=17cm]{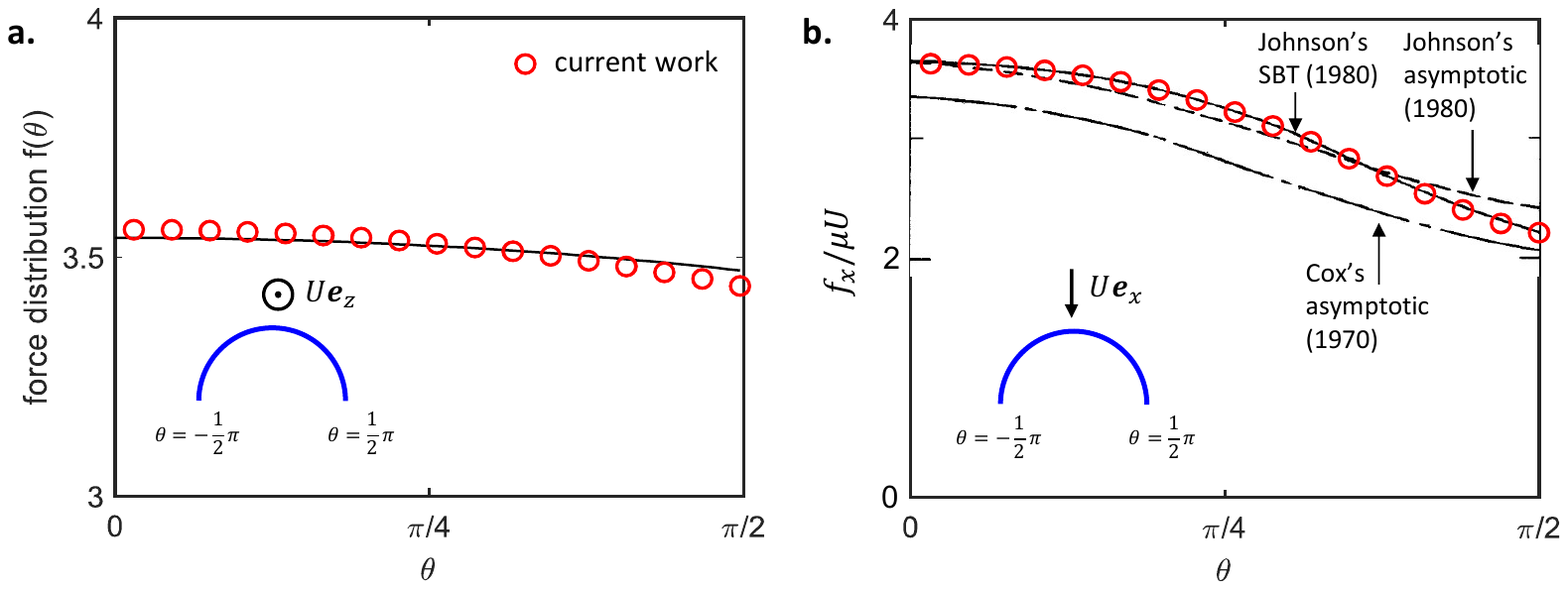}
	\caption{Tests for a semi-circular filament. (a) Verifying our implementation of SBT (red circles) against Johnson's asymptotic solution (solid line) from Eq.~(36) in Ref.~\cite{Johnson1980}. The quantity being plotted is the vertical component of the force density exerted by a horizontal semi-circular filament, with slenderness $\epsilon = 0.1$, translating vertically as shown in the inset. (b) Verifying our implementation of SBT (red circles) against Johnson's SBT computations (solid line) and two asymptotic solutions (dashed lines) by Cox \cite{Cox1970} and Johnson \cite{Johnson1980}. The quantity being plotted is the drag force per unit length on a semi-circular filament, with slenderness $\epsilon = 0.1$, translating along its axis of symmetry as shown in the inset. For both (a) and (b), we use a truncation of $N_{\mathrm{Legendre}} = 10$ in our implementation of SBT. In (b), our results are overlaid onto Fig.~4 from Ref.~\cite{Johnson1980}.} 
	\label{fig:app-arc}
\end{figure}

\begin{figure}
	\landscapetrim{17cm}{8.242cm}
	\includegraphics[trim={{.5\cutwidth} {.5\cutheight} {.5\cutwidth} {.5\cutheight}},clip,width=17cm]{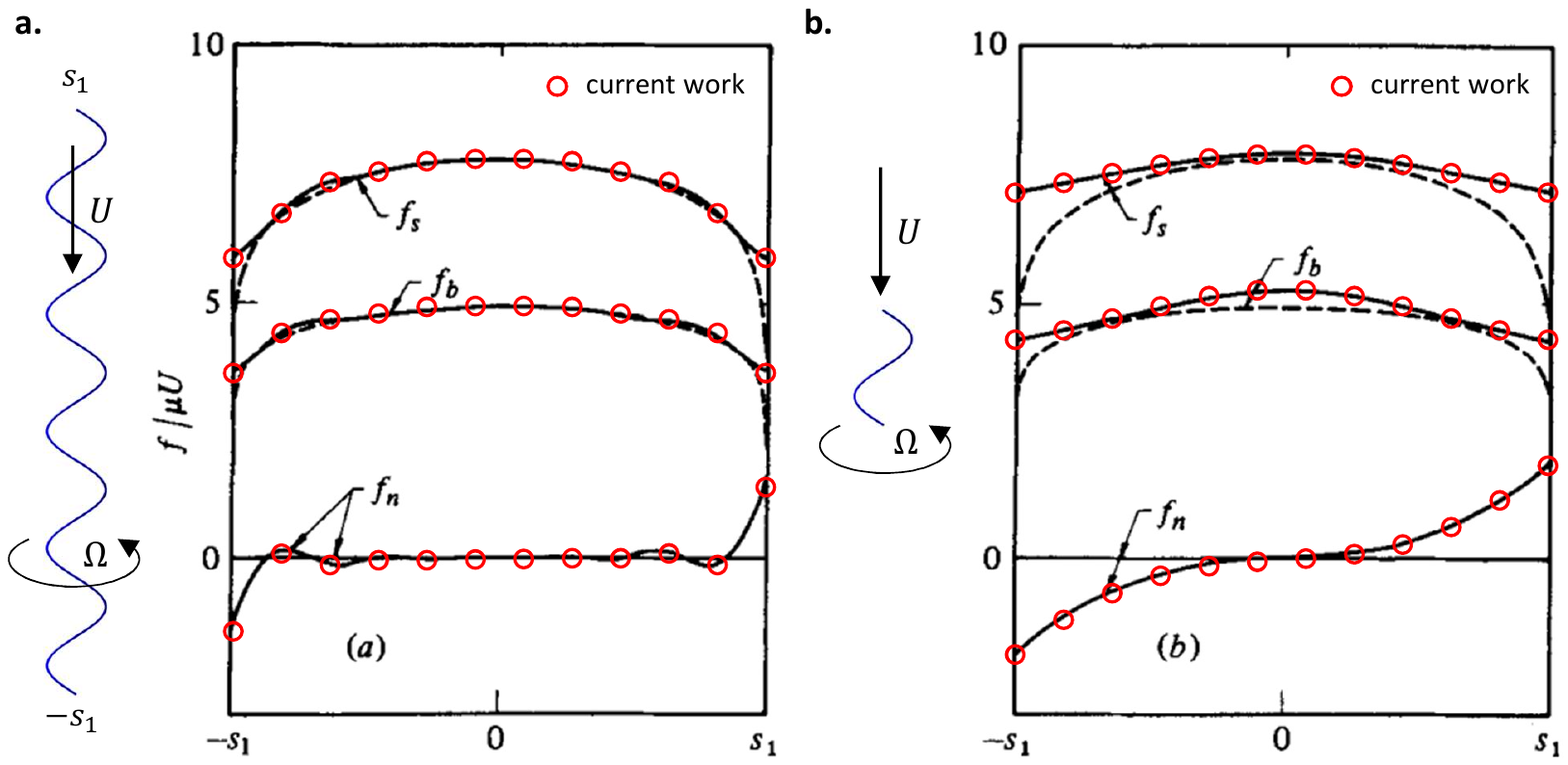}
	\caption{Tests for a helical filament. Verifying our implementation of SBT (red circles) against Lighthill's exact solution for infinite helices (dashed line) \cite{Lighthill1976} and Johnson's SBT (solid line) \cite{Johnson1980}. The three data series represent the tangential ($f_s$), normal ($f_n$) and binormal ($f_b$) components of the force density along a helix (with a prolate spheroidal cross-section) that is translating and rotating about its axis of symmetry such that the net force along the axis is zero. Helix parameters: (a) five helical turns, $\epsilon = 0.0021$, $\psi = 1.0039$ rad; (b) one helical turn, $\epsilon = 0.0107$, $\psi = 1.0039$ rad. In our implementation of SBT, we truncate the solution to $N_{\mathrm{Legendre}} = 25$ Legendre polynomial modes. Our results are overlaid onto Fig.~6 from Ref.~\cite{Johnson1980}.}
	\label{fig:app-helix}
\end{figure}

\begin{figure}
	\landscapetrim{17cm}{8.5cm}
	\includegraphics[trim={{.5\cutwidth} {.5\cutheight} {.5\cutwidth} {.5\cutheight}},clip,width=17cm]{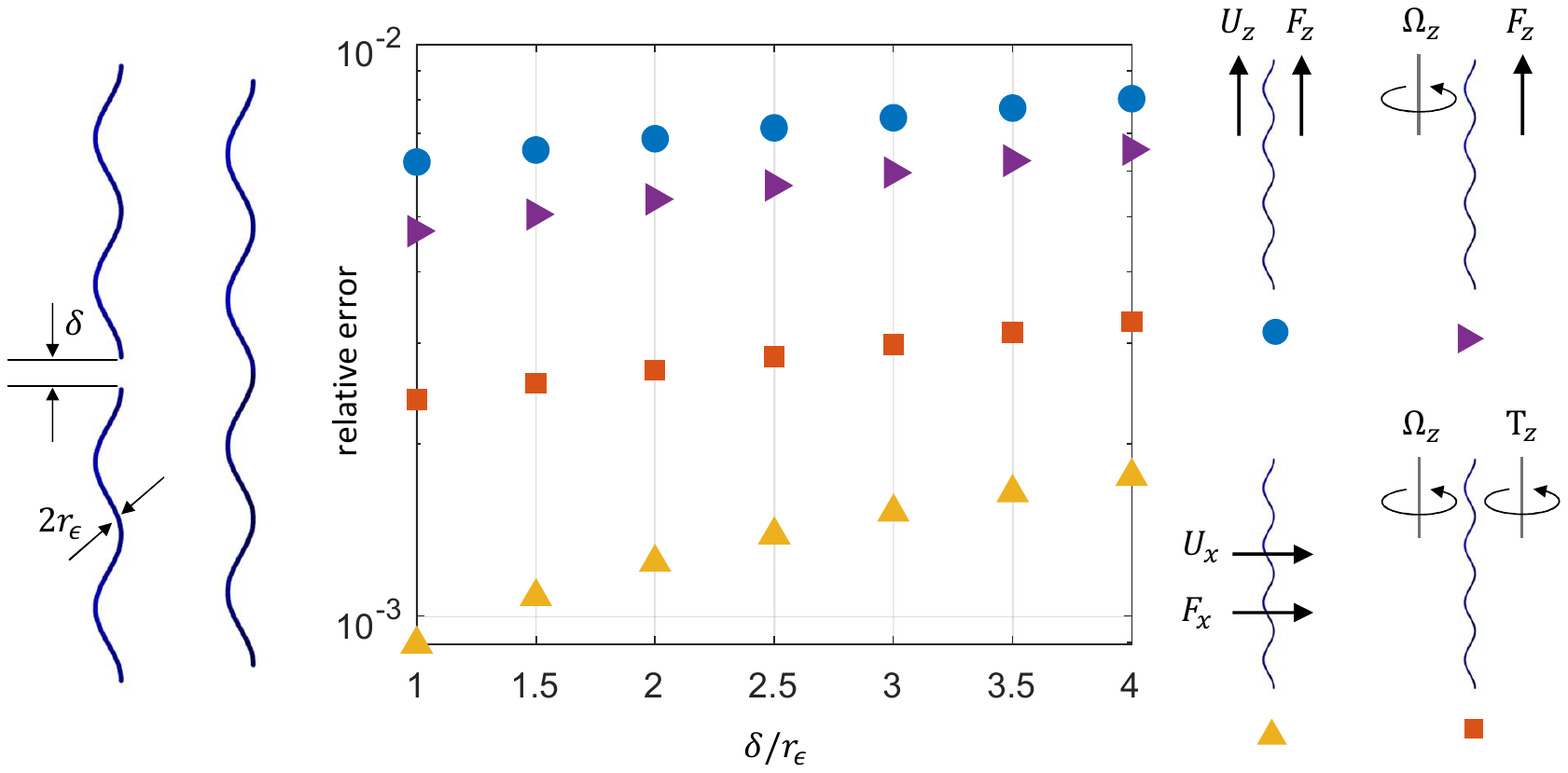}
	\caption{Tests for cross-filament HIs. We compare the hydrodynamic resistance of two interacting helical filaments, having two helical turns each and placed head-to-head (H2H), with that of one helical filament that has four helical turns (ONE). In the limit $\delta \to 0$, the two problems are identical. We plot (on a log-linear scale) the relative errors in the coefficients of the resistance matrix $A_{33}=F_z/U_z$ (blue circles), $B_{33} = F_z/\Omega_z$ (purple rightward triangles), $A_{11} = F_x/U_x$ (yellow upward triangles) and $D_{33} = T_z/\Omega_z$ (orange squares). The relative error is defined as $|1-\textrm{H2H}/\textrm{ONE}|$ and decays with decreasing separation $\delta$, thus validating our implementation of cross-filament HIs. Helix parameters: $\psi = 0.5043$ rad, $\epsilon =  0.0024$ (for $N=4$ helical turns) and $\epsilon =  0.0048$ (for $N=2$ helical turns).}
	\label{fig:app-twovsone}
\end{figure}

In order to validate our implementation of slender-body theory (SBT), we carry out four different tests. For the implementation of single-filament dynamics (i.e.~standard SBT), we verify our computations against classical results for prolate spheroids (Fig.~\ref{fig:app-spheroid}), semi-circular arcs (Fig.~\ref{fig:app-arc}) and helices (Fig.~\ref{fig:app-helix}). In all three cases, our implementation of SBT is in excellent agreement with results published in the literature. To test cross-filament interactions, we compare the hydrodynamic resistance of two helical filaments placed head-to-head and that of a single helical filament with twice the length (Fig.~\ref{fig:app-twovsone}).  The relative error between the two setups decays with decreasing distance between the two half-filaments, thus validating our implementation of cross-filament HIs as well.

\begin{figure}
	\landscapetrim{17cm}{8.5cm}
	\includegraphics[trim={{.5\cutwidth} {.5\cutheight} {.5\cutwidth} {.5\cutheight}},clip,width=17cm]{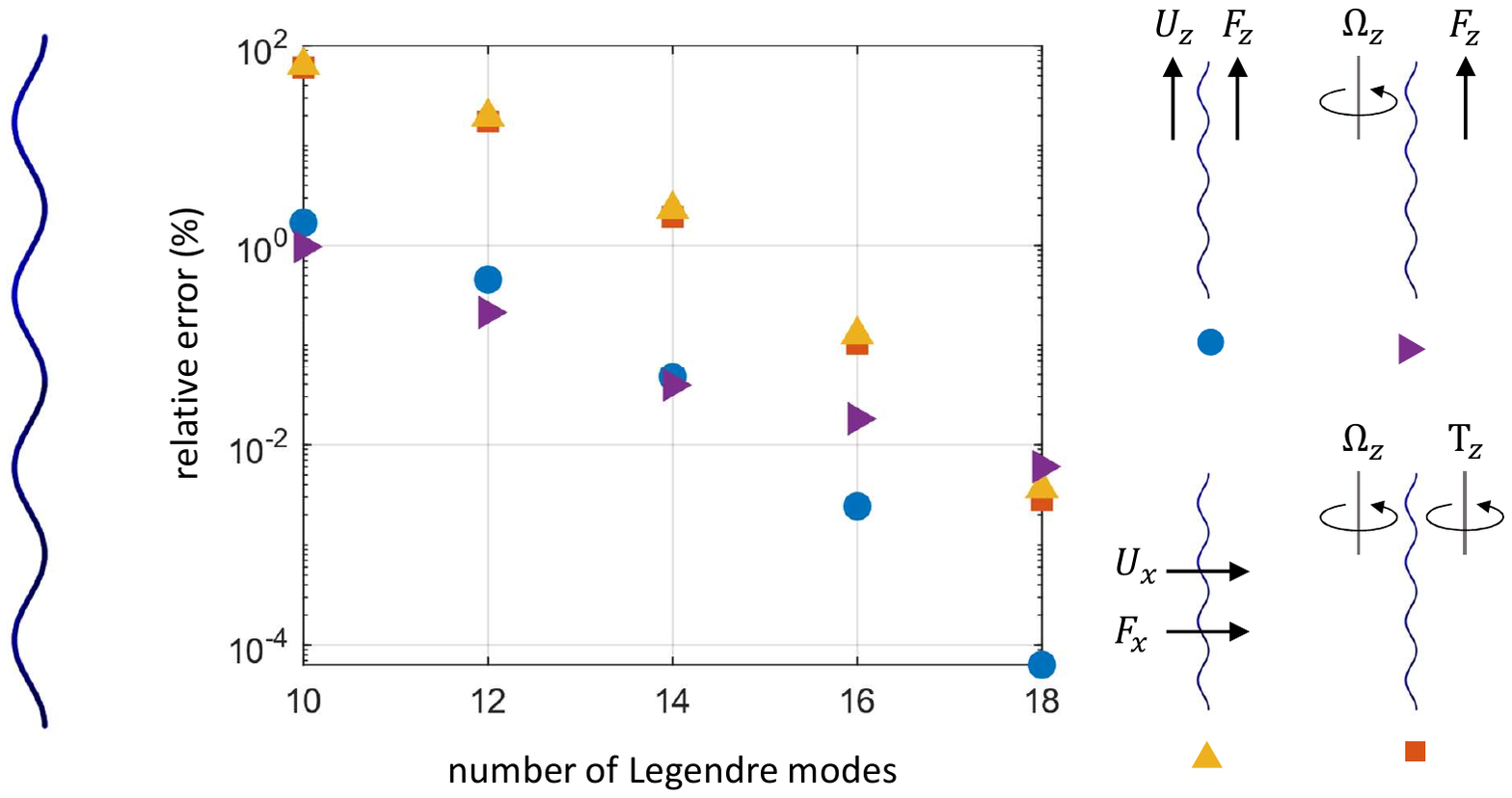}
	\caption{Self-convergence test for a single helical filament. The percentage errors in $A_{33}=F_z/U_z$ (blue circles), $B_{33} = F_z/\Omega_z$ (purple rightward triangles), $A_{11} = F_x/U_x$ (yellow upward triangles) and $D_{33} = T_z/\Omega_z$ (orange squares) are calculated at each number of Legendre modes relative to the most accurate numerical solution available, in this case $N_{\mathrm{Legendre}} = 20$. Helix parameters: $\psi = 0.5043$ rad, $\epsilon =0.0038$, $N=4$ helical turns.}
	\label{fig:app-accuracy}
\end{figure}

To determine the appropriate level of truncation in our spectral method implementation of SBT, we perform self-convergence tests for a single helical filament. In Fig.~\ref{fig:app-accuracy}, we vary the number of Legendre polynomial modes from $10$ to $20$, and we compare the resistance matrix at a given number of modes, $N_{\mathrm{Legendre}}$, with the most refined numerical solution available, i.e.~$N_{\mathrm{Legendre}} = 20$. The results of the self-convergence test for a helix with four helical turns suggest that a truncation level of $N_{\mathrm{Legendre}} = 15$ is sufficient to obtain 99\% accuracy. Unless otherwise stated, this is the level of truncation used for the simulations presented in this paper.

\bibliography{mybibliography}

\begin{thebibliography}{69}%
\makeatletter
\providecommand \@ifxundefined [1]{%
 \@ifx{#1\undefined}
}%
\providecommand \@ifnum [1]{%
 \ifnum #1\expandafter \@firstoftwo
 \else \expandafter \@secondoftwo
 \fi
}%
\providecommand \@ifx [1]{%
 \ifx #1\expandafter \@firstoftwo
 \else \expandafter \@secondoftwo
 \fi
}%
\providecommand \natexlab [1]{#1}%
\providecommand \enquote  [1]{``#1''}%
\providecommand \bibnamefont  [1]{#1}%
\providecommand \bibfnamefont [1]{#1}%
\providecommand \citenamefont [1]{#1}%
\providecommand \href@noop [0]{\@secondoftwo}%
\providecommand \href [0]{\begingroup \@sanitize@url \@href}%
\providecommand \@href[1]{\@@startlink{#1}\@@href}%
\providecommand \@@href[1]{\endgroup#1\@@endlink}%
\providecommand \@sanitize@url [0]{\catcode `\\12\catcode `\$12\catcode
  `\&12\catcode `\#12\catcode `\^12\catcode `\_12\catcode `\%12\relax}%
\providecommand \@@startlink[1]{}%
\providecommand \@@endlink[0]{}%
\providecommand \url  [0]{\begingroup\@sanitize@url \@url }%
\providecommand \@url [1]{\endgroup\@href {#1}{\urlprefix }}%
\providecommand \urlprefix  [0]{URL }%
\providecommand \Eprint [0]{\href }%
\providecommand \doibase [0]{https://doi.org/}%
\providecommand \selectlanguage [0]{\@gobble}%
\providecommand \bibinfo  [0]{\@secondoftwo}%
\providecommand \bibfield  [0]{\@secondoftwo}%
\providecommand \translation [1]{[#1]}%
\providecommand \BibitemOpen [0]{}%
\providecommand \bibitemStop [0]{}%
\providecommand \bibitemNoStop [0]{.\EOS\space}%
\providecommand \EOS [0]{\spacefactor3000\relax}%
\providecommand \BibitemShut  [1]{\csname bibitem#1\endcsname}%
\let\auto@bib@innerbib\@empty
\bibitem [{\citenamefont {Darnton}\ \emph {et~al.}(2010)\citenamefont
  {Darnton}, \citenamefont {Turner}, \citenamefont {Rojevsky},\ and\
  \citenamefont {Berg}}]{Darnton2010}%
  \BibitemOpen
  \bibfield  {author} {\bibinfo {author} {\bibfnamefont {N.~C.}\ \bibnamefont
  {Darnton}}, \bibinfo {author} {\bibfnamefont {L.}~\bibnamefont {Turner}},
  \bibinfo {author} {\bibfnamefont {S.}~\bibnamefont {Rojevsky}},\ and\
  \bibinfo {author} {\bibfnamefont {H.~C.}\ \bibnamefont {Berg}},\ }\bibfield
  {title} {\bibinfo {title} {{Dynamics of bacterial swarming}},\ }\href
  {https://doi.org/https://doi.org/10.1016/j.bpj.2010.01.053} {\bibfield
  {journal} {\bibinfo  {journal} {Biophys. J.}\ }\textbf {\bibinfo {volume}
  {98}},\ \bibinfo {pages} {2082} (\bibinfo {year} {2010})}\BibitemShut
  {NoStop}%
\bibitem [{\citenamefont {Brumley}\ \emph {et~al.}(2014)\citenamefont
  {Brumley}, \citenamefont {Wan}, \citenamefont {Polin},\ and\ \citenamefont
  {Goldstein}}]{Brumley2014}%
  \BibitemOpen
  \bibfield  {author} {\bibinfo {author} {\bibfnamefont {D.~R.}\ \bibnamefont
  {Brumley}}, \bibinfo {author} {\bibfnamefont {K.~Y.}\ \bibnamefont {Wan}},
  \bibinfo {author} {\bibfnamefont {M.}~\bibnamefont {Polin}},\ and\ \bibinfo
  {author} {\bibfnamefont {R.~E.}\ \bibnamefont {Goldstein}},\ }\bibfield
  {title} {\bibinfo {title} {Flagellar synchronization through direct
  hydrodynamic interactions},\ }\href {https://doi.org/10.7554/eLife.02750}
  {\bibfield  {journal} {\bibinfo  {journal} {eLife}\ }\textbf {\bibinfo
  {volume} {3}},\ \bibinfo {pages} {e02750} (\bibinfo {year}
  {2014})}\BibitemShut {NoStop}%
\bibitem [{\citenamefont {Wan}\ and\ \citenamefont
  {Goldstein}(2016)}]{Wan2016}%
  \BibitemOpen
  \bibfield  {author} {\bibinfo {author} {\bibfnamefont {K.~Y.}\ \bibnamefont
  {Wan}}\ and\ \bibinfo {author} {\bibfnamefont {R.~E.}\ \bibnamefont
  {Goldstein}},\ }\bibfield  {title} {\bibinfo {title} {Coordinated beating of
  algal flagella is mediated by basal coupling},\ }\href
  {https://doi.org/10.1073/pnas.1518527113} {\bibfield  {journal} {\bibinfo
  {journal} {Proc. Natl. Acad. Sci. USA}\ }\textbf {\bibinfo {volume} {113}},\
  \bibinfo {pages} {E2784} (\bibinfo {year} {2016})}\BibitemShut {NoStop}%
\bibitem [{\citenamefont {Guo}\ \emph {et~al.}(2021)\citenamefont {Guo},
  \citenamefont {Man}, \citenamefont {Wan},\ and\ \citenamefont
  {Kanso}}]{Kanso2021}%
  \BibitemOpen
  \bibfield  {author} {\bibinfo {author} {\bibfnamefont {H.}~\bibnamefont
  {Guo}}, \bibinfo {author} {\bibfnamefont {Y.}~\bibnamefont {Man}}, \bibinfo
  {author} {\bibfnamefont {K.~Y.}\ \bibnamefont {Wan}},\ and\ \bibinfo {author}
  {\bibfnamefont {E.}~\bibnamefont {Kanso}},\ }\bibfield  {title} {\bibinfo
  {title} {Intracellular coupling modulates biflagellar synchrony},\ }\href
  {https://doi.org/10.1098/rsif.2020.0660} {\bibfield  {journal} {\bibinfo
  {journal} {J. R. Soc. Interface}\ }\textbf {\bibinfo {volume} {18}},\
  \bibinfo {pages} {20200660} (\bibinfo {year} {2021})}\BibitemShut {NoStop}%
\bibitem [{\citenamefont {Ishikawa}(2009)}]{Ishikawa2009}%
  \BibitemOpen
  \bibfield  {author} {\bibinfo {author} {\bibfnamefont {T.}~\bibnamefont
  {Ishikawa}},\ }\bibfield  {title} {\bibinfo {title} {Suspension biomechanics
  of swimming microbes},\ }\href {https://doi.org/10.1098/rsif.2009.0223}
  {\bibfield  {journal} {\bibinfo  {journal} {J. R. Soc. Interface}\ }\textbf
  {\bibinfo {volume} {6}},\ \bibinfo {pages} {815} (\bibinfo {year}
  {2009})}\BibitemShut {NoStop}%
\bibitem [{\citenamefont {Shaqfeh}\ and\ \citenamefont
  {Fredrickson}(1990)}]{Shaqfeh1990}%
  \BibitemOpen
  \bibfield  {author} {\bibinfo {author} {\bibfnamefont {E.~S.}\ \bibnamefont
  {Shaqfeh}}\ and\ \bibinfo {author} {\bibfnamefont {G.~H.}\ \bibnamefont
  {Fredrickson}},\ }\bibfield  {title} {\bibinfo {title} {{The hydrodynamic
  stress in a suspension of rods}},\ }\href {https://doi.org/10.1063/1.857683}
  {\bibfield  {journal} {\bibinfo  {journal} {Phys. Fluids A}\ }\textbf
  {\bibinfo {volume} {2}},\ \bibinfo {pages} {7} (\bibinfo {year}
  {1990})}\BibitemShut {NoStop}%
\bibitem [{\citenamefont {Mackaplow}\ and\ \citenamefont
  {Shaqfeh}(1996)}]{Mackaplow1996}%
  \BibitemOpen
  \bibfield  {author} {\bibinfo {author} {\bibfnamefont {M.~B.}\ \bibnamefont
  {Mackaplow}}\ and\ \bibinfo {author} {\bibfnamefont {E.~S.}\ \bibnamefont
  {Shaqfeh}},\ }\bibfield  {title} {\bibinfo {title} {{A numerical study of the
  rheological properties of suspensions of rigid, non-Brownian fibres}},\
  }\href {https://doi.org/10.1017/S0022112096008889} {\bibfield  {journal}
  {\bibinfo  {journal} {J. Fluid Mech.}\ }\textbf {\bibinfo {volume} {329}},\
  \bibinfo {pages} {155} (\bibinfo {year} {1996})}\BibitemShut {NoStop}%
\bibitem [{\citenamefont {Guazzelli}\ and\ \citenamefont
  {Hinch}(2011)}]{Guazelli2011}%
  \BibitemOpen
  \bibfield  {author} {\bibinfo {author} {\bibfnamefont {{\'{E}}.}~\bibnamefont
  {Guazzelli}}\ and\ \bibinfo {author} {\bibfnamefont {J.}~\bibnamefont
  {Hinch}},\ }\bibfield  {title} {\bibinfo {title} {Fluctuations and
  instability in sedimentation},\ }\href
  {https://doi.org/10.1146/annurev-fluid-122109-160736} {\bibfield  {journal}
  {\bibinfo  {journal} {Annu. Rev. Fluid Mech.}\ }\textbf {\bibinfo {volume}
  {43}},\ \bibinfo {pages} {97} (\bibinfo {year} {2011})}\BibitemShut {NoStop}%
\bibitem [{\citenamefont {du~Roure}\ \emph {et~al.}(2019)\citenamefont
  {du~Roure}, \citenamefont {Lindner}, \citenamefont {Nazockdast},\ and\
  \citenamefont {Shelley}}]{Shelley2019}%
  \BibitemOpen
  \bibfield  {author} {\bibinfo {author} {\bibfnamefont {O.}~\bibnamefont
  {du~Roure}}, \bibinfo {author} {\bibfnamefont {A.}~\bibnamefont {Lindner}},
  \bibinfo {author} {\bibfnamefont {E.~N.}\ \bibnamefont {Nazockdast}},\ and\
  \bibinfo {author} {\bibfnamefont {M.~J.}\ \bibnamefont {Shelley}},\
  }\bibfield  {title} {\bibinfo {title} {Dynamics of flexible fibers in viscous
  flows and fluids},\ }\href
  {https://doi.org/10.1146/annurev-fluid-122316-045153} {\bibfield  {journal}
  {\bibinfo  {journal} {Annu. Rev. Fluid Mech.}\ }\textbf {\bibinfo {volume}
  {51}},\ \bibinfo {pages} {539} (\bibinfo {year} {2019})}\BibitemShut
  {NoStop}%
\bibitem [{\citenamefont {Ramaswamy}(2010)}]{Ramaswamy2010}%
  \BibitemOpen
  \bibfield  {author} {\bibinfo {author} {\bibfnamefont {S.}~\bibnamefont
  {Ramaswamy}},\ }\bibfield  {title} {\bibinfo {title} {The mechanics and
  statistics of active matter},\ }\href
  {https://doi.org/10.1146/annurev-conmatphys-070909-104101} {\bibfield
  {journal} {\bibinfo  {journal} {Annu. Rev. Condens. Matter Phys.}\ }\textbf
  {\bibinfo {volume} {1}},\ \bibinfo {pages} {323} (\bibinfo {year}
  {2010})}\BibitemShut {NoStop}%
\bibitem [{\citenamefont {Marchetti}\ \emph {et~al.}(2013)\citenamefont
  {Marchetti}, \citenamefont {Joanny}, \citenamefont {Ramaswamy}, \citenamefont
  {Liverpool}, \citenamefont {Prost}, \citenamefont {Rao},\ and\ \citenamefont
  {Simha}}]{Marchetti2013}%
  \BibitemOpen
  \bibfield  {author} {\bibinfo {author} {\bibfnamefont {M.~C.}\ \bibnamefont
  {Marchetti}}, \bibinfo {author} {\bibfnamefont {J.~F.}\ \bibnamefont
  {Joanny}}, \bibinfo {author} {\bibfnamefont {S.}~\bibnamefont {Ramaswamy}},
  \bibinfo {author} {\bibfnamefont {T.~B.}\ \bibnamefont {Liverpool}}, \bibinfo
  {author} {\bibfnamefont {J.}~\bibnamefont {Prost}}, \bibinfo {author}
  {\bibfnamefont {M.}~\bibnamefont {Rao}},\ and\ \bibinfo {author}
  {\bibfnamefont {R.~A.}\ \bibnamefont {Simha}},\ }\bibfield  {title} {\bibinfo
  {title} {Hydrodynamics of soft active matter},\ }\href
  {https://doi.org/10.1103/RevModPhys.85.1143} {\bibfield  {journal} {\bibinfo
  {journal} {Rev. Mod. Phys.}\ }\textbf {\bibinfo {volume} {85}},\ \bibinfo
  {pages} {1143} (\bibinfo {year} {2013})}\BibitemShut {NoStop}%
\bibitem [{\citenamefont {Sharifi-Mood}\ \emph {et~al.}(2016)\citenamefont
  {Sharifi-Mood}, \citenamefont {Mozaffari},\ and\ \citenamefont
  {C{\'{o}}rdova-Figueroa}}]{Sharifi2016}%
  \BibitemOpen
  \bibfield  {author} {\bibinfo {author} {\bibfnamefont {N.}~\bibnamefont
  {Sharifi-Mood}}, \bibinfo {author} {\bibfnamefont {A.}~\bibnamefont
  {Mozaffari}},\ and\ \bibinfo {author} {\bibfnamefont {U.~M.}\ \bibnamefont
  {C{\'{o}}rdova-Figueroa}},\ }\bibfield  {title} {\bibinfo {title} {Pair
  interaction of catalytically active colloids: from assembly to escape},\
  }\href {https://doi.org/10.1017/jfm.2016.317} {\bibfield  {journal} {\bibinfo
   {journal} {J. Fluid Mech.}\ }\textbf {\bibinfo {volume} {798}},\ \bibinfo
  {pages} {910–954} (\bibinfo {year} {2016})}\BibitemShut {NoStop}%
\bibitem [{\citenamefont {Varma}\ \emph {et~al.}(2018)\citenamefont {Varma},
  \citenamefont {Montenegro-Johnson},\ and\ \citenamefont
  {Michelin}}]{Varma2018}%
  \BibitemOpen
  \bibfield  {author} {\bibinfo {author} {\bibfnamefont {A.}~\bibnamefont
  {Varma}}, \bibinfo {author} {\bibfnamefont {T.~D.}\ \bibnamefont
  {Montenegro-Johnson}},\ and\ \bibinfo {author} {\bibfnamefont
  {S.}~\bibnamefont {Michelin}},\ }\bibfield  {title} {\bibinfo {title}
  {{Clustering-induced self-propulsion of isotropic autophoretic particles}},\
  }\href {https://doi.org/10.1039/c8sm00690c} {\bibfield  {journal} {\bibinfo
  {journal} {Soft Matter}\ }\textbf {\bibinfo {volume} {14}},\ \bibinfo {pages}
  {7155} (\bibinfo {year} {2018})}\BibitemShut {NoStop}%
\bibitem [{\citenamefont {Varma}\ and\ \citenamefont
  {Michelin}(2019)}]{Varma2019}%
  \BibitemOpen
  \bibfield  {author} {\bibinfo {author} {\bibfnamefont {A.}~\bibnamefont
  {Varma}}\ and\ \bibinfo {author} {\bibfnamefont {S.}~\bibnamefont
  {Michelin}},\ }\bibfield  {title} {\bibinfo {title} {{Modeling
  chemo-hydrodynamic interactions of phoretic particles: A unified
  framework}},\ }\href {https://doi.org/10.1103/PhysRevFluids.4.124204}
  {\bibfield  {journal} {\bibinfo  {journal} {Phys. Rev. Fluids}\ }\textbf
  {\bibinfo {volume} {4}},\ \bibinfo {pages} {124204} (\bibinfo {year}
  {2019})}\BibitemShut {NoStop}%
\bibitem [{\citenamefont {Saha}\ \emph {et~al.}(2019)\citenamefont {Saha},
  \citenamefont {Ramaswamy},\ and\ \citenamefont {Golestanian}}]{Saha2019}%
  \BibitemOpen
  \bibfield  {author} {\bibinfo {author} {\bibfnamefont {S.}~\bibnamefont
  {Saha}}, \bibinfo {author} {\bibfnamefont {S.}~\bibnamefont {Ramaswamy}},\
  and\ \bibinfo {author} {\bibfnamefont {R.}~\bibnamefont {Golestanian}},\
  }\bibfield  {title} {\bibinfo {title} {Pairing, waltzing and scattering of
  chemotactic active colloids},\ }\href
  {https://doi.org/10.1088/1367-2630/ab20fd} {\bibfield  {journal} {\bibinfo
  {journal} {New J. Phys.}\ }\textbf {\bibinfo {volume} {21}},\ \bibinfo
  {pages} {063006} (\bibinfo {year} {2019})}\BibitemShut {NoStop}%
\bibitem [{\citenamefont {Vicsek}\ and\ \citenamefont
  {Zafeiris}(2012)}]{Vicsek2012}%
  \BibitemOpen
  \bibfield  {author} {\bibinfo {author} {\bibfnamefont {T.}~\bibnamefont
  {Vicsek}}\ and\ \bibinfo {author} {\bibfnamefont {A.}~\bibnamefont
  {Zafeiris}},\ }\bibfield  {title} {\bibinfo {title} {Collective motion},\
  }\href {https://doi.org/https://doi.org/10.1016/j.physrep.2012.03.004}
  {\bibfield  {journal} {\bibinfo  {journal} {Phys. Rep.}\ }\textbf {\bibinfo
  {volume} {517}},\ \bibinfo {pages} {71 } (\bibinfo {year}
  {2012})}\BibitemShut {NoStop}%
\bibitem [{\citenamefont {Elgeti}\ \emph {et~al.}(2015)\citenamefont {Elgeti},
  \citenamefont {Winkler},\ and\ \citenamefont {Gompper}}]{Elgeti2015}%
  \BibitemOpen
  \bibfield  {author} {\bibinfo {author} {\bibfnamefont {J.}~\bibnamefont
  {Elgeti}}, \bibinfo {author} {\bibfnamefont {R.}~\bibnamefont {Winkler}},\
  and\ \bibinfo {author} {\bibfnamefont {G.}~\bibnamefont {Gompper}},\
  }\bibfield  {title} {\bibinfo {title} {Physics of
  microswimmers{\textemdash}single particle motion and collective behavior: a
  review},\ }\href {https://doi.org/10.1088/0034-4885/78/5/056601} {\bibfield
  {journal} {\bibinfo  {journal} {Rep. Prog. Phys.}\ }\textbf {\bibinfo
  {volume} {78}},\ \bibinfo {pages} {056601} (\bibinfo {year}
  {2015})}\BibitemShut {NoStop}%
\bibitem [{\citenamefont {Li}\ \emph {et~al.}(2016)\citenamefont {Li},
  \citenamefont {Ostace},\ and\ \citenamefont {Ardekani}}]{Arezoo2016}%
  \BibitemOpen
  \bibfield  {author} {\bibinfo {author} {\bibfnamefont {G.}~\bibnamefont
  {Li}}, \bibinfo {author} {\bibfnamefont {A.}~\bibnamefont {Ostace}},\ and\
  \bibinfo {author} {\bibfnamefont {A.~M.}\ \bibnamefont {Ardekani}},\
  }\bibfield  {title} {\bibinfo {title} {Hydrodynamic interaction of swimming
  organisms in an inertial regime},\ }\href
  {https://doi.org/10.1103/PhysRevE.94.053104} {\bibfield  {journal} {\bibinfo
  {journal} {Phys. Rev. E}\ }\textbf {\bibinfo {volume} {94}},\ \bibinfo
  {pages} {053104} (\bibinfo {year} {2016})}\BibitemShut {NoStop}%
\bibitem [{\citenamefont {Weihs}(1973)}]{Weihs1973}%
  \BibitemOpen
  \bibfield  {author} {\bibinfo {author} {\bibfnamefont {D.}~\bibnamefont
  {Weihs}},\ }\bibfield  {title} {\bibinfo {title} {{Hydromechanics of Fish
  Schooling}},\ }\href {https://doi.org/https://doi.org/10.1038/241290a0}
  {\bibfield  {journal} {\bibinfo  {journal} {Nature}\ }\textbf {\bibinfo
  {volume} {241}},\ \bibinfo {pages} {290–291} (\bibinfo {year}
  {1973})}\BibitemShut {NoStop}%
\bibitem [{\citenamefont {Dai}\ \emph {et~al.}(2018)\citenamefont {Dai},
  \citenamefont {He}, \citenamefont {Zhang},\ and\ \citenamefont
  {Zhang}}]{Dai2018}%
  \BibitemOpen
  \bibfield  {author} {\bibinfo {author} {\bibfnamefont {L.}~\bibnamefont
  {Dai}}, \bibinfo {author} {\bibfnamefont {G.}~\bibnamefont {He}}, \bibinfo
  {author} {\bibfnamefont {X.}~\bibnamefont {Zhang}},\ and\ \bibinfo {author}
  {\bibfnamefont {X.}~\bibnamefont {Zhang}},\ }\bibfield  {title} {\bibinfo
  {title} {Stable formations of self-propelled fish-like swimmers induced by
  hydrodynamic interactions},\ }\href {https://doi.org/10.1098/rsif.2018.0490}
  {\bibfield  {journal} {\bibinfo  {journal} {J. R. Soc. Interface}\ }\textbf
  {\bibinfo {volume} {15}},\ \bibinfo {pages} {20180490} (\bibinfo {year}
  {2018})}\BibitemShut {NoStop}%
\bibitem [{\citenamefont {Pan}\ and\ \citenamefont {Dong}(2020)}]{Pan2020}%
  \BibitemOpen
  \bibfield  {author} {\bibinfo {author} {\bibfnamefont {Y.}~\bibnamefont
  {Pan}}\ and\ \bibinfo {author} {\bibfnamefont {H.}~\bibnamefont {Dong}},\
  }\bibfield  {title} {\bibinfo {title} {Computational analysis of hydrodynamic
  interactions in a high-density fish school},\ }\href
  {https://doi.org/10.1063/5.0028682} {\bibfield  {journal} {\bibinfo
  {journal} {Phys. Fluids}\ }\textbf {\bibinfo {volume} {32}},\ \bibinfo
  {pages} {121901} (\bibinfo {year} {2020})}\BibitemShut {NoStop}%
\bibitem [{\citenamefont {Jeffery}(1915)}]{Jeffery1915}%
  \BibitemOpen
  \bibfield  {author} {\bibinfo {author} {\bibfnamefont {G.~B.}\ \bibnamefont
  {Jeffery}},\ }\bibfield  {title} {\bibinfo {title} {On the steady rotation of
  a solid of revolution in a viscous fluid},\ }\href
  {https://doi.org/https://doi.org/10.1112/plms/s2_14.1.327} {\bibfield
  {journal} {\bibinfo  {journal} {Proc. Lond. Math. Soc.}\ }\textbf {\bibinfo
  {volume} {14}},\ \bibinfo {pages} {327} (\bibinfo {year} {1915})}\BibitemShut
  {NoStop}%
\bibitem [{\citenamefont {Stimson}\ and\ \citenamefont
  {Jeffery}(1926)}]{StimsonJeffery1926}%
  \BibitemOpen
  \bibfield  {author} {\bibinfo {author} {\bibfnamefont {M.}~\bibnamefont
  {Stimson}}\ and\ \bibinfo {author} {\bibfnamefont {G.~B.}\ \bibnamefont
  {Jeffery}},\ }\bibfield  {title} {\bibinfo {title} {{The motion of two
  spheres in a viscous fluid}},\ }\href@noop {} {\bibfield  {journal} {\bibinfo
   {journal} {Proc. R. Soc. Lond. A}\ }\textbf {\bibinfo {volume} {111}},\
  \bibinfo {pages} {110} (\bibinfo {year} {1926})}\BibitemShut {NoStop}%
\bibitem [{\citenamefont {Goddard}\ \emph {et~al.}(2020)\citenamefont
  {Goddard}, \citenamefont {Mills-Williams},\ and\ \citenamefont
  {Sun}}]{Goddard2020}%
  \BibitemOpen
  \bibfield  {author} {\bibinfo {author} {\bibfnamefont {B.~D.}\ \bibnamefont
  {Goddard}}, \bibinfo {author} {\bibfnamefont {R.~D.}\ \bibnamefont
  {Mills-Williams}},\ and\ \bibinfo {author} {\bibfnamefont {J.}~\bibnamefont
  {Sun}},\ }\bibfield  {title} {\bibinfo {title} {The singular hydrodynamic
  interactions between two spheres in stokes flow},\ }\href
  {https://doi.org/https://doi.org/10.1063/5.0009053} {\bibfield  {journal}
  {\bibinfo  {journal} {Phys. Fluids}\ }\textbf {\bibinfo {volume} {32}},\
  \bibinfo {pages} {062001} (\bibinfo {year} {2020})}\BibitemShut {NoStop}%
\bibitem [{\citenamefont {Goldman}\ \emph {et~al.}(1966)\citenamefont
  {Goldman}, \citenamefont {Cox},\ and\ \citenamefont {Brenner}}]{Goldman1966}%
  \BibitemOpen
  \bibfield  {author} {\bibinfo {author} {\bibfnamefont {A.}~\bibnamefont
  {Goldman}}, \bibinfo {author} {\bibfnamefont {R.}~\bibnamefont {Cox}},\ and\
  \bibinfo {author} {\bibfnamefont {H.}~\bibnamefont {Brenner}},\ }\bibfield
  {title} {\bibinfo {title} {The slow motion of two identical arbitrarily
  oriented spheres through a viscous fluid},\ }\href
  {https://doi.org/https://doi.org/10.1016/0009-2509(66)85036-4} {\bibfield
  {journal} {\bibinfo  {journal} {Chem. Eng. Sci}\ }\textbf {\bibinfo {volume}
  {21}},\ \bibinfo {pages} {1151} (\bibinfo {year} {1966})}\BibitemShut
  {NoStop}%
\bibitem [{\citenamefont {Wakiya}(1967)}]{Wakiya1967}%
  \BibitemOpen
  \bibfield  {author} {\bibinfo {author} {\bibfnamefont {S.}~\bibnamefont
  {Wakiya}},\ }\bibfield  {title} {\bibinfo {title} {Slow motions of a viscous
  fluid around two spheres},\ }\href {https://doi.org/10.1143/JPSJ.22.1101}
  {\bibfield  {journal} {\bibinfo  {journal} {J. Phys. Soc. Japan}\ }\textbf
  {\bibinfo {volume} {22}},\ \bibinfo {pages} {1101} (\bibinfo {year}
  {1967})}\BibitemShut {NoStop}%
\bibitem [{\citenamefont {Dabro\'{s}}(1985)}]{Dabros1985}%
  \BibitemOpen
  \bibfield  {author} {\bibinfo {author} {\bibfnamefont {T.}~\bibnamefont
  {Dabro\'{s}}},\ }\bibfield  {title} {\bibinfo {title} {A singularity method
  for calculating hydrodynamic forces and particle velocities in
  low-{R}eynolds-number flows},\ }\href
  {https://doi.org/https://doi.org/10.1017/S0022112085001951} {\bibfield
  {journal} {\bibinfo  {journal} {J. Fluid Mech.}\ }\textbf {\bibinfo {volume}
  {156}},\ \bibinfo {pages} {1–21} (\bibinfo {year} {1985})}\BibitemShut
  {NoStop}%
\bibitem [{\citenamefont {Kim}\ and\ \citenamefont {Mifflin}(1985)}]{Kim1985}%
  \BibitemOpen
  \bibfield  {author} {\bibinfo {author} {\bibfnamefont {S.}~\bibnamefont
  {Kim}}\ and\ \bibinfo {author} {\bibfnamefont {R.~T.}\ \bibnamefont
  {Mifflin}},\ }\bibfield  {title} {\bibinfo {title} {The resistance and
  mobility functions of two equal spheres in low‐{R}eynolds‐number flow},\
  }\href {https://doi.org/10.1063/1.865384} {\bibfield  {journal} {\bibinfo
  {journal} {Phys. Fluids}\ }\textbf {\bibinfo {volume} {28}},\ \bibinfo
  {pages} {2033} (\bibinfo {year} {1985})}\BibitemShut {NoStop}%
\bibitem [{\citenamefont {Yoon}\ and\ \citenamefont {Kim}(1987)}]{YoonKim1987}%
  \BibitemOpen
  \bibfield  {author} {\bibinfo {author} {\bibfnamefont {B.~J.}\ \bibnamefont
  {Yoon}}\ and\ \bibinfo {author} {\bibfnamefont {S.}~\bibnamefont {Kim}},\
  }\bibfield  {title} {\bibinfo {title} {Note on the direct calculation of
  mobility functions for two equal-sized spheres in {S}tokes flow},\ }\href
  {https://doi.org/https://doi.org/10.1017/S0022112087003240} {\bibfield
  {journal} {\bibinfo  {journal} {J. Fluid Mech.}\ }\textbf {\bibinfo {volume}
  {185}},\ \bibinfo {pages} {437–446} (\bibinfo {year} {1987})}\BibitemShut
  {NoStop}%
\bibitem [{\citenamefont {Felderhof}(1977)}]{Felderhof1977}%
  \BibitemOpen
  \bibfield  {author} {\bibinfo {author} {\bibfnamefont {B.}~\bibnamefont
  {Felderhof}},\ }\bibfield  {title} {\bibinfo {title} {Hydrodynamic
  interaction between two spheres},\ }\href
  {https://doi.org/https://doi.org/10.1016/0378-4371(77)90111-X} {\bibfield
  {journal} {\bibinfo  {journal} {Physica A}\ }\textbf {\bibinfo {volume}
  {89}},\ \bibinfo {pages} {373 } (\bibinfo {year} {1977})}\BibitemShut
  {NoStop}%
\bibitem [{\citenamefont {Cichocki}\ \emph {et~al.}(1988)\citenamefont
  {Cichocki}, \citenamefont {Felderhof},\ and\ \citenamefont
  {Schmitz}}]{Cichocki1988}%
  \BibitemOpen
  \bibfield  {author} {\bibinfo {author} {\bibfnamefont {B.}~\bibnamefont
  {Cichocki}}, \bibinfo {author} {\bibfnamefont {B.}~\bibnamefont
  {Felderhof}},\ and\ \bibinfo {author} {\bibfnamefont {R.}~\bibnamefont
  {Schmitz}},\ }\bibfield  {title} {\bibinfo {title} {Hydrodynamic interactions
  between two spherical particles},\ }\href
  {https://www.researchgate.net/publication/257160907_Hydrodynamic_interactions_between_two_spherical_particles}
  {\bibfield  {journal} {\bibinfo  {journal} {Physico Chem. Hyd.}\ }\textbf
  {\bibinfo {volume} {10}},\ \bibinfo {pages} {383} (\bibinfo {year}
  {1988})}\BibitemShut {NoStop}%
\bibitem [{\citenamefont {Jayaweera}\ \emph {et~al.}(1964)\citenamefont
  {Jayaweera}, \citenamefont {Mason},\ and\ \citenamefont
  {Slack}}]{Jayaweera1964}%
  \BibitemOpen
  \bibfield  {author} {\bibinfo {author} {\bibfnamefont {K.~O. L.~F.}\
  \bibnamefont {Jayaweera}}, \bibinfo {author} {\bibfnamefont {B.~J.}\
  \bibnamefont {Mason}},\ and\ \bibinfo {author} {\bibfnamefont {G.~W.}\
  \bibnamefont {Slack}},\ }\bibfield  {title} {\bibinfo {title} {{The behaviour
  of clusters of spheres falling in a viscous fluid Part 1. Experiment}},\
  }\href {https://doi.org/https://10.1017/S0022112064001069} {\bibfield
  {journal} {\bibinfo  {journal} {J. Fluid Mech.}\ }\textbf {\bibinfo {volume}
  {20}},\ \bibinfo {pages} {121–} (\bibinfo {year} {1964})}\BibitemShut
  {NoStop}%
\bibitem [{\citenamefont {Cichocki}\ \emph {et~al.}(1994)\citenamefont
  {Cichocki}, \citenamefont {Felderhof}, \citenamefont {Hinsen}, \citenamefont
  {Wajnryb},\ and\ \citenamefont {B\l{}awzdziewicz}}]{Cichocki1994}%
  \BibitemOpen
  \bibfield  {author} {\bibinfo {author} {\bibfnamefont {B.}~\bibnamefont
  {Cichocki}}, \bibinfo {author} {\bibfnamefont {B.~U.}\ \bibnamefont
  {Felderhof}}, \bibinfo {author} {\bibfnamefont {K.}~\bibnamefont {Hinsen}},
  \bibinfo {author} {\bibfnamefont {E.}~\bibnamefont {Wajnryb}},\ and\ \bibinfo
  {author} {\bibfnamefont {J.}~\bibnamefont {B\l{}awzdziewicz}},\ }\bibfield
  {title} {\bibinfo {title} {Friction and mobility of many spheres in {S}tokes
  flow},\ }\href {https://doi.org/10.1063/1.466366} {\bibfield  {journal}
  {\bibinfo  {journal} {J. Chem. Phys.}\ }\textbf {\bibinfo {volume} {100}},\
  \bibinfo {pages} {3780} (\bibinfo {year} {1994})}\BibitemShut {NoStop}%
\bibitem [{\citenamefont {Hocking}(1964)}]{Hocking1964}%
  \BibitemOpen
  \bibfield  {author} {\bibinfo {author} {\bibfnamefont {L.~M.}\ \bibnamefont
  {Hocking}},\ }\bibfield  {title} {\bibinfo {title} {{The behaviour of
  clusters of spheres falling in a viscous fluid Part 2. Slow motion theory}},\
  }\href {https://doi.org/https://doi.org/10.1017/S0022112064001070} {\bibfield
   {journal} {\bibinfo  {journal} {J. Fluid Mech.}\ }\textbf {\bibinfo {volume}
  {20}},\ \bibinfo {pages} {129–} (\bibinfo {year} {1964})}\BibitemShut
  {NoStop}%
\bibitem [{\citenamefont {Nasseri}\ and\ \citenamefont
  {Phan-Thien}(1997)}]{Nasseri1997}%
  \BibitemOpen
  \bibfield  {author} {\bibinfo {author} {\bibfnamefont {S.}~\bibnamefont
  {Nasseri}}\ and\ \bibinfo {author} {\bibfnamefont {N.}~\bibnamefont
  {Phan-Thien}},\ }\bibfield  {title} {\bibinfo {title} {{Hydrodynamic
  interaction between two nearby swimming micromachines}},\ }\href
  {https://doi.org/https://doi.org/10.1007/s004660050275} {\bibfield  {journal}
  {\bibinfo  {journal} {Comp. Mech.}\ }\textbf {\bibinfo {volume} {20}},\
  \bibinfo {pages} {551} (\bibinfo {year} {1997})}\BibitemShut {NoStop}%
\bibitem [{\citenamefont {Ishikawa}\ \emph {et~al.}(2007)\citenamefont
  {Ishikawa}, \citenamefont {Sekiya}, \citenamefont {Imai},\ and\ \citenamefont
  {Yamaguchi}}]{Ishikawa2007}%
  \BibitemOpen
  \bibfield  {author} {\bibinfo {author} {\bibfnamefont {T.}~\bibnamefont
  {Ishikawa}}, \bibinfo {author} {\bibfnamefont {G.}~\bibnamefont {Sekiya}},
  \bibinfo {author} {\bibfnamefont {Y.}~\bibnamefont {Imai}},\ and\ \bibinfo
  {author} {\bibfnamefont {T.}~\bibnamefont {Yamaguchi}},\ }\bibfield  {title}
  {\bibinfo {title} {{Hydrodynamic interactions between two swimming
  bacteria}},\ }\href
  {https://doi.org/https://doi.org/10.1529/biophysj.107.110254} {\bibfield
  {journal} {\bibinfo  {journal} {Biophys. J.}\ }\textbf {\bibinfo {volume}
  {93}},\ \bibinfo {pages} {2217} (\bibinfo {year} {2007})}\BibitemShut
  {NoStop}%
\bibitem [{\citenamefont {Ishikawa}\ \emph {et~al.}(2020)\citenamefont
  {Ishikawa}, \citenamefont {Pedley}, \citenamefont {Drescher},\ and\
  \citenamefont {Goldstein}}]{Ishikawa2020}%
  \BibitemOpen
  \bibfield  {author} {\bibinfo {author} {\bibfnamefont {T.}~\bibnamefont
  {Ishikawa}}, \bibinfo {author} {\bibfnamefont {T.~J.}\ \bibnamefont
  {Pedley}}, \bibinfo {author} {\bibfnamefont {K.}~\bibnamefont {Drescher}},\
  and\ \bibinfo {author} {\bibfnamefont {R.~E.}\ \bibnamefont {Goldstein}},\
  }\bibfield  {title} {\bibinfo {title} {{Stability of dancing
  \textit{Volvox}}},\ }\href {https://doi.org/10.1017/jfm.2020.613} {\bibfield
  {journal} {\bibinfo  {journal} {J. Fluid. Mech.}\ }\textbf {\bibinfo {volume}
  {903}},\ \bibinfo {pages} {A11} (\bibinfo {year} {2020})}\BibitemShut
  {NoStop}%
\bibitem [{\citenamefont {Gyrya}\ \emph {et~al.}(2010)\citenamefont {Gyrya},
  \citenamefont {Aranson}, \citenamefont {Berlyand},\ and\ \citenamefont
  {Karpeev}}]{Gyrya2010}%
  \BibitemOpen
  \bibfield  {author} {\bibinfo {author} {\bibfnamefont {V.}~\bibnamefont
  {Gyrya}}, \bibinfo {author} {\bibfnamefont {I.~S.}\ \bibnamefont {Aranson}},
  \bibinfo {author} {\bibfnamefont {L.~V.}\ \bibnamefont {Berlyand}},\ and\
  \bibinfo {author} {\bibfnamefont {D.}~\bibnamefont {Karpeev}},\ }\bibfield
  {title} {\bibinfo {title} {A model of hydrodynamic interaction between
  swimming bacteria},\ }\href
  {https://doi.org/https://doi.org/10.1007/s11538-009-9442-6} {\bibfield
  {journal} {\bibinfo  {journal} {Bull. Math. Biol.}\ }\textbf {\bibinfo
  {volume} {72}},\ \bibinfo {pages} {148} (\bibinfo {year} {2010})}\BibitemShut
  {NoStop}%
\bibitem [{\citenamefont {G\"{o}tze}\ and\ \citenamefont
  {Gompper}(2010)}]{Goetze2010}%
  \BibitemOpen
  \bibfield  {author} {\bibinfo {author} {\bibfnamefont {I.~O.}\ \bibnamefont
  {G\"{o}tze}}\ and\ \bibinfo {author} {\bibfnamefont {G.}~\bibnamefont
  {Gompper}},\ }\bibfield  {title} {\bibinfo {title} {Mesoscale simulations of
  hydrodynamic squirmer interactions},\ }\href
  {https://doi.org/10.1103/PhysRevE.82.041921} {\bibfield  {journal} {\bibinfo
  {journal} {Phys. Rev. E}\ }\textbf {\bibinfo {volume} {82}},\ \bibinfo
  {pages} {041921} (\bibinfo {year} {2010})}\BibitemShut {NoStop}%
\bibitem [{\citenamefont {Molina}\ \emph {et~al.}(2013)\citenamefont {Molina},
  \citenamefont {Nakayama},\ and\ \citenamefont {Yamamoto}}]{Molina2013}%
  \BibitemOpen
  \bibfield  {author} {\bibinfo {author} {\bibfnamefont {J.~J.}\ \bibnamefont
  {Molina}}, \bibinfo {author} {\bibfnamefont {Y.}~\bibnamefont {Nakayama}},\
  and\ \bibinfo {author} {\bibfnamefont {R.}~\bibnamefont {Yamamoto}},\
  }\bibfield  {title} {\bibinfo {title} {Hydrodynamic interactions of
  self-propelled swimmers},\ }\href {https://doi.org/10.1039/C3SM00140G}
  {\bibfield  {journal} {\bibinfo  {journal} {Soft Matter}\ }\textbf {\bibinfo
  {volume} {9}},\ \bibinfo {pages} {4923} (\bibinfo {year} {2013})}\BibitemShut
  {NoStop}%
\bibitem [{\citenamefont {Liao}\ \emph {et~al.}(2007)\citenamefont {Liao},
  \citenamefont {Subramanian}, \citenamefont {DeLisa}, \citenamefont {Koch},\
  and\ \citenamefont {Wu}}]{Liao2007}%
  \BibitemOpen
  \bibfield  {author} {\bibinfo {author} {\bibfnamefont {Q.}~\bibnamefont
  {Liao}}, \bibinfo {author} {\bibfnamefont {G.}~\bibnamefont {Subramanian}},
  \bibinfo {author} {\bibfnamefont {M.}~\bibnamefont {DeLisa}}, \bibinfo
  {author} {\bibfnamefont {D.}~\bibnamefont {Koch}},\ and\ \bibinfo {author}
  {\bibfnamefont {M.}~\bibnamefont {Wu}},\ }\bibfield  {title} {\bibinfo
  {title} {Pair velocity correlations among swimming \textit{{E}scherichia
  coli} bacteria are determined by force-quadrupole hydrodynamic
  interactions},\ }\href {https://doi.org/https://doi.org/10.1063/1.2742423}
  {\bibfield  {journal} {\bibinfo  {journal} {Phys. Fluids}\ }\textbf {\bibinfo
  {volume} {19}},\ \bibinfo {pages} {061701} (\bibinfo {year}
  {2007})}\BibitemShut {NoStop}%
\bibitem [{\citenamefont {Drescher}\ \emph {et~al.}(2009)\citenamefont
  {Drescher}, \citenamefont {Leptos}, \citenamefont {Tuval}, \citenamefont
  {Ishikawa}, \citenamefont {Pedley},\ and\ \citenamefont
  {Goldstein}}]{Drescher2009}%
  \BibitemOpen
  \bibfield  {author} {\bibinfo {author} {\bibfnamefont {K.}~\bibnamefont
  {Drescher}}, \bibinfo {author} {\bibfnamefont {K.~C.}\ \bibnamefont
  {Leptos}}, \bibinfo {author} {\bibfnamefont {I.}~\bibnamefont {Tuval}},
  \bibinfo {author} {\bibfnamefont {T.}~\bibnamefont {Ishikawa}}, \bibinfo
  {author} {\bibfnamefont {T.~J.}\ \bibnamefont {Pedley}},\ and\ \bibinfo
  {author} {\bibfnamefont {R.~E.}\ \bibnamefont {Goldstein}},\ }\bibfield
  {title} {\bibinfo {title} {Dancing \textit{{V}olvox}: Hydrodynamic bound
  states of swimming algae},\ }\href
  {https://doi.org/10.1103/PhysRevLett.102.168101} {\bibfield  {journal}
  {\bibinfo  {journal} {Phys. Rev. Lett.}\ }\textbf {\bibinfo {volume} {102}},\
  \bibinfo {pages} {1168101} (\bibinfo {year} {2009})}\BibitemShut {NoStop}%
\bibitem [{\citenamefont {Kim}\ and\ \citenamefont {Powers}(2004)}]{Kim2004b}%
  \BibitemOpen
  \bibfield  {author} {\bibinfo {author} {\bibfnamefont {M.}~\bibnamefont
  {Kim}}\ and\ \bibinfo {author} {\bibfnamefont {T.~R.}\ \bibnamefont
  {Powers}},\ }\bibfield  {title} {\bibinfo {title} {{Hydrodynamic interactions
  between rotating helices}},\ }\href
  {https://doi.org/10.1103/PhysRevE.69.061910} {\bibfield  {journal} {\bibinfo
  {journal} {Phys. Rev. E}\ }\textbf {\bibinfo {volume} {69}},\ \bibinfo
  {pages} {061910} (\bibinfo {year} {2004})}\BibitemShut {NoStop}%
\bibitem [{\citenamefont {Reigh}\ \emph {et~al.}(2012)\citenamefont {Reigh},
  \citenamefont {Winkler},\ and\ \citenamefont {Gompper}}]{Reigh2012}%
  \BibitemOpen
  \bibfield  {author} {\bibinfo {author} {\bibfnamefont {S.~Y.}\ \bibnamefont
  {Reigh}}, \bibinfo {author} {\bibfnamefont {R.~G.}\ \bibnamefont {Winkler}},\
  and\ \bibinfo {author} {\bibfnamefont {G.}~\bibnamefont {Gompper}},\
  }\bibfield  {title} {\bibinfo {title} {Synchronization and bundling of
  anchored bacterial flagella},\ }\href
  {https://doi.org/https://doi.org/10.1039/c2sm07378a} {\bibfield  {journal}
  {\bibinfo  {journal} {Soft Matter}\ }\textbf {\bibinfo {volume} {8}},\
  \bibinfo {pages} {4363} (\bibinfo {year} {2012})}\BibitemShut {NoStop}%
\bibitem [{\citenamefont {Reigh}\ \emph {et~al.}(2013)\citenamefont {Reigh},
  \citenamefont {Winkler},\ and\ \citenamefont {Gompper}}]{Reigh2013}%
  \BibitemOpen
  \bibfield  {author} {\bibinfo {author} {\bibfnamefont {S.~Y.}\ \bibnamefont
  {Reigh}}, \bibinfo {author} {\bibfnamefont {R.~G.}\ \bibnamefont {Winkler}},\
  and\ \bibinfo {author} {\bibfnamefont {G.}~\bibnamefont {Gompper}},\
  }\bibfield  {title} {\bibinfo {title} {Synchronization, slippage, and
  unbundling of driven helical flagella},\ }\href
  {https://doi.org/https://doi.org/10.1371/journal.pone.0070868} {\bibfield
  {journal} {\bibinfo  {journal} {PLoS ONE}\ }\textbf {\bibinfo {volume} {8}},\
  \bibinfo {pages} {e70868} (\bibinfo {year} {2013})}\BibitemShut {NoStop}%
\bibitem [{\citenamefont {Chakrabarti}\ and\ \citenamefont
  {Saintillan}(2019)}]{Chakrabarti2019}%
  \BibitemOpen
  \bibfield  {author} {\bibinfo {author} {\bibfnamefont {B.}~\bibnamefont
  {Chakrabarti}}\ and\ \bibinfo {author} {\bibfnamefont {D.}~\bibnamefont
  {Saintillan}},\ }\bibfield  {title} {\bibinfo {title} {Hydrodynamic
  synchronization of spontaneously beating filaments},\ }\href
  {https://doi.org/https://doi.org/10.1103/PhysRevLett.123.208101} {\bibfield
  {journal} {\bibinfo  {journal} {Phys. Rev. Lett.}\ }\textbf {\bibinfo
  {volume} {123}},\ \bibinfo {pages} {208101} (\bibinfo {year}
  {2019})}\BibitemShut {NoStop}%
\bibitem [{\citenamefont {Man}\ and\ \citenamefont {Kanso}(2020)}]{Man2020}%
  \BibitemOpen
  \bibfield  {author} {\bibinfo {author} {\bibfnamefont {Y.}~\bibnamefont
  {Man}}\ and\ \bibinfo {author} {\bibfnamefont {E.}~\bibnamefont {Kanso}},\
  }\bibfield  {title} {\bibinfo {title} {Multisynchrony in active
  microfilaments},\ }\href
  {https://doi.org/https://doi.org/10.1103/PhysRevLett.125.148101} {\bibfield
  {journal} {\bibinfo  {journal} {Phys. Rev. Lett.}\ }\textbf {\bibinfo
  {volume} {125}},\ \bibinfo {pages} {148101} (\bibinfo {year}
  {2020})}\BibitemShut {NoStop}%
\bibitem [{\citenamefont {Guirao}\ and\ \citenamefont
  {Joanny}(2007)}]{Joanny2007mcw}%
  \BibitemOpen
  \bibfield  {author} {\bibinfo {author} {\bibfnamefont {B.}~\bibnamefont
  {Guirao}}\ and\ \bibinfo {author} {\bibfnamefont {J.-F.}\ \bibnamefont
  {Joanny}},\ }\bibfield  {title} {\bibinfo {title} {Spontaneous creation of
  macroscopic flow and metachronal waves in an array of cilia},\ }\href
  {https://doi.org/https://doi.org/10.1529/biophysj.106.084897} {\bibfield
  {journal} {\bibinfo  {journal} {Biophys. J.}\ }\textbf {\bibinfo {volume}
  {92}},\ \bibinfo {pages} {1900 } (\bibinfo {year} {2007})}\BibitemShut
  {NoStop}%
\bibitem [{\citenamefont {Elgeti}\ and\ \citenamefont
  {Gompper}(2013)}]{Elgeti2013}%
  \BibitemOpen
  \bibfield  {author} {\bibinfo {author} {\bibfnamefont {J.}~\bibnamefont
  {Elgeti}}\ and\ \bibinfo {author} {\bibfnamefont {G.}~\bibnamefont
  {Gompper}},\ }\bibfield  {title} {\bibinfo {title} {Emergence of metachronal
  waves in cilia arrays},\ }\href {https://doi.org/10.1073/pnas.1218869110}
  {\bibfield  {journal} {\bibinfo  {journal} {Proc. Natl. Acad. Sci. USA}\
  }\textbf {\bibinfo {volume} {110}},\ \bibinfo {pages} {4470} (\bibinfo {year}
  {2013})}\BibitemShut {NoStop}%
\bibitem [{\citenamefont {Nguyen}\ and\ \citenamefont
  {Graham}(2018)}]{Nguyen2018}%
  \BibitemOpen
  \bibfield  {author} {\bibinfo {author} {\bibfnamefont {F.~T.~M.}\
  \bibnamefont {Nguyen}}\ and\ \bibinfo {author} {\bibfnamefont {M.~D.}\
  \bibnamefont {Graham}},\ }\bibfield  {title} {\bibinfo {title} {Impacts of
  multiflagellarity on stability and speed of bacterial locomotion},\ }\href
  {https://doi.org/https://doi.org/10.1103/PhysRevE.98.042419} {\bibfield
  {journal} {\bibinfo  {journal} {Phys. Rev. E}\ }\textbf {\bibinfo {volume}
  {98}},\ \bibinfo {pages} {042419} (\bibinfo {year} {2018})}\BibitemShut
  {NoStop}%
\bibitem [{\citenamefont {Man}\ \emph {et~al.}(2016)\citenamefont {Man},
  \citenamefont {Koens},\ and\ \citenamefont {Lauga}}]{Man2016}%
  \BibitemOpen
  \bibfield  {author} {\bibinfo {author} {\bibfnamefont {Y.}~\bibnamefont
  {Man}}, \bibinfo {author} {\bibfnamefont {L.}~\bibnamefont {Koens}},\ and\
  \bibinfo {author} {\bibfnamefont {E.}~\bibnamefont {Lauga}},\ }\bibfield
  {title} {\bibinfo {title} {{Hydrodynamic interactions between nearby slender
  filaments}},\ }\href {https://doi.org/10.1209/0295-5075/116/24002} {\bibfield
   {journal} {\bibinfo  {journal} {Europhys. Lett.}\ }\textbf {\bibinfo
  {volume} {116}},\ \bibinfo {pages} {24002} (\bibinfo {year}
  {2016})}\BibitemShut {NoStop}%
\bibitem [{\citenamefont {Kim}\ \emph {et~al.}(2003)\citenamefont {Kim},
  \citenamefont {Bird}, \citenamefont {Van~Parys}, \citenamefont {Breuer},\
  and\ \citenamefont {Powers}}]{Kim2003}%
  \BibitemOpen
  \bibfield  {author} {\bibinfo {author} {\bibfnamefont {M.~J.}\ \bibnamefont
  {Kim}}, \bibinfo {author} {\bibfnamefont {J.~C.}\ \bibnamefont {Bird}},
  \bibinfo {author} {\bibfnamefont {A.~J.}\ \bibnamefont {Van~Parys}}, \bibinfo
  {author} {\bibfnamefont {K.~S.}\ \bibnamefont {Breuer}},\ and\ \bibinfo
  {author} {\bibfnamefont {T.~R.}\ \bibnamefont {Powers}},\ }\bibfield  {title}
  {\bibinfo {title} {A macroscopic scale model of bacterial flagellar
  bundling},\ }\href {https://doi.org/https://doi.org/10.1073/pnas.2633596100}
  {\bibfield  {journal} {\bibinfo  {journal} {Proc. Natl. Acad. Sci. USA}\
  }\textbf {\bibinfo {volume} {100}},\ \bibinfo {pages} {15481} (\bibinfo
  {year} {2003})}\BibitemShut {NoStop}%
\bibitem [{\citenamefont {Kim}\ \emph {et~al.}(2004)\citenamefont {Kim},
  \citenamefont {Kim}, \citenamefont {Bird}, \citenamefont {Park},
  \citenamefont {Powers},\ and\ \citenamefont {Breuer}}]{Kim2004a}%
  \BibitemOpen
  \bibfield  {author} {\bibinfo {author} {\bibfnamefont {M.~J.}\ \bibnamefont
  {Kim}}, \bibinfo {author} {\bibfnamefont {M.~J.}\ \bibnamefont {Kim}},
  \bibinfo {author} {\bibfnamefont {J.~C.}\ \bibnamefont {Bird}}, \bibinfo
  {author} {\bibfnamefont {J.}~\bibnamefont {Park}}, \bibinfo {author}
  {\bibfnamefont {T.~R.}\ \bibnamefont {Powers}},\ and\ \bibinfo {author}
  {\bibfnamefont {K.~S.}\ \bibnamefont {Breuer}},\ }\bibfield  {title}
  {\bibinfo {title} {Particle image velocimetry experiments on a macro-scale
  model for bacterial flagellar bundling},\ }\href
  {https://doi.org/10.1007/s00348-004-0848-5} {\bibfield  {journal} {\bibinfo
  {journal} {Exp. Fluids}\ }\textbf {\bibinfo {volume} {37}},\ \bibinfo {pages}
  {782} (\bibinfo {year} {2004})}\BibitemShut {NoStop}%
\bibitem [{\citenamefont {Hancock}(1953)}]{Hancock1953}%
  \BibitemOpen
  \bibfield  {author} {\bibinfo {author} {\bibfnamefont {G.}~\bibnamefont
  {Hancock}},\ }\bibfield  {title} {\bibinfo {title} {{The self-propulsion of
  microscopic organisms through liquids}},\ }\href
  {https://doi.org/https://doi.org/10.1098/rspa.1953.0048} {\bibfield
  {journal} {\bibinfo  {journal} {Proc. R. Soc. A}\ }\textbf {\bibinfo {volume}
  {217}},\ \bibinfo {pages} {96} (\bibinfo {year} {1953})}\BibitemShut
  {NoStop}%
\bibitem [{\citenamefont {Gray}\ and\ \citenamefont
  {Hancock}(1955)}]{Gray1955}%
  \BibitemOpen
  \bibfield  {author} {\bibinfo {author} {\bibfnamefont {J.}~\bibnamefont
  {Gray}}\ and\ \bibinfo {author} {\bibfnamefont {G.~J.}\ \bibnamefont
  {Hancock}},\ }\bibfield  {title} {\bibinfo {title} {The propulsion of
  sea-urchin spermatozoa},\ }\href
  {https://jeb.biologists.org/content/32/4/802} {\bibfield  {journal} {\bibinfo
   {journal} {J. Exp. Biol.}\ }\textbf {\bibinfo {volume} {32}},\ \bibinfo
  {pages} {802} (\bibinfo {year} {1955})}\BibitemShut {NoStop}%
\bibitem [{\citenamefont {Lighthill}(1996)}]{Lighthill1996_helical}%
  \BibitemOpen
  \bibfield  {author} {\bibinfo {author} {\bibfnamefont {J.}~\bibnamefont
  {Lighthill}},\ }\bibfield  {title} {\bibinfo {title} {Helical distributions
  of stokeslets},\ }\href {https://doi.org/https://doi.org/10.1007/BF00118823}
  {\bibfield  {journal} {\bibinfo  {journal} {J. Eng. Math.}\ }\textbf
  {\bibinfo {volume} {30}},\ \bibinfo {pages} {35} (\bibinfo {year}
  {1996})}\BibitemShut {NoStop}%
\bibitem [{\citenamefont {Cox}(1970)}]{Cox1970}%
  \BibitemOpen
  \bibfield  {author} {\bibinfo {author} {\bibfnamefont {R.~G.}\ \bibnamefont
  {Cox}},\ }\bibfield  {title} {\bibinfo {title} {{The motion of long slender
  bodies in a viscous fluid Part 1. General theory}},\ }\href
  {https://doi.org/10.1017/S002211207000215X} {\bibfield  {journal} {\bibinfo
  {journal} {J. Fluid Mech.}\ }\textbf {\bibinfo {volume} {44}},\ \bibinfo
  {pages} {791–810} (\bibinfo {year} {1970})}\BibitemShut {NoStop}%
\bibitem [{\citenamefont {Lighthill}(1976)}]{Lighthill1976}%
  \BibitemOpen
  \bibfield  {author} {\bibinfo {author} {\bibfnamefont {J.}~\bibnamefont
  {Lighthill}},\ }\bibfield  {title} {\bibinfo {title} {Flagellar
  hydrodynamics---{The John von Neumann} lecture, 1975},\ }\href
  {https://www.jstor.org/stable/2028784} {\bibfield  {journal} {\bibinfo
  {journal} {SIAM Rev.}\ }\textbf {\bibinfo {volume} {18}},\ \bibinfo {pages}
  {161} (\bibinfo {year} {1976})}\BibitemShut {NoStop}%
\bibitem [{\citenamefont {Johnson}(1980)}]{Johnson1980}%
  \BibitemOpen
  \bibfield  {author} {\bibinfo {author} {\bibfnamefont {R.~E.}\ \bibnamefont
  {Johnson}},\ }\bibfield  {title} {\bibinfo {title} {{An improved slender-body
  theory for Stokes flow}},\ }\href {https://doi.org/10.1017/S0022112080000687}
  {\bibfield  {journal} {\bibinfo  {journal} {J. Fluid Mech.}\ }\textbf
  {\bibinfo {volume} {99}},\ \bibinfo {pages} {411} (\bibinfo {year}
  {1980})}\BibitemShut {NoStop}%
\bibitem [{\citenamefont {Tornberg}\ and\ \citenamefont
  {Shelley}(2004)}]{Tornberg2004}%
  \BibitemOpen
  \bibfield  {author} {\bibinfo {author} {\bibfnamefont {A.~K.}\ \bibnamefont
  {Tornberg}}\ and\ \bibinfo {author} {\bibfnamefont {M.~J.}\ \bibnamefont
  {Shelley}},\ }\bibfield  {title} {\bibinfo {title} {{Simulating the dynamics
  and interactions of flexible fibers in Stokes flows}},\ }\href
  {https://doi.org/10.1016/j.jcp.2003.10.017} {\bibfield  {journal} {\bibinfo
  {journal} {J. Comp. Phys.}\ }\textbf {\bibinfo {volume} {196}},\ \bibinfo
  {pages} {8} (\bibinfo {year} {2004})}\BibitemShut {NoStop}%
\bibitem [{\citenamefont {Maxian}\ \emph {et~al.}(2021)\citenamefont {Maxian},
  \citenamefont {Mogilner},\ and\ \citenamefont {Donev}}]{Maxian2021}%
  \BibitemOpen
  \bibfield  {author} {\bibinfo {author} {\bibfnamefont {O.}~\bibnamefont
  {Maxian}}, \bibinfo {author} {\bibfnamefont {A.}~\bibnamefont {Mogilner}},\
  and\ \bibinfo {author} {\bibfnamefont {A.}~\bibnamefont {Donev}},\ }\bibfield
   {title} {\bibinfo {title} {Integral-based spectral method for inextensible
  slender fibers in stokes flow},\ }\href
  {https://doi.org/10.1103/PhysRevFluids.6.014102} {\bibfield  {journal}
  {\bibinfo  {journal} {Phys. Rev. Fluids}\ }\textbf {\bibinfo {volume} {6}},\
  \bibinfo {pages} {014102} (\bibinfo {year} {2021})}\BibitemShut {NoStop}%
\bibitem [{\citenamefont {Koens}(2016)}]{thesisKoens}%
  \BibitemOpen
  \bibfield  {author} {\bibinfo {author} {\bibfnamefont {L.~M.}\ \bibnamefont
  {Koens}},\ }\emph {\bibinfo {title} {{The Hydrodynamics of Complex
  Microswimmers: An exploration of slender filaments and ribbons}}},\
  \href@noop {} {Ph.D. thesis},\ \bibinfo  {school} {University of Cambridge}
  (\bibinfo {year} {2016})\BibitemShut {NoStop}%
\bibitem [{\citenamefont {G\"{o}tz}(2000)}]{thesisGotz}%
  \BibitemOpen
  \bibfield  {author} {\bibinfo {author} {\bibfnamefont {T.}~\bibnamefont
  {G\"{o}tz}},\ }\emph {\bibinfo {title} {Interactions of fibers and flow:
  asymptotics, theory and numerics}},\ \href@noop {} {Ph.D. thesis},\ \bibinfo
  {school} {University of Kaiserslautern} (\bibinfo {year} {2000})\BibitemShut
  {NoStop}%
\bibitem [{\citenamefont {Darnton}\ \emph {et~al.}(2004)\citenamefont
  {Darnton}, \citenamefont {Turner}, \citenamefont {Breuer},\ and\
  \citenamefont {Berg}}]{Darnton2004}%
  \BibitemOpen
  \bibfield  {author} {\bibinfo {author} {\bibfnamefont {N.}~\bibnamefont
  {Darnton}}, \bibinfo {author} {\bibfnamefont {L.}~\bibnamefont {Turner}},
  \bibinfo {author} {\bibfnamefont {K.}~\bibnamefont {Breuer}},\ and\ \bibinfo
  {author} {\bibfnamefont {H.~C.}\ \bibnamefont {Berg}},\ }\bibfield  {title}
  {\bibinfo {title} {Moving fluid with bacterial carpets},\ }\href
  {https://doi.org/10.1016/S0006-3495(04)74253-8} {\bibfield  {journal}
  {\bibinfo  {journal} {Biophys. J.}\ }\textbf {\bibinfo {volume} {86}},\
  \bibinfo {pages} {1863} (\bibinfo {year} {2004})}\BibitemShut {NoStop}%
\bibitem [{\citenamefont {Kim}\ and\ \citenamefont {Breuer}(2008)}]{Kim2008}%
  \BibitemOpen
  \bibfield  {author} {\bibinfo {author} {\bibfnamefont {M.}~\bibnamefont
  {Kim}}\ and\ \bibinfo {author} {\bibfnamefont {K.}~\bibnamefont {Breuer}},\
  }\bibfield  {title} {\bibinfo {title} {Microfluidic pump powered by
  self-organizing bacteria},\ }\href {https://doi.org/10.1002/smll.200700641}
  {\bibfield  {journal} {\bibinfo  {journal} {Small}\ }\textbf {\bibinfo
  {volume} {4}},\ \bibinfo {pages} {111} (\bibinfo {year} {2008})}\BibitemShut
  {NoStop}%
\bibitem [{\citenamefont {Martindale}\ and\ \citenamefont
  {Fu}(2017)}]{Martindale2017}%
  \BibitemOpen
  \bibfield  {author} {\bibinfo {author} {\bibfnamefont {J.~D.}\ \bibnamefont
  {Martindale}}\ and\ \bibinfo {author} {\bibfnamefont {H.~C.}\ \bibnamefont
  {Fu}},\ }\bibfield  {title} {\bibinfo {title} {Autonomously responsive
  pumping by a bacterial flagellar forest: A mean-field approach},\ }\href
  {https://doi.org/10.1103/PhysRevE.96.033107} {\bibfield  {journal} {\bibinfo
  {journal} {Phys. Rev. E}\ }\textbf {\bibinfo {volume} {96}},\ \bibinfo
  {pages} {033107} (\bibinfo {year} {2017})}\BibitemShut {NoStop}%
\bibitem [{\citenamefont {Dauparas}\ \emph {et~al.}(2018)\citenamefont
  {Dauparas}, \citenamefont {Das},\ and\ \citenamefont {Lauga}}]{Dauparas2018}%
  \BibitemOpen
  \bibfield  {author} {\bibinfo {author} {\bibfnamefont {J.}~\bibnamefont
  {Dauparas}}, \bibinfo {author} {\bibfnamefont {D.}~\bibnamefont {Das}},\ and\
  \bibinfo {author} {\bibfnamefont {E.}~\bibnamefont {Lauga}},\ }\bibfield
  {title} {\bibinfo {title} {{Helical micropumps near surfaces}},\ }\href
  {https://doi.org/10.1063/1.5012070} {\bibfield  {journal} {\bibinfo
  {journal} {Biomicrofluidics}\ }\textbf {\bibinfo {volume} {12}},\ \bibinfo
  {pages} {014108} (\bibinfo {year} {2018})}\BibitemShut {NoStop}%
\bibitem [{\citenamefont {Buchmann}\ \emph {et~al.}(2018)\citenamefont
  {Buchmann}, \citenamefont {Fauci}, \citenamefont {Leiderman}, \citenamefont
  {Strawbridge},\ and\ \citenamefont {Zhao}}]{Buchmann2018}%
  \BibitemOpen
  \bibfield  {author} {\bibinfo {author} {\bibfnamefont {A.}~\bibnamefont
  {Buchmann}}, \bibinfo {author} {\bibfnamefont {L.~J.}\ \bibnamefont {Fauci}},
  \bibinfo {author} {\bibfnamefont {K.}~\bibnamefont {Leiderman}}, \bibinfo
  {author} {\bibfnamefont {E.}~\bibnamefont {Strawbridge}},\ and\ \bibinfo
  {author} {\bibfnamefont {L.}~\bibnamefont {Zhao}},\ }\bibfield  {title}
  {\bibinfo {title} {{Mixing and pumping by pairs of helices in a viscous
  fluid}},\ }\href {https://doi.org/10.1103/PhysRevE.97.023101} {\bibfield
  {journal} {\bibinfo  {journal} {Phys. Rev. E}\ }\textbf {\bibinfo {volume}
  {97}},\ \bibinfo {pages} {023101} (\bibinfo {year} {2018})}\BibitemShut
  {NoStop}%
\bibitem [{\citenamefont {Chwang}\ and\ \citenamefont {Wu}(1975)}]{Chwang1975}%
  \BibitemOpen
  \bibfield  {author} {\bibinfo {author} {\bibfnamefont {A.~T.}\ \bibnamefont
  {Chwang}}\ and\ \bibinfo {author} {\bibfnamefont {T.~Y.-T.}\ \bibnamefont
  {Wu}},\ }\bibfield  {title} {\bibinfo {title} {{Hydromechanics of
  low-Reynolds-number flow. Part 2. Singularity method for Stokes flows.}},\
  }\href {https://doi.org/10.1017/S0022112075000614} {\bibfield  {journal}
  {\bibinfo  {journal} {J. Fluid Mech.}\ }\textbf {\bibinfo {volume} {67}},\
  \bibinfo {pages} {787} (\bibinfo {year} {1975})}\BibitemShut {NoStop}%
\end{thebibliography}%


%

\end{document}